\documentclass[12pt]{article}
\usepackage{fullpage,amsmath,amssymb,mathtools,natbib,hyperref}
\usepackage{titling,enumitem}
\usepackage[skip=0pt]{caption} 

\usepackage{setspace}
\usepackage{datetime,bigints}
\usepackage[dvipsnames]{xcolor}
\usepackage[most]{tcolorbox}
\usdate
\usepackage{multirow,algorithm} 
\usepackage{float}
\usepackage{algorithm,bbm}
\usepackage[]{algpseudocode}
\usepackage{arydshln}
\usepackage[section]{placeins}
\usepackage[titletoc,toc,title]{appendix}
\usepackage{fancyvrb} 
\usepackage{listings}
\usepackage{bm}
\usepackage{xcolor,colortbl} 
\newcommand{\notes}[1]{%
    \linespread{0.2}\vspace{0.1em}%
    \captionsetup{justification=justified}%
    \caption*{\footnotesize #1}%
}

\definecolor{lightgray}{gray}{0.9}
\xdefinecolor{gray}{rgb}{0.8,0.8,0.8}
\xdefinecolor{blue}{RGB}{58,95,205}
\DeclareCaptionFormat{listing}{\rule{\dimexpr\textwidth+17pt\relax}{0.4pt}\vskip1pt#1#2#3}
\newcommand{\mc}[2]{\multicolumn{#1}{c}{#2}}
\definecolor{gray}{gray}{0.85}
\definecolor{LightCyan}{rgb}{0.88,1,1}

\newcolumntype{a}{>{\columncolor{gray}}c}
\newcolumntype{b}{>{\columncolor{white}}c}

\usepackage{xr,caption}

\renewenvironment{thebibliography}[1]{
  \begin{oldthebibliography}{#1}
    \setlength{\itemsep}{0em}
    \setlength{\parskip}{0em}
}
{
  \end{oldthebibliography}
}

\usepackage{thmtools} 
\usepackage{amsthm}
{
      \theoremstyle{plain}
      
      \newtheorem{theorem}{Theorem}
      \newtheorem{example}{Example}
      \newtheorem{proposition}{Proposition}
      \newtheorem{lemma}{Lemma}
      \newtheorem{corollary}{Corollary}
      \newtheorem{assumption}{Assumption}
      
  }

\renewcommand{\arraystretch}{1.5}

\makeatletter
\renewenvironment{abstract}{%
    \if@twocolumn
      \section*{\abstractname}%
    \else 
      \begin{center}%
        {\bfseries \normalsize\abstractname\vspace{\z@}}
      \end{center} \vspace{-0.5cm}%
      \quotation
    \fi}
    {\if@twocolumn\else\endquotation\fi}
\makeatother

\begin{document}

  \title{Occasionally Misspecified}
  \author{ Jean-Jacques Forneron\thanks{Department of Economics, Boston University, 270 Bay State Road, Boston, MA 02215 USA.\newline Email: \href{mailto:jjmf@bu.edu}{jjmf@bu.edu}, Website: \href{http://jjforneron.com}{http://jjforneron.com}.\\
  I would like to thank Richard Crump, Claire Labonne, participants at the CREST/PSE/Science Po and Wisconsin Econometrics seminars, and the BC/BU Greenline workshop for useful comments and suggestions.}} 
  \date{\today}
  \maketitle 

  \begin{abstract}  
    When fitting a particular Economic model on a sample of data, the model may turn out to be heavily misspecified for some observations. This can happen because of unmodelled idiosyncratic events, such as an abrupt but short-lived change in policy. These outliers can significantly alter estimates and inferences. A robust estimation is desirable to limit their influence. For skewed data, this induces another bias which can also invalidate the estimation and inferences. This paper proposes a robust GMM estimator with a simple bias correction that does not degrade robustness significantly. The paper provides finite-sample robustness bounds, and asymptotic uniform equivalence with an oracle that discards all outliers. Consistency and asymptotic normality ensue from that result. An application to the ``Price-Puzzle,'' which finds inflation increases when monetary policy tightens, illustrates the concerns and the method. The proposed estimator finds the intuitive result: tighter monetary policy leads to a decline in inflation.
  \end{abstract}
  

  \baselineskip=18.0pt
  \thispagestyle{empty}
  \setcounter{page}{0}
  
\newpage

\section{Introduction}
Empirical data is routinely used to fit and test Economic models or predictions. Although the model may explain much of the variation in the data, it may also turn out to be particularly misspecified for some observations. This can result from sudden, yet temporary, changes in policy. To illustrate: monetary policy is often measured via changes in interest rates. Between 1979 and 1982, the Federal Reserve no longer fixed the Federal Funds Rate as a policy tool, targetting monetary aggregates instead \citep[p3]{coibion2012}. Sharp changes in interest rates during that period generate significant identifying power on the effects of monetary policy. Yet, misspecification threatens the validity of the resulting estimates and inferences. Other factors that can cause occasional misspecification include imperfect data matching, or some rare - but significant - prediction errors when generating regressors. 

A robust estimator is desirable in these scenarios: being less sensitive to influential outliers. However, robust estimates can be biased and inconsistent when the underlying data is asymmetric. To illustrate: the sample median is more robust than the mean; however, it estimates a different quantity when the data is skewed. This is relevant as many economic variables -- income, prices, and quantity, to name a few -- tend to be skewed. When symmetric data is contaminated asymmetrically, both the mean and median are biased. Further, in a linear regression context, \citet{hamilton1992} stresses that robust M-estimators are  ``designed for protection against wild errors or y-outliers. x-outliers are its Achilles' heel.'' Leverage characterizes x-outliers, which is bounded for ordinary least-squares. In the example above: sharp changes in interest rates imply high leverage around 1979-1982. The issue is even more pronounced in non-linear regressions where leverage is not necessarily bounded \citep{st1992}. This superleverage can further exacerbate the influence of outliers. 

This paper proposes a robust Generalized Method of Moments (GMM) estimator with a simple bias-correction step. Building on \citet{ronchetti2001}, the sample moments are estimated robustly; here using a penalized student log-likelihood criterion. The particular choice of criterion makes the asymptotic asymmetry bias tractable. A linear combination, known as Richardson extrapolation, of two robust moment estimates is asymptotically unbiased. The bias, which depends on higher-order moments, is not estimated. The correction does not degrade robustness significantly.  Also, in linear regressions, robust GMM estimates are robust against x-outlier, unlike M-estimates which only screen for large residuals. Given these moment estimates, the model is estimated in the same fashion as a standard GMM.

Finite and large sample results describe the properties of the method against adversarial contamination. First, uniform finite-sample exponential bounds, for cross-sections and mixing time-series, measure how robust moment estimates deviate from their biased target. This provides a worst-case global robustness guarantee for a given level of data contamination. The combination of the student likelihood, which is neither convex nor bounded but has a bounded influence function, with the particular choice of penalty is key for this result.

The large-sample results require the number of outliers to increase more slowly than the sample size. Their influence can grow rapidly: non-robust estimates may be inconsistent, or diverge. This captures the finite-sample setting where a few observations overwhelm the estimation. The bias-corrected robust moment and parameter estimates are shown to be first-order equivalent to an oracle which discards all outliers. Asymptotic normality follows from standard regularity conditions on the oracle. For linear models, the robust GMM estimates can be expressed as weighted least-squares or weighted two-stage least-squares. The weights are easy to compute and report, highlighting which observations were downweighted in the process. This should reduce concerns about black-box results.

Simulations illustrate the small sample properties of the proposed estimator in the presence of x-outliers, which have high leverage. OLS is very sensitive. A robust M-estimator packaged in R is biased and sensitive. Without correction, the procedure is more robust but biased. Bias correction reduces estimation error and improves coverage of t-tests. As the proportion of outliers increases, its performance degrades but remains better than the benchmarks. Undersmoothing, sometimes suggested in the literature, is also less robust than bias correction. Three empirical applications illustrate the relevance of the procedure. 

The first estimates the effect of a monetary policy shock on inflation using a structural Vector Autoregressive (VAR) model as in \citet{stock2001}. OLS estimates a “Price-Puzzle:” predicting an inflation increase when monetary policy tightens. Two historical sub-periods of unusual monetary policy -- including 1979-1982 -- significantly influence this result. The proposed estimates find the intuitive result: a negative impact on inflation. The weights reveal that the two historical subperiods are downweighted to get this result. Robust estimates overweight some observations. Bias correction re-adjusts towards equal weighting.

Recently, \citet{young2022} found that many instrumental variable (IV) results involve highly leveraged regressions, and are very sensitive to outliers. Two applications illustrate the methodology in this setting. The first considers the relationship between trade openness and inflation \citep{romer1993}. The second is about the effect of segregation on the quality of government \citep{alesina2011}. Both regressions are highly influenced by a few observations. Robust estimates have significantly smaller standard errors, producing more precise inferences. Bias correction reveals non-negligible bias in robust estimates.




\paragraph{Structure of the paper.} Section \ref{sec:motivating} motivates the paper with the Price Puzzle example. Section \ref{sec:lit} surveys the existing literature. Section \ref{sec:modsamest} introduces the setting, sampling assumptions, and the estimator. Derivations for a simplified estimator give insights for the finite and large sample results. Section \ref{sec:theory} provides finite-sample bounds and asymptotic results. Simulated and empirical applications are in Section \ref{sec:apps}. Appendices \ref{apx:prelim}, \ref{apx:proofs} give the proofs for the main results and preliminary ones. Supplemental Appendices \ref{apx:proof_prelim}, \ref{apx:lev_outliers}, \ref{apx:lev_IV}, \ref{apx:extra_mc}, \ref{apx:extra_emp}, \ref{apx:algos} provide proofs for the preliminary results, simple derivations with leveraged outliers, derivations for influence and leverage in IV regressions, additional simulation and empirical results, and detailed numerical Algorithms to perform the estimation.

\newcommand{\vardbtilde}[1]{\tilde{\raisebox{0pt}[0.85\height]{$\tilde{#1}$}}}

\section{Motivating Example: the Price Puzzle} \label{sec:motivating}
To illustrate the issues considered in this paper, consider estimating the impact of monetary policy with a recursive vector autoregressive (VAR) model as in \citet{stock2001}. There are three variables: inflation ($\pi_t$), unemployment rate $(u_t)$, and the federal funds rate ($R_t$). The VAR is estimated by OLS with four lags on U.S. data from 1960Q1 to 2000Q4. 

\begin{figure}[ht] \caption{Recursive VAR: Impulse Response, Diagnostics} \label{fig:VAR1}
  \includegraphics[scale = 0.5]{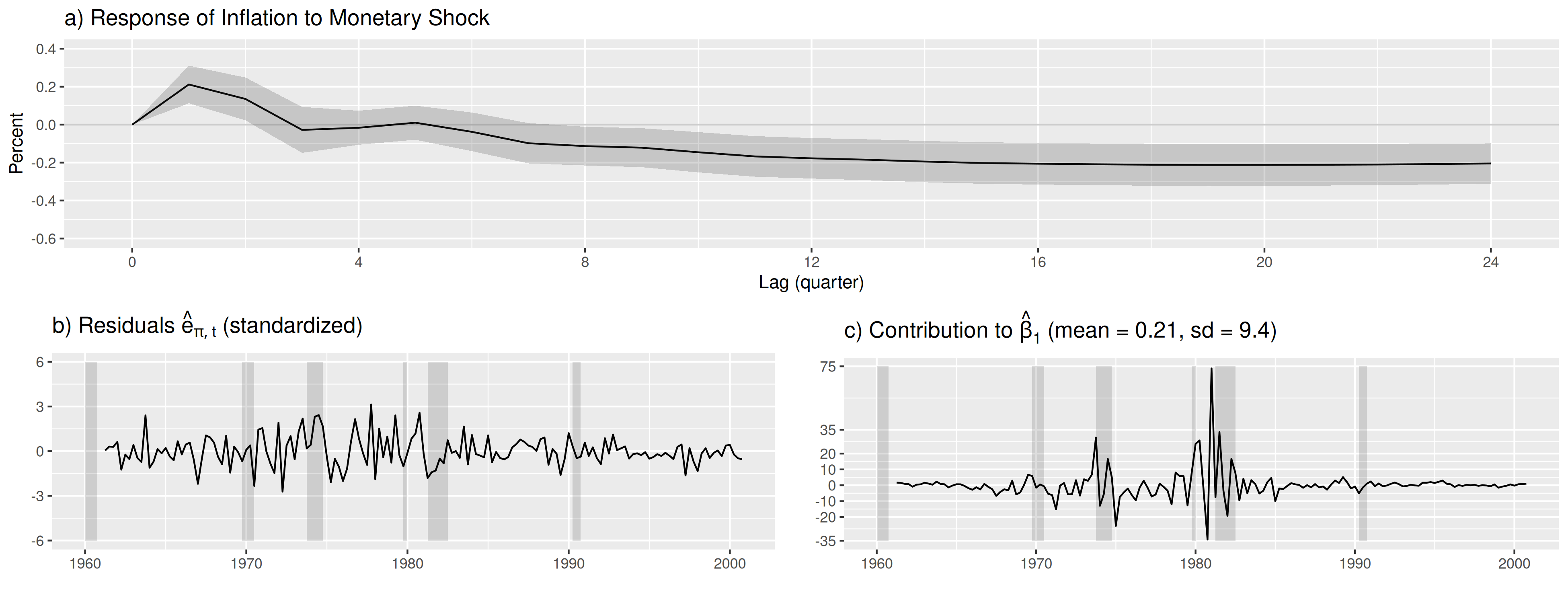}\\
  {\footnotesize \textbf{Note:} a) Estimated response of inflation $\pi$ to a unit increase in interest rate $R$, shaded = estimates $\pm$ one standard error, b) Standardized Residuals = $\hat{e}_{\pi,t}/\hat{\sigma}_{\hat{e}_\pi}$, c) Contribution of observation $t$ to $\hat{\beta}_n$ measured by $(X^\prime X/n)^{-1} x_t \pi_t$, $x_t$ is the vector of regressors. b,c) Shaded vertical bars = NBER recession dates.}
\end{figure}

Panel a) in Figure \ref{fig:VAR1} plots the estimated response of inflation to a  unit increase in $R_t$. It shows a positive and significant increase in inflation for nearly four consecutive quarters. This was first observed by \citet{sims1992} and immediately coined as a `Price Puzzle' by \citet{Eichenbaum1992}. It has since been studied extensively. \citet{rusnak2013} performed a meta-analysis of $1000$ estimates and put forward several potential forms of model misspecification to explain the puzzle. The number of specifications they explore is several times greater than the sample size so there should be some concerns about overfitting, however. 

The following presents some simple diagnostics that indicate two time periods strongly influence the estimates. The puzzle begins with a positive and significant initial impact. It is measured by $\beta_1$ in the regression:
\begin{align}
  \pi_t &= \beta_0 + \beta_1 R_{t-1} + \beta_2 u_{t-1} + \beta_3 \pi_{t-1} + \dots + \beta_{10} R_{t-4} + \beta_{11} u_{t-4} + \beta_{12} \pi_{t-4} + e_{\pi,t}. \label{eq:VAR1}
\end{align}
Figure \ref{fig:VAR1} investigates this regression more closely. 
Panel a) plots the residuals $\hat{e}_{\pi,t}$ over time. Besides some increased volatility between 1970-1982, there are no obvious outliers in the series. In fact, the skewness and kurtosis are $0.36$ and $3.78$, respectively, not far from a normal distribution. Panel c) approximates the contribution of each $t$ to $\hat{\beta}_1$. Since $\hat{\beta}_n = \sum_{t=1}^n (X^\prime X /n )^{-1} x_t y_t /n$ is a sample mean, $(X^\prime X /n )^{-1} x_t y_t$ approximates the contribution of each $t$ to the mean. Some observations stand out: for instance, 1981Q1 alone positively contributes $\approx 75/n = 0.47$ to $\hat{\beta}_1 = 0.21$, about $3.5$ standard errors.\footnote{Most coefficients in (\ref{eq:VAR1}) are strongly influenced by a few observations as shown in Table \ref{tab:VAR1}. The contribution reported here is related to Cook's distance which measures changes in predicted values $\hat{y}_t$ when observation $t$ is excluded in the estimation \citep{cook1977}. Here, the effect of observation $t$ on the estimated regression coefficients is the object of interest -- this will be referred to as contribution.}

Panels b,c) show that, although none of the residuals $\hat{e}_{\pi,t}$ are particularly large, two time periods, around 1974-1975 and 1979-1982, have a disproportionate influence on the results. The latter has historical significance: the Federal Reserve changed to non-borrowed reserves targeting where the interest rate $R_t$ was no longer a fixed policy instrument, as discussed in the introduction.
 Richmond FED President, Robert P. Black, summarized the tactical change during the October 1979 FOMC meeting as follows:

\begin{quote}
  ``I often think of our position as being analogous to that of a monopolist in the sense that we control the money supply. A monopolist has a choice of controlling either price or quantity but he can never control both. I believe we’ve been trying to control the quantity of money by setting the price and we have misjudged. We’ve jiggled the price, in terms of the federal funds rate, one way or the other, and we‘ve usually met with less than complete success in judging what quantity of money will be forthcoming from that.'' \citep[][p23]{FOMC100679}
\end{quote}

This has several implications for the VAR estimates. First, $R_t$ was no longer a direct measure of monetary policy: the recursive VAR may not correctly identify monetary shocks during that time period. Importantly, this goes beyond parameter instability. Time-varying parameters, regime-switching, or structural break models would still require $R_t$ to provide a measure of monetary policy shocks. As emphasized by Robert Black, monetary policy was conducted on monetary aggregates at that time, not interest rates.  
Second, interest rates were significantly more volatile with the policy change;\footnote{This was anticipated and monitored by board members as shown by FOMC Transcripts of 1979-1982.} producing significant regression leverage. This, as highlighted in Figure \ref{fig:VAR1}, gives excess influence to these observations. 

Misspecification arises because the central bank relies on multiple policy instruments, the VAR only uses $R_t$. \citet{friedman1963} argued that well-known historical events clearly identify large monetary shocks. This narrative approach was popularized by \citet{romer1989}, \citet{romer2004}. Narrative and VAR estimates can differ when the central bank relies on different instruments throughout the sample \citep{coibion2012,monnet2014}. Narrative estimates, however, aggregate multiple types of monetary policies; results cannot be interpreted as e.g. an interest rate shock.

To identify the effect of an interest rate shock, a robust estimation is desirable. However, as noted in the introduction robust M-estimates may be biased and may not be robust to these x-outliers. Because residuals are small, robust M-estimates with Huber loss and high-breakdown MM estimates (\textit{rlm}, \textit{lmRob} in R) are nearly identical to Figure \ref{fig:VAR1} (not reported). 

Diagnostics, as presented above, are useful to assess whether the estimation might present some irregularities. A robust estimation, presented below, is meant to reduce the influence of abnormal observations. The two are complementary, see \citet[Ch1.2.4]{huber2011} for further discussion.  

\begin{figure}[ht] \caption{Recursive VAR, IRF: OLS, Robust and Bias-Corrected Estimates} \label{fig:VAR2}
  \includegraphics[scale = 0.5]{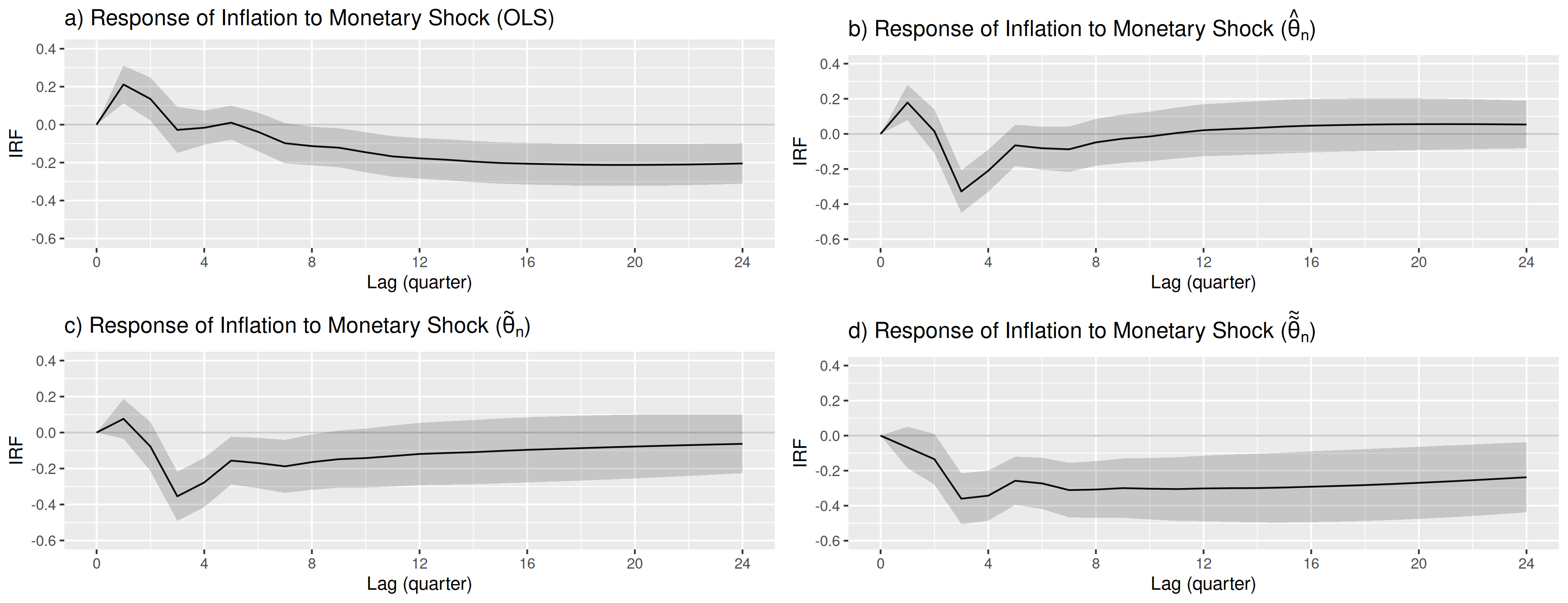}\\
  {\footnotesize \textbf{Note:} a) OLS estimates, b) $\hat{\theta}_{n}$ robust estimates without bias correction, c) $\tilde{\theta}_{n}$ robust estimates with bias correction, d) $\vardbtilde{\theta}_{n}$ robust estimates with repeated bias correction. b,c,d) Estimates computed with tuning parameter $\hat{\nu}_n = 8.99$. Results for other $\nu$ in Appendix \ref{apx:extra_emp}. Bands: estimates $\pm$ one standard error.}
\end{figure}

Figure \ref{fig:VAR2} re-estimates the effect on the same data, with the same model specification: using OLS (panel a), the proposed robust estimator without bias correction (panel b), with bias correction (panel c), with bias correction and a small sample correction (panel d). Without bias correction, the price puzzle remains -- but does not last 4 quarters anymore. With bias correction, the price puzzle disappears; the initial effect is not significant. With the additional adjustment, the effect is qualitatively larger and negative.  

\begin{figure}[ht] \caption{Recursive VAR, Estimation Weights: OLS, Robust, and Bias-Corrected Estimates } \label{fig:VARw}
  \includegraphics[scale = 0.5]{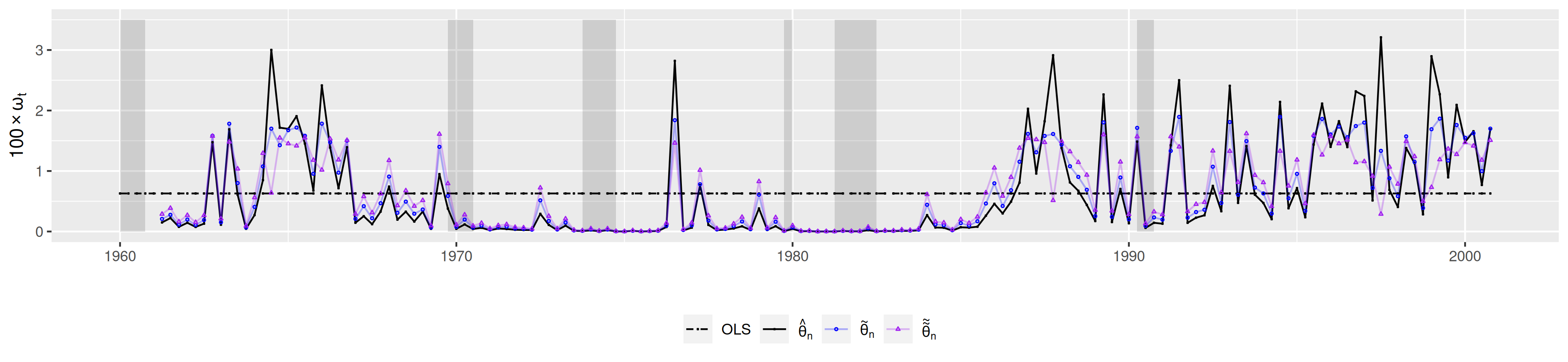}\\
  \includegraphics[scale = 0.5]{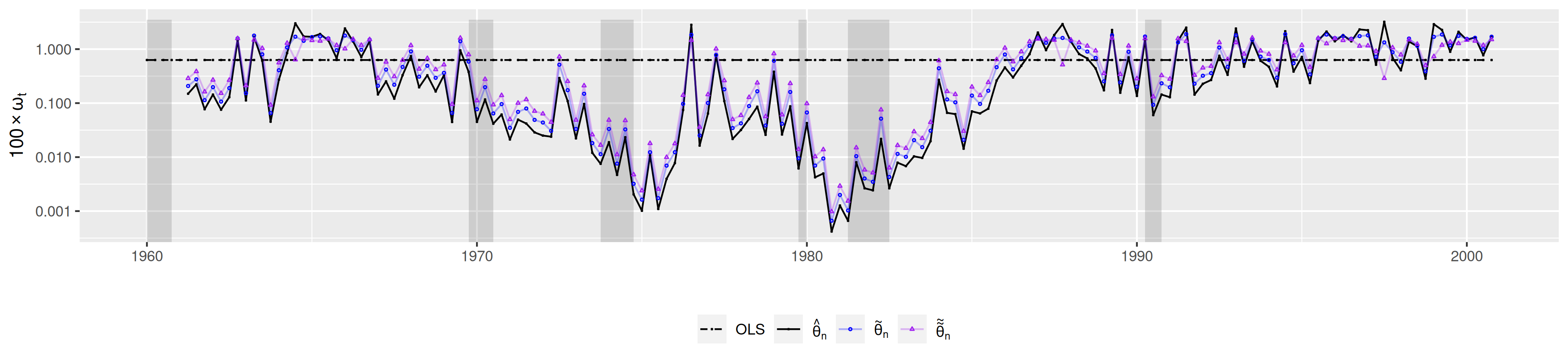}\\
  {\footnotesize \textbf{Note:} Top and bottom panels: levels and log scale, respectively. Estimation weights $\omega_t$, implicitly used to estimate $\theta$. OLS (dashed/black): $\omega_t = 1/n$. Robust estimates $\hat{\theta}_n$ (solid/black). Bias-corrected robust estimates $\tilde{\theta}_n$ (solid/circle/blue).  Repeated bias-corrected robust estimates $\vardbtilde{\theta}_n$ (solid/triangle/purple). Shaded vertical bars = NBER recession dates.}
\end{figure}

As discussed above, the estimates can be seen as weighted least-squares. Figure \ref{fig:VARw} compares the weights, for each time period, used by each method on a regular and a log-scale (resp. top, bottom). OLS uses equal weighting (black/dashed). Without bias correction, robust estimates downweigh the leveraged outliers, especially 1979-1982, but overweigh other periods (black/solid). Bias correction re-adjusts towards equal weighting (blue/dot). The small sample adjustment further re-adjusts in that direction (purple/triangle).

\section{Related Literature} \label{sec:lit} The paper is mainly related to the literature on robust estimation, mostly developed in statistics. Textbook references such as \citet{huber2011} and \citet{maronna2019} survey a wide range of estimators and their properties. To focus the discussion, consider a linear regression: $y_t = x_t^\prime \theta + e_t$. Robust M-estimators minimize the loss $\sum_{t=1}^n\psi( y_t - x_t^\prime \theta )$ over $\theta$. While OLS uses a quadratic $\psi$, least-absolute deviation (LAD), and the \citet{Huber1964} loss are non-quadratic. They increase linearly with large residuals $|y_t - x_t^\prime \theta|$. This reduces the influence of y-outliers. Winsorizing and trimming are popular alternatives. \citet[p80]{Huber1964} notes that trimming can be sensitive around the cutoffs. The first-order condition implies the solution $\hat{\theta}_n$ satifies $\sum_{t=1}^n x_t \psi^\prime(y_t - x_t^\prime \hat{\theta}_n)$. Large residuals $\hat{e}_t = y_t - x_t^\prime \hat{\theta}_n$ are handled by $\psi^\prime$. However, x-outliers with a large $x_t$, are not screened by $\psi^\prime$.\footnote{Mallows type estimators separately screen for leverage, see e.g. \citet{carroll1988}.} When the distribution of $e_t$ is symmetric and the sample is contaminated symmetrically, robust estimates are consistent and asymptotically normal under regularity conditions. Symmetry is critical. \citet{jaeckel1971} derived, for estimating a location parameter, with asymmetric contamination of symmetric data, an asymptotic bias of order $n^{-1/2}$ when the proportion of outliers is $O(n^{-1/2})$ -- i.e. $n_o = O(n^{1/2})$. $n_o$ is the number of outliers in the sample of size $n$. Recently, \citet{dalalyan2022} proposed an attractive robust location estimator for multivariate Gaussian or sub-Gaussian data with a high-breakdown point - i.e. robust to a large fraction of outliers in the sample. Here, the finite-sample results are derived using only finite second moment conditions. Also, the focus here is on settings where data is asymmetric, contaminated by a small number of highly influential outliers.

For asymmetric data, the estimator may not be consistent, see \citet{carroll1988} for linear regressions. Quasi-Maximum Likelihood estimation, with a student distribution for the errors, is commonly used to estimate volatility models. \citet{newey1997} show that the estimates may not be consistent without symmetry conditions. In a parametric setup, \citet{cantoni2001} provide analytical bias formulas for generalized linear models, used to correct the first-order condition of the M-estimation. 
Here, parametric assumptions are not required. \citet{zhou2018} derive bias bounds and exponential inequalities for linear regressions with the Huber loss when $e_t$ has finite variance. They do not consider sample contamination and require sub-gaussian regressors - i.e. no x-outliers. These two issues are particularly relevant for the Price Puzzle.  Another approach to robustness is to bound the asymptotic bias in a local neighborhood of the model using the influence curve (IC) of \citet{hampel1974}, see e.g. \citet[Ch4.9]{huber2011}. \citet{andrews1986} relates the IC to the stability of estimators. Recently, several papers have used the IC to study and bound local misspecification bias for GMM, e.g. \citet{andrews2017}, \citet{armstrong2021}, \citet{bonhomme2022}. Under these local asymptotics, the estimator remains consistent and asymptotically normal with a bias proportional to sampling uncertainty. In this paper, the model is grossly misspecified, but only for $1 \leq n_o \ll n$ outliers. Non-robust estimates can be inconsistent, or diverge: a robust estimation is required. \citet{christensen2023} propose global sensitivity analyses on distributional assumptions, the model is otherwise correctly specified. It is common in Economics to apply more robust testing to non-robust estimates, assuming consistency, asymptotic normality -- unlike here.  One can adjust standard errors \citep[e.g.][]{mackinnon2012}, critical values \citep[e.g.][]{muller2020,potscher2023}, or both. \citet{sasaki2023} propose a test for finite moments at a point, as required for consistency and central limit theory. \citet{cowell1996} and \citet{cowell2007} consider the robustness properties of inequality measures, e.g. Gini coefficient. Surveying a large number of empirical results, \citet{young2022} finds that many IV regressions are highly leveraged and sensitive to a few observations, or clusters of observations.

For GMM estimation, \citet{ronchetti2001} proposed a robust estimator that is locally asymptotically robust, using the IC criteria. \citet{hill2010}, \citet{vcivzek2016} consider trimming in GMM estimation. \citet{rohatgi2022} use a \textsc{filter} algorithm to screen out outliers in GMM estimation. The median-of-means is popular in prediction problems, which could also be considered here: the dataset is split into $K \geq 2$ subsamples of $m = n/K$ observations. $K$ sample means are computed. The median of the $K$ means is the estimator. The estimate is robust for up to $n_o \leq K/2-1$ outliers, see e.g. \citet{lecue2020}, \citet{laforgue2021}. To accommodate an increasing $n_o$, having $K \to \infty$ as $n \to \infty$ is necessary. This introduces a bias, bounded above by $\sigma/\sqrt{m} = \sigma \sqrt{K/n}$.\footnote{For any distribution, the median and the mean differ by at most: $|\text{median}(X) - \mathbb{E}(X)| \leq \sigma(X)$.} Even for $K$ fixed, an asymptotic bias can arise. Without a tractable expression for the bias, it is not clear how one would correct the asymptotic bias. Here, the choice of loss function makes the asymptotic bias tractable. An alternative is undersmoothing where the tuning parameter diverges fast enough that the bias is asymptotically negligible. It only requires to bound the asymptotic bias. Section \ref{sec:mc} illustrates that it is less robust than bias-correction.  


\section{Models, Sample, Estimator} \label{sec:modsamest}
This paper considers estimations from unconditional moment restrictions:
\begin{align} \mathbb{E}_P \left[g\left(z_t;\theta\right) \right] = 0 \Leftrightarrow \theta = \theta_0, \label{eq:moments} \end{align}
where $z_t \overset{d}{\sim} P$ and the solution $\theta_0 \in \Theta$, a compact subset of $\mathbb{R}^k$. OLS regressions correspond to $g(z_t;\theta) =  x_t(y_t - x_t^\prime \theta)$ where $z_t = (y_t,x_t)$ collects the dependent variable and the regressors. For instrumental variable regressions, take $g(z_t;\theta) =  w_t(y_t - x_t^\prime \theta)$ where $z_t = (y_t,x_t,w_t)$ collects the dependent variable, the regressors and the instruments. Non-linear estimations also fit into this framework. Concave Likelihood maximization, such as Probit or Logit, would set (\ref{eq:moments}) to be the first-order condition. The main examples are linear.

The dataset consists of $n$ observations but $z_t \sim P$ may not hold for all $t = 1,\dots,n$. This is presented in the following Assumption.
\begin{assumption}[Sample] \label{as:sample}
There are $n = n_P + n_o$ observations such that
\begin{itemize}
  \item[i)]  for $t \in \{1,\dots,n_P\}$, $z_t \sim P$ for which (\ref{eq:moments}) holds, are either iid or strictly stationary, $\beta$-mixing with rate $\beta_{m} \leq a \exp(- b m)$ for $0 < a,b < \infty$; 
  \item[ii)] for $t \in \{n_P+1,\dots,n \}$ and $0 < A,\alpha < \infty$:
  \begin{align} z_t \in \mathcal{O}_n := \{ z \text{ s.t. } \sup_{\theta \in \Theta} \|g(z;\theta)\|^2 \leq  A n^{\alpha} \}. \label{eq:On} \end{align}
\end{itemize}
\end{assumption}
The first $n_P$ observations are such that (\ref{eq:moments}) holds. However, the last $n_o$ observations, or outliers, can be arbitrary in $\mathcal{O}_n$. The ordering between observations simplifies notation and, for time-series, preserves the dependence structure of the good $n_P$ observations. The mixing condition typically holds for stationary VAR models, as in the motivating example. In practice, the user does not know which observations are drawn from $P$ and those that are not. The $n_o$ outliers could be allocated anywhere within the sample. The outliers will be chosen in an adversarial fashion, looking at the least-favorable collection $(z_{n_{P}+1},\dots,z_n) \in \mathcal{O}_n$ for each $\theta$, without restrictions on dependence.

The goal here is to derive finite-sample robustness properties against the \textit{worst-case realization} of the $n_o$ outliers. Ex-ante, if the $n_o$ outliers are randomly distributed, such that $\mathbb{P}(\sup_{\theta \in \Theta}\|g(z_t;\theta)\| > t) \leq t^{-\varepsilon}$ for some $\varepsilon > 0$. Then $\mathbb{P}( z_t \in \mathcal{O}_n \text{ for each }t=n_{P}+1,\dots,n  ) \geq 1-A^{-\varepsilon} n_o n^{-\alpha \varepsilon}$, can be made arbitrarily close to $1$ setting $\alpha$ large enough. In practice, the user does not specify $(A,\alpha)$. For random data contamination, Assumption \ref{as:sample} can be interpreted as conditioning on a realization with $n_o$ outliers in the set $\mathcal{O}_n$ which has arbitrarily high-probability given an appropriate choice of $n_o$, $A$ and $\alpha$.\footnote{See also Remark 1 in \citet{laforgue2021}.} 

Outliers can take many forms in (\ref{eq:On}). Figure \ref{fig:VAR1} illustrates that residuals $y_t - x_t^\prime \theta$ are not the only source of influence, captured here by $x_t(y_t - x_t^\prime \theta)$. High leverage observations are only influential if $|y_t - x_t^\prime\theta| \gg 0$. Likewise, $x_t(y_t - x_t^\prime \theta)$ can be large when neither $y_t - x_t^\prime \theta$ nor $x_t$ are individually large but their product is non-negligible. This implies that screening residuals and regressors separately, as suggested in \citet{hamilton1992}, can be insufficient. The influence of a single observation can also vary depending on the model specification: a regression that is linear in $x_t$ is typically less leveraged than in a quadratic specification with $(x_t,x_t^2)$ as regressors. Collinearity also plays a role on influence, as $(X^\prime X/n)^{-1} x_t y_t$, reported in Figure \ref{fig:VAR1}, can be greatly inflated by the collinearity factor $(X^\prime X/n)^{-1}$. In the motivating example, the regressors are lagged variables which are autocorrelated, i.e. collinear. A rotation invariance property is important to ensure robustness when there are multiple regressors. For instrumental variable regressions, the relevant quantity $w_t(y_t - x_t^\prime \theta)$ involves the instruments $w_t$ and the residual. In the context of time-series, one concern would be innovation outliers associated with a large shock $y_t - x_t^\prime \theta$. Another, similar to the description in the motivating example, would be additive outliers. Here the effect is isolated, as in a different regime that occurs only once within the sample.


The main concern here is that the sample mean $\overline{g}_n(\theta) = 1/n \sum_{t=1}^n g(z_t;\theta)$ is not a consistent estimator for $\mathbb{E}_P[g(z_t;\theta)]$ when $(n_o n^{\alpha})/n  \not \to 0$. This allows to capture the concern that a minority of observations has significant influence, even as the sample size $n$ increases. For $n_o =1$, $\alpha = 1/2$ the estimates are consistent but asymptotically biased, standard error estimates are also affected.\footnote{This is illustrated in Appendix \ref{apx:lev_outliers}.} For $n_o = 1$, $\alpha = 1$ estimates are inconsistent. They diverge when $\alpha >1$. Mild outliers are also problematic: for $n_o = n^{1/4}$ and $\alpha = 1/4$ estimates are asymptotically biased. 

To handle contaminated samples, \citet{ronchetti2001} showed that a robust estimate of $\mathbb{E}_P[g(z_t;\theta)]$ is required. The following first computes a robust estimate of $\mu(\theta) = \mathbb{E}_P[g(z_t;\theta)]$, then corrects the first-order asymptotic bias, and finally solves for $\mu(\theta)=0$.

\paragraph{Step 1.} For each $\theta \in \Theta$, find $\hat{\psi}_n(\theta;\nu)$ which minimizes the sample criterion:
\begin{align}
  Q_n(\psi;\theta) = \frac{\nu + p}{n}\sum_{t=1}^n \log\left(1+ \frac{\|g(z_t;\theta)-\mu\|^2_{\Sigma^{-1}}}{\nu}\right) + \log|\Sigma| +  \frac{\kappa_1}{\nu}\|\mu\|^2_{\Sigma^{-1}} + \frac{\kappa_2}{\nu} \text{trace}(\Sigma), \label{eq:tloss}
\end{align}
where $\psi = (\mu,\Sigma)$ and $p = \text{dim}(g(z_t;\theta))$. The location and scale parameters are estimated jointly to ensure the first is invariant to rotation and less sensitive to re-scaling. The loss $Q_n$ consists of a student quasi-likelihood plus two penalization terms. The tuning parameter $\nu > 0$ controls the robustness of the estimates. Here, it is not estimated and acts as a critical value. For observations such that $\|g(z_t;\theta)-\mu\|^2_{\Sigma} \ll \nu$, the loss is approximately quadratic and approximates the Gaussian log-likelihood. In contrast, for observations such that $\|g(z_t;\theta)-\mu\|^2_{\Sigma^{-1}} \gg \nu$ the loss is approximately logarithmic. Large values for $g(z_t;\theta)$ have a lesser impact compared to the Gaussian likelihood. 

To fully capture robustness, the parameter space $\Psi$ for $\psi$ is unbounded:
\[ \Psi = \{ (\mu,\Sigma), \, \mu \in \mathbb{R}^p, 0 < s_0 \leq \lambda_{\min}(\Sigma) \leq \lambda_{\max}(\Sigma) \leq +\infty \},\]
where $s_0$ is such that $s_0 \leq \lambda_{\min}(\text{var}_P[g(z_t;\theta)]) < +\infty$ for all $\theta \in \Theta$. In the presence of outliers, the main concern is in estimating a large $\hat{\mu}_n$ and/or $\hat{\Sigma}_n$. Here, setting $s_0 >0$ simplifies some derivations to focus on finite-sample upper bounds.

The robustness of the student log-likelihood has some downsides numerically. 
Without regularization ($\kappa_1=\kappa_2 = 0$), the derivative $\partial_{\mu} Q_n(\psi;\theta) = 0$ for $\|\mu\| = +\infty$ and any $\Sigma$. The student likelihood becomes flat for larger values of $\|\mu\|$. With non-zero penalties, i.e. $\kappa_1$ and $\kappa_2 \neq 0$, $\partial_{\mu} Q_n(\psi;\theta) \to \infty$ when $\|\mu\| \to \infty$. The combination of the student log-likelihood, which has bounded influence, with this choice of penalty implies the estimates $\hat\psi_n(\theta;\nu)$ are bounded, as shown in the next Section. The self-normalization $\|\mu\|_{\Sigma^{-1}}$ is invariant to rotations of the moments and less sensitive to scale. At the solution $\theta = \theta_0$, $\mu(\theta_0)=\mathbb{E}_P[g(z_t;\theta_0)] = 0$ holds. This motivates penalizing towards zero in this particular setting. 

Simultaneously estimating the location and scale parameters can seem problematic. A large $\hat{\Sigma}_n$ is effectively similar to using a large $\nu$, leading to less robust location estimates $\hat{\mu}_n$. The second penalty $\text{trace}(\Sigma)$ is important in that regard, as it ensures $\hat{\Sigma}_n$ cannot be too large in finite samples. This is shown in the next Section.

\paragraph{Step 2.} For each $\theta \in \Theta$, compute: 
\begin{align}\tilde{\mu}_n(\theta;\nu) = 2 \hat{\mu}_n(\theta;\nu) - \hat{\mu}_n(\theta;\nu/2). \label{eq:BC}\end{align} 
This type of adjustment is known as Richardson extrapolation in numerical analysis.
Unlike the sample mean, the estimator $\hat{\mu}_n(\theta;\nu)$ is typically biased for $\nu < +\infty$. Taking $\nu \to \infty$ with $n \to\infty$ at an appropriate rate, the adjustment $2 \hat{\mu}_n(\theta;\nu) - \hat{\mu}_n(\theta;\nu/2)$ corrects the first-order asymptotic bias. The bias depends on higher-order moments (see below). Estimating this bias is not straightforward: robustly estimating the first moment is already a challenge in this setting. The correction (\ref{eq:BC}) is simple to implement and widely applicable.

\paragraph{Step 3.} Find $\tilde\theta_{n}$ such that:
\begin{align} \|\tilde{\mu}_n(\tilde\theta_{n};\nu)\|^2_{W_n} \leq  \inf_{\theta \in \Theta}\|\tilde{\mu}_n(\theta;\nu)\|^2_{W_n} + o_p(n^{-1}). \end{align}
The estimated $\tilde\theta_{n}$ inherits the asymptotic bias properties of the bias corrected moments $\tilde{\mu}_n$. 

Step 1. continuously updates both $\mu$ and $\Sigma$ with $\theta$. The scaling $\hat{\Sigma}_n(\theta;\nu)$ used to normalize the estimation of $\hat{\mu}_n(\theta;\nu)$ adapts to the value of $\theta$. Appendix \ref{apx:algos} gives generic Algorithms \ref{algo:minQn}, \ref{algo:minGMM} used to compute $\hat{\psi}_n$, $\tilde{\theta}_n$ in the applications. $\tilde{\mu}_n(\theta;\nu)$ is as smooth as $g(z_t;\theta)$ -- cf. implicit function Theorem. Gradient-based optimizers, e.g. gradient-descent or Gauss-Newton, can be used. They are globally convergent under rank conditions \citep[][Th1,2]{forneron2023}. Unlike trimmed moments, the estimated $\tilde{\mu}_n(\theta;\nu)$ varies continuously with $\nu$. This implies that the estimates $\tilde{\theta}_n$ can be less sensitive to small changes in tuning parameters. Figures \ref{fig:VAR2b}-\ref{fig:VAR2bbb} reproduce Figure \ref{fig:VAR2} with larger values of $\nu$, illustrating that the estimated impulse response function changes continuously with $\nu$.

Numerical software typically proceeds iteratively, see e.g. \citet[Ch7.8]{huber2011}. Fix a tuning parameter and fit an initial regression $\hat{\theta}_n^1$. Then, update the scale parameter - here $\hat{\Sigma}_n^1$, re-estimate the regression $\hat{\theta}_n^2$, re-estimate the scale parameter, and repeat until convergence. The same scaling is applied for all $\theta$ at each stage. For least-squares, \textit{rreg} in Stata and \textit{rlm} in R proceed this way. Stata's \textit{rreg} is initialized with a non-robust OLS estimate. The properties of the estimates after many iterations are not easy to derive, especially as scale estimates are less robust than those of location.
Here, uniform-in-$\theta$ non-asymptotic concentration inequalities for the joint parameter $\hat\psi_n(\theta;\nu)$ are derived. This gives some finite-sample guarantees for step 1. above.  


\paragraph{Intuition for the results.} To better understand the role of the tuning parameter $\nu$ and the bias-correction step, consider estimating a scalar parameter $\theta_0 = \mathbb{E}_P(z_t)$ using:
\[ \hat{\mu}_n(\nu) = \frac{1}{n} \sum_{t=1}^n \frac{ z_t }{ 1+|z_t|^2/\nu }, \]
which simplifies the first-order condition of $Q_n$ with respect to $\mu$.\footnote{The first-order condition $\partial_\mu Q_n = 0$ reads $\frac{\nu + p}{\nu n} \sum_{t=1}^n \frac{ z_t -\mu }{1+\|z_t - \mu\|^2_{\Sigma^{-1}}/\nu} + \frac{\kappa_1 \mu}{\nu} = 0$.} For any $z$, $\frac{|z|}{1+|z|^2/\nu} \leq \frac{\sqrt{\nu}}{2}$ bounds the influence of a single observation. Let $\mu(\nu) = \mathbb{E}_P \left( z_t/(1+|z_t|^2/\nu) \right)$. If $z_t$ are iid for $t \in \{1,\dots,n_P\}$, regardless of the remaining $n_o$ observations:\footnote{This inequality implies $\mathbb{P}( \sup_{z_t \in \mathcal{O}_n,t>n_P}|\hat{\mu}_n(\nu)-\mu(\nu)| \geq \frac{\sqrt{\nu} n_o}{n} + C\frac{n_P}{n}[\sqrt{\frac{x}{n_P}} + \frac{x}{n_P}] ) \leq 2 \exp ( - x )$ for some constant $C$. This is the form used in a later Theorem. }
\[ \mathbb{P}\left( \sup_{z_t \in \mathcal{O}_n,t>n_P}|\hat{\mu}_n(\nu)-\mu(\nu)| \geq \frac{\sqrt{\nu} n_o}{n} + \frac{n_P}{n}\frac{x}{\sqrt{n_P}} \right) \leq 2 \exp \left( - \frac{x^2}{2\sigma^2_\nu + \frac{2}{3}\sqrt{\frac{\nu}{n_P}}x} \right), \]
using Bernstein's inequality, with $\sigma^2_\nu = \text{var}_P\left(\frac{z_t}{1+|z_t|^2/\nu}\right) \to \text{var}_P(z_t)$ as $\nu \to \infty$. The right-hand-side is approximately sub-Gaussian for $x \ll \sqrt{n_P/\nu}$ and sub-exponential for $x \gg \sqrt{n_P/\nu}$. The factor $\sqrt{\nu/n_P}$ indicates the rate at which the estimator becomes sub-Gaussian.

As expected, outliers introduce a bias. The worst-case bias is at most $\sqrt{\nu} n_o/n$. Consistency of $\hat{\mu}_n$ requires $(\sqrt{\nu}/n)n_o = o(1)$ and asymptotic normality $(\sqrt{\nu/n})n_o = o(1)$. More contamination $n_o$ requires a smaller $\nu$ to compensate. The same $\nu$ introduces another bias:
\[ \mu(\nu) = \theta_0 - \frac{1}{\nu}\mathbb{E}_P\left( \frac{z_t^3}{1+z_t^2/\nu} \right), \]
as measured by the last term. It is typically non-zero when the distribution is not symmetric around $0$. The bias is at most $\mathbb{E}_P(|z_t|^3)/\nu$ or $\mathbb{E}_P(|z_t|^2)/(2 \sqrt{\nu})$ if, respectively, the third or second moment is finite. Consistency requires $\nu \to \infty$ and asymptotic normality $\sqrt{n}/\nu = o(1)$. There is some tradeoff between the outlier bias $\sqrt{\nu}n_o/n$, which mandates a smaller $\nu$, and this robustness bias, which compels using a larger $\nu$. A bias reduction that does not significantly degrade robustness can be achieved using $\tilde{\mu}_n = 2 \hat{\mu}_n(\nu) - \hat{\mu}_n(\nu/2)$, since:
\[ \tilde{\mu}(\nu) = 2\mu(\nu) - \mu(\nu/2) = \theta_0 - \frac{1}{\nu^2}\mathbb{E}_P\left( \frac{z_t^5}{(1+z_t^2/\nu)(1+2z_t^2/\nu)} \right). \]
Now the bias is at most $\mathbb{E}_P(|z_t|^5)/\nu^2$ or $\mathbb{E}_P(|z_t|^4)/\nu^{3/2}$ if, respectively, the fifth or fourth moment is finite. For the former, asymptotic normality only requires $\sqrt{n}/\nu^2 = o(1)$. The effect of a single observation on the estimate $\tilde{\mu}_n$ is no more than $\sqrt{2 \nu} + \sqrt{\nu}/2$, compared to $\sqrt{\nu/2}$ for the non-corrected $\hat{\mu}_n$. The bias correction does require more regularity from the uncontaminated data in terms of moments - 5 instead of 3 finite ones. 

Higher-order Richardson extrapolation could further reduce the order of the asymptotic bias. Simulations suggest the following can give better results in small samples.
Applying the correction once more using $\vardbtilde{\mu}_n = 2\tilde{\mu}_n(\nu) - \tilde{\mu}_n(\nu/2)$ flips the sign of the asymptotic bias and can have some small sample effects:
\[ \vardbtilde{\mu}(\nu) = \theta_0 + \frac{2}{\nu^2} \mathbb{E}_P \left( \frac{z_t^5(1 - 4z_t^4/\nu^2)}{(1+z_t^2/\nu)(1+2z_t^2/\nu)(1+2z_t^2/\nu)(1+4z_t^2/\nu)} \right). \] 
To illustrate, take $z_t = \theta_0$ constant. Then $\tilde{\mu}(\nu) = \theta_0$ if, and only if, $\theta_0 = 0$ whereas $\vardbtilde{\mu}(\nu) = \theta_0$ if $\theta_0 \in \{\theta_0,-\sqrt{\nu/2},\sqrt{\nu/2}\}$. For finite $\nu$, the bias of $\vardbtilde{\mu}$ has two additional roots. Simulations in Section \ref{sec:mc} indicate small-sample improvements for estimation and inference.\footnote{Note that averaging $2/3 \tilde{\mu}(\nu) + 1/3  \vardbtilde{\mu}(\nu) = \theta_0 + o(\nu^{-2})$ can reduce the asymptotic bias, by the dominated convergence Theorem. This is not pursued here.}

\section{Properties of the Estimator} \label{sec:theory}

\subsection{Finite Sample Bounds} \label{sec:finite_sample}

The following Lemma shows the importance of the penalization $\kappa_1,\kappa_2$  in (\ref{eq:tloss}) which effectively bounds the parameter space $\Psi$.

\begin{lemma} \label{lem:bounds}
  For any $\theta \in \Theta$ and $\nu >0$, the minimizer $\hat\psi_n = (\hat{\mu}_n,\hat{\Sigma}_n)$ of (\ref{eq:tloss}) over $\Psi$ satisfies:
  \begin{align}
    \|\hat{\Sigma}_n^{-1/2}\hat{\mu}_n\| \leq \frac{\nu^{3/2}(1+p/\nu)}{2\kappa_1}, \quad 
    \text{trace}(\hat{\Sigma}_n) \leq \frac{\nu^2 (1+p/\nu)}{\kappa_2} + \frac{\nu^4(1+p/\nu)^2}{4 \kappa_1\kappa_2} + \frac{p \nu}{\kappa_2}. \label{eq:bounds}
  \end{align}
\end{lemma}
The dependence of $\hat\psi_n$ on $\theta,\nu$ is omitted to simplify notation.
Lemma \ref{lem:bounds} implies $\|\hat{\mu}_n\| \leq \nu^{7/2}$ and $\hat{\Sigma}_n \leq \nu^4$, up to constants. Although $\Psi$ is unbounded, the estimates are bounded with probability $1$. In the following, $\Psi$ will be replaced with:
\[ \Psi_n = \{ (\mu,\Sigma) \in \Psi \text{ s.t. } (\ref{eq:bounds}) \text{ holds} \},\]
without loss of generality. The upper bounds increase rapidly. With Lemma \ref{lem:Lip}, they imply an envelope function of size $\nu^{17}$ which diverges too quickly to directly apply standard empirical process results, e.g. \citet[Th2.14.1]{van1996}. Instead, the results directly rely on the functional form of (\ref{eq:tloss}) and the following assumption to derive exponential inequalities under cross-sectional and time-series dependence (Lemma \ref{lem:Concentration1}).

\begin{assumption} \label{as:moments} $z_t \sim P$, a distribution such that for two $0 \leq M_2,M_4 < \infty$:\\ i. $\sup_{\theta \in \Theta}\mathbb{E}_{P}(\|g(z_t;\theta)\|^2) \leq M_2$, ii. for all $(\theta_1,\theta_2) \in \Theta$, $\|g(z_t;\theta_1) - g(z_t;\theta_2)\| \leq G_t \|\theta_1-\theta_2\|$ with $\mathbb{E}_P(\|G_t\|^2) \leq M_2$, iii. $\sup_{\theta \in \Theta}\mathbb{E}_{P}(\|g(z_t;\theta)\|^4) \leq M_4$. In ii. $G_t = G(z_t)$ is either iid or strictly stationary and mixing with rate $\beta_m$ found in Assumption \ref{as:sample} i. 
\end{assumption}

Let $Q_\nu = \mathbb{E}_P(Q_{n})$ be the population analog of $Q_n$ without any contamination: \[Q_\nu(\psi;\theta) = \mathbb{E}_P\left[(\nu+p)\log\left(1+\|g(z_t;\theta)-\mu\|^2_{\Sigma^{-1}}/\nu\right)\right] + \log|\Sigma| +  \frac{\kappa_1}{\nu}\|\mu\|^2_{\Sigma^{-1}} + \frac{\kappa_2}{\nu} \text{trace}(\Sigma).\] 

\begin{proposition} \label{prop:unif} Take $x \geq 0$ and $1 \leq \nu \leq n$, suppose Assumptions \ref{as:sample} and \ref{as:moments} i-ii hold with $z_t$ iid for $t \in \{1,\dots,n_P\}$. For each $\theta \in \Theta$, let $\hat{\psi}_n(\theta;\nu)$ be the minimizer of (\ref{eq:tloss}) and $\psi(\theta;\nu)$ the minimizer of $Q_\nu$ on $\Psi$. Set $C_n = 1 + (k + 2p^2 )[\log(p) + \log(\nu) + \log(n_P)],$ with $p = \text{dim}(g)$ and $k = \text{dim}(\theta)$ then:
  \begin{align*}
    \mathbb{P} \Bigg( \sup_{\theta\in\Theta} &\sup_{z_t \in \mathcal{O}_n, t>n_P} \left\{ Q_{\nu}(\hat\psi_n(\theta;\nu);\theta)-Q_{\nu}(\psi(\theta;\nu);\theta) \right\} \geq C_{\mathcal{O}} \frac{n_o(\nu+p)}{n}[1 + \log(n) ]  \\ &+ L\frac{n_P}{n} (\nu+p)\log(1+\nu p)\left[ \sqrt{\frac{x}{n_P}} + \frac{x}{n_P} + \sqrt{\frac{C_n}{n_P}} + \frac{C_n}{n_P} \right]  \Bigg) \leq 4 \exp(-x),
  \end{align*}
  for a constant $L$ which depends on $s_0,\kappa_1,\kappa_2,M_2$ and $C_{\mathcal{O}}$ depends on $s_0,M_2,\kappa_1,A,\alpha$.
  If $z_t$ is strictly stationary and $\beta$-mixing for $t \in \{1,\dots,n_P\}$, then:
  \begin{align*}
    \mathbb{P} \Bigg( \sup_{\theta\in\Theta} &\sup_{z_t \in \mathcal{O}_n, t>n_P}\left\{ Q_{\nu}(\hat\psi_n(\theta;\nu);\theta)-Q_{\nu}(\psi(\theta;\nu);\theta) \right\} \geq C_{\mathcal{O}} \frac{n_o(\nu+p)}{n}[1 + \log(n) ] \\ &+ \tilde{L}\frac{n_P}{n} (\nu+p)\log(1+\nu p)\left[ \sqrt{\frac{(x+C_n)x}{n_P}} + \frac{(x+C_n)x}{n_P} + \sqrt{\frac{C_n}{n_P}} + \frac{C_n}{n_P} \right]  \Bigg) \leq 12 \exp(-x),
  \end{align*}
  for $\tilde{L}$ which additionally depends on the mixing coefficients $a,b$.
\end{proposition}

Because $\psi(\theta;\nu)$ is a minimizer, $Q_{\nu}(\hat\psi_n(\theta;\nu);\theta)-Q_{\nu}(\psi(\theta;\nu);\theta) \geq 0$ always holds. Proposition \ref{prop:unif} gives exponential inequalities for deviations from the biased solutions $\psi(\theta;\nu)$, uniformly in both parameters $\theta$ and outliers $z_t \in \mathcal{O}_n$, with respect to the loss $Q_\nu$. The bounds only require finite second moments, allowing for heavy tails under $P$. This is important in macroeconomic and financial applications since $P$ typically does not have sub-exponential, or Gaussian, tails.\footnote{Heavy-tailed distributions, unlike the exponential and Gaussian distributions, may not have all finite moments. Student and Pareto are both heavy-tailed distributions.} The worst-case contamination bias is of order $n_0(\nu+p)/n[1+\log(n)]$ which depends on the proportion of outliers $n_o/n$ and the tuning parameter $\nu$. It differs from the $(\sqrt{\nu}/n)n_o$ term for the simple estimator above. The proofs indicate that $(\nu/n)n_o$ corresponds to the influence of outliers when estimating $\Sigma$. 

For iid data, similar to Bernstein's inequality, the tails are thin: approximately sub-Gaussian for small $x \ll \sqrt{n_P}$ and sub-exponential for large $x \gg \sqrt{n_P}$.\footnote{The inequality $\mathbb{P}(Z \geq \sqrt{x/n} + x/n + a_n) \leq 4\exp(-x)$ implies $\mathbb{P}(Z \geq u/\sqrt{n} + a_n) \leq 4\min[\exp(-u^2),\exp(-\sqrt{n}u)]$ which is sub-Gaussian for $u \ll \sqrt{n}$ and sub-exponential for $u \gg \sqrt{n}$.} For time-series data, the tails are thicker: approximately sub-Gaussian for $x \ll C_n$, sub-exponential for  $C_n \ll x \ll \sqrt{n_P}$ and sub-Weibull for $x \gg \sqrt{n_P}$ with tail parameter $1/2$ \citep{vladimirova2020}. This is comparable to Bernstein inequalities for sample means of bounded $\beta$-mixing processes in  \citet{doukhan1994}.

Estimating both $\mu$ and $\Sigma$ consistently requires $(\nu/n)\log(n) n_o \to 0$. This is more restrictive than $(\sqrt{\nu}/n)n_o \to 0$ which appears under local asymptotics for $\mu$. This is related to the discussion above on iterative procedures and joint estimation of $\psi$. The dependence on the number of moment conditions $p$ is made explicit to show how it affects the bounds. The $p^2$ term in $C_n$ comes from estimating $p(p+1)/2$ coefficients in $\Sigma$. For the large sample results below, the number of parameters $k$ and moments $p$ will be assumed to be fixed and finite.

\subsection{Asymptotic Properties} \label{sec:large_sample}
The following builds on Proposition \ref{prop:unif} to derive uniform consistency and then oracle equivalence results which involve the amount of contamination $n_o$ and the bias. The large-sample results can be used to compute standard errors and compute confidence intervals the usual way (i.e. reporting $\tilde{\theta}_n \pm 1.96 \text{se}(\tilde{\theta}_n)$).
\begin{corollary} \label{corr:unif} Suppose the conditions for Proposition \ref{prop:unif}, Assumption \ref{as:moments} iii hold, and:
  \[ n_o = o\left( \frac{n}{\nu\log(n)}\right), \, \nu\log(\nu) = o\left( \sqrt{\frac{n}{\log(n)}} \right).  \]
  Let $\psi(\theta;\infty)$ denote the pair $\mu(\theta;\infty) = \mathbb{E}_P[g(z_t;\theta)]$, $\Sigma(\theta;\infty) = \text{var}_P[g(z_t;\theta)]$, then:
  \[ \sup_{\theta \in \Theta} \left( \sup_{ z_t \in \mathcal{O}_n, t > n_P } \|\hat{\psi}_n(\theta;\nu) - \psi(\theta;\infty)\| \right) = o_p(1). \]
\end{corollary}
Proposition \ref{prop:unif} and the following two bounds: $|Q_\nu(\psi;\theta) - Q_\infty(\psi;\theta)| \leq O(\nu^{-1})$ and $\|\psi(\theta;\nu)-\psi(\theta;\infty)\| \leq O(\nu^{-1})$, uniformly in $\theta$, imply the uniform consistency result above. Taking the supremum over $\mathcal{O}_n$ ensures the result is robust against the least favorable outliers.  
\begin{proposition} \label{prop:equiv} Suppose the conditions of Corollary \ref{corr:unif} hold. Let $\max[\mathbb{E}_P(\|g(z_t;\theta_0)\|^{r + \delta}),\mathbb{E}_P(|G_t|^{r + \delta})] := M_{r,\delta}$ for $r \geq 1$ and $\delta >0$.
  Let $\overline{g}_{n_P}(\theta) = \frac{1}{n_P} \sum_{t=1}^{n_P} g(z_t;\theta)$, if $M_{3,\delta}$ is finite for some $\delta>0$:
  \begin{align*}
    \sup_{\theta \in \Theta} \left( \sup_{ z_t \in \mathcal{O}_n, t > n_P } \| \hat{\mu}_{n}(\theta;\nu) - \overline{g}_{n_P}(\theta) \| \right) = O_p\left(\max\left[\frac{1}{\nu},\frac{\sqrt{\nu}n_o}{n}\right] \right)
  \end{align*}
  If, in addition $M_{5,\delta}$ is finite for some $\delta>0$:
  \begin{align*}
    \sup_{\theta \in \Theta} \left( \sup_{ z_t \in \mathcal{O}_n, t > n_P } \| \tilde{\mu}_{n}(\theta;\nu) - \overline{g}_{n_P}(\theta) \| \right) = O_p\left(\max\left[\frac{1}{\nu^2},\frac{\sqrt{\nu}n_o}{n}\right]\right).
  \end{align*}  
\end{proposition}
Using the same two inequalities, and a bound on the score, Proposition \ref{prop:equiv} shows that the robust and bias-corrected estimates are uniformly close to an oracle that computes the sample mean using only the good $n_P$ observations. An empirical researcher might want to trim out outliers without altering, as much as possible, the rest of the sample. This oracle result precisely states this property. In that sense, it gives a more desirable characterization than limit theorems for $\hat{\mu}_n(\theta;\nu) - \mu(\theta;\infty)$ and $\tilde{\mu}_n(\theta;\nu) - \mu(\theta;\infty)$.

Similar to non-parametric regressions which derive bias from smoothness, stronger moment conditions are needed to derive faster rates of convergence. Without outliers, OLS estimates are asymptotically normal for iid data when $\mathbb{E}_P(\|x_t y_t\|^2),\mathbb{E}_P(\|x_t \|^4) < \infty$. Here the condition is more restrictive, it reads $\mathbb{E}_P(\|x_t y_t\|^{5+\delta}),\mathbb{E}_P(\|x_t \|^{10 + 2\delta}) < \infty$.

The worst-case impact of outliers is of order $(\sqrt{\nu}/n)n_o$, with and without bias correction. Note that the estimator $\hat{\mu}_n$ is ``redescending.'' The maximal influence of a single observation $z$ given by $(\sqrt{\nu/2})/n$, is attained at $\|g(z;\theta)-\mu\|_{\Sigma^{-1}} = \sqrt{\nu}$ and then monotonically declines to zero as $\|g(z;\theta)-\mu\|_{\Sigma^{-1}}$ increases.\footnote{This is also discussed in \citet[p432]{mcdonald1988}, \citet[Ch4.8]{huber2011}.}  The result requires $\hat{\Sigma}_n(\theta;\nu)$ uniformly convergent. Importantly, the influence function is not redescending for $\Sigma$: it is strictly increasing and bounded above by $\nu > \sqrt{\nu}$. Hence, consistency of $\hat{\Sigma}_n$ is more restrictive: $(\nu/n) n_o \to 0$.


\begin{assumption} \label{as:momcond}
  i. $\mathbb{E}_P[g(z_t;\cdot)]$ is continuously differentiable in $\theta \in \Theta$, ii. $\mathbb{E}_P[g(z_t;\theta)]=0$ if, and only if, $\theta = \theta_0 \in \text{int}(\Theta)$, iii. $G(\theta_0) := \partial_\theta \mathbb{E}_P[g(z_t;\theta_0)]$ has full rank, iv. for any $\delta_{n_P} \to 0$, $\sup_{\|\theta - \theta_0\|\leq \delta_{n_P}}\sqrt{n_P}\|\overline{g}_{n_P}(\theta)-\overline{g}_{n_P}(\theta_0) - \partial_\theta \mathbb{E}_P[g(z_t;\theta_0)](\theta-\theta_0)\|/[1+\sqrt{n_P}\|\theta-\theta_0\|] = o_p(1)$, v. $\sqrt{n_P}\overline{g}_{n_P}(\theta_0) \overset{d}{\to} \mathcal{N}(0,\Sigma_0)$, vi. $W_n \overset{p}{\to} W$ positive definite.
\end{assumption}

Assumption \ref{as:moments} repeats conditions from \citet{newey1994}, only for the good $n_P$ observations. They imply consistency and asymptotic normality of $\hat{\theta}_{n_P}$, an oracle estimator which uses only the good $n_P$ datapoints.  

\begin{theorem} \label{th:asym} Suppose Assumption \ref{as:momcond} and the conditions of Proposition \ref{prop:equiv} hold with $M_{5,\delta}$ finite for some $\delta >0$. Suppose $n_o$ and $\nu$ are such that:  \[ \frac{\sqrt{n}}{\nu^{2}} = o(1), \text{ and } \sqrt{\frac{\nu}{n}}n_o = o(1).\]
  Let $\hat{\theta}_{n_P} = \text{argmin}_{\theta \in \Theta} \|\overline{g}_{n_P}(
\theta)\|_{W_n}$, the estimator $\tilde{\theta}_n$ satisfies:
  \begin{align*} 
    &\sup_{z_t \in \mathcal{O}_n, t > n_P}\|\sqrt{n_P}(\tilde{\theta}_n - \hat{\theta}_{n_P})\| = o_p(1), \quad \text{and} \quad
    \sqrt{n_P}(\tilde{\theta}_n - \theta_0) \overset{d}{\to} \mathcal{N}(0,V)
  \end{align*}
  for any sequence $z_t \in \mathcal{O}_n$, $t=n_P+1,\dots,n$, where $V = (G^\prime W G)^{-1} G^\prime W \Sigma_0 W G (G^\prime W G)^{-1}$, $G = \partial_\theta \mathbb{E}_P[g(z_t;\theta_0)]$.
\end{theorem}

Theorem \ref{th:asym} presents the main result: the bias-corrected estimates are asymptotically equivalent to the oracle $\hat{\theta}_{n_P}$. They inherit its asymptotic properties. The supremum over $\mathcal{O}_n$ ensures robustness against least favorable outliers. The bias is asymptotically negligible if $\nu^2 = o(\sqrt{n})$. If $n_o$ were known, setting $\nu \asymp (n/n_o)^{2/5}$ would achieve the optimal rate in Proposition \ref{prop:equiv}. For this choice of $\nu$, the condition $\sqrt{n}/\nu^2 = o(1)$ reads $n_o = o(n^{3/8})$. Setting $\nu = O(n^{1/4} \log(n))$ is nearly optimal when $n_o$ becomes arbitrarily close to this bound as it requires implies $n_o = o( n^{3/8}/\sqrt{\log(n)} )$. A data-driven rule is given below to select $\nu$ in practice while enforcing this rate. The large sample properties of $\vardbtilde{\mu}_n(\theta;\nu) = 2 \tilde{\mu}_n(\theta;\nu) - \tilde{\mu}_n(\theta;\nu/2)$, from Section \ref{sec:modsamest}, and the resulting $\vardbtilde{\theta}_n$ follow from those of $\tilde{\mu}(\theta;\nu)$, $\tilde{\theta}_n$.

Assumption \ref{as:momcond} requires $\theta$ to be strongly identified using the good $n_P$ observations. Given that Proposition \ref{prop:equiv} does not restrict identification status, one should compute an identification-robust test statistic - e.g. Anderson-Rubin (AR) - from the robust bias-corrected moment estimates. This is related to \citet{klooster2023} who consider AR test statistics with bounded influence curve.



With the oracle result (Proposition \ref{prop:equiv}) and the regularity conditions (Assumption \ref{as:momcond}), further results could be derived. One could consider two-step GMM with robust weighting $W_n = \hat{\Sigma}_n(\tilde{\theta}_n;\nu)^{-1}$ in the second step, robust overidentifying restrictions, quasi-Likelihood Ratio and Lagrange multiplier tests, etc. This is not pursued here.
\begin{proposition} \label{prop:HC} Suppose the assumptions for Theorem \ref{th:asym} hold.
  For each $\theta \in \Theta$ and $\nu >0$, the estimates $\hat{\mu}_n(\theta;\nu)$, $\tilde{\mu}_n(\theta;\nu)$ satisfy:
  \[ \hat{\mu}_n(\theta;\nu) = \sum_{t=1}^n \omega_t(\theta;\nu) g(z_t;\theta), \quad \tilde{\mu}_n(\theta;\nu) = \sum_{t=1}^n \tilde{\omega}_t(\theta;\nu) g(z_t;\theta) \]
  where the weights are given by $\omega_t(\theta;\nu) = \frac{ (1+p/\nu)/n[1+q_t(\theta;\nu)/\nu]^{-1}}{(1+p/\nu)/n\sum_{t=1}^n [1+q_t(\theta;\nu)/\nu]^{-1} + \kappa_1/\nu }$ and $\tilde{\omega}_t(\theta;\nu) = 2\omega_t(\theta;\nu) - \omega_t(\theta;\nu/2)$ using $q_t(\theta;\nu) =  \|g(z_t;\theta)-\hat{\mu}_n\|^2_{\hat{\Sigma}_n^{-1}}$. 
  
  Let $\hat{\varepsilon}_t(\theta) = g(z_t;\theta) - \hat{\mu}_n(\theta;\nu)$, $\tilde{\varepsilon}_t(\theta) = g(z_t;\theta) - \tilde{\mu}_n(\theta;\nu)$. The following weighted variance estimators are consistent:
  \begin{align*} &\hat{\Sigma}_{n,\omega}(\theta) = \sum_{t=1}^n \omega_t(\theta;\nu) \hat{\varepsilon}_t(\theta) \hat{\varepsilon}_t(\theta)^\prime \overset{p}{\to} \Sigma(\theta), \quad \tilde{\Sigma}_{n,\omega}(\theta) = \sum_{t=1}^n \tilde{\omega}_t(\theta;\nu) \tilde{\varepsilon}_t(\theta) \tilde{\varepsilon}_t(\theta)^\prime \overset{p}{\to} \Sigma(\theta),\end{align*}
  where $\Sigma(\theta) = \text{var}_P( g(z_t;\theta) )$ here denotes the short-run variance under $P$.
\end{proposition}

For cross-sections and serially uncorrelated moments, $\hat{\Sigma}_n(\hat{\theta}_n,\nu)$ can be used to estimate $\Sigma_0$ in Theorem \ref{th:asym}. Because of the penalty $\kappa_2 >0$ on $\Sigma$, it tends to be downward biased. An alternative is to use the same weights as $\tilde{\mu}_n$ to match the properties of the estimator more closely. Proposition \ref{prop:HC} above shows that such an estimator is also consistent for the short-run variance. Long-run variance estimates, required for serially correlated moments, are not considered here. 

The weighted average representation further implies, for linear models, that $\tilde{\theta}_n$ are weighted least-squares estimates since $\tilde{\mu}_n(\tilde{\theta}_n;\nu) = \sum_{t=1}^n \tilde{\omega}(\tilde{\theta}_n;\nu) x_t( y_t - x_t^\prime \tilde{\theta}_n ) = 0 \Rightarrow \tilde{\theta}_n  = ( \sum_{t=1}^n \tilde{\omega}(\tilde{\theta}_n;\nu) x_t x_t^\prime )^{-1} \sum_{t=1}^n \tilde{\omega}(\tilde{\theta}_n;\nu) x_t y_t$. The weighting can be used to interpret the results.

\paragraph{Data-driven choice of tuning parameter $\nu$.} The following describes a data-driven procedure to select the tuning parameter $\nu$. Take $0<a_0<a_1<\dots<a_J$ and $\nu_j = a_j n^{s}$ with $1/4 < s < 1/2$ or $\nu_j = a_j n^{s} \log(n)$ with $1/4 \leq s < 1/2$. The simulated and empirical examples use $s = 1/4$ and $0.5 = \log(a_0) < \dots < \log(a_J) = 35$ so that each $\nu_j = O(n^{1/4}\log(n))$ satisfies the requirements for Theorem \ref{th:asym}.

Using $\nu = \nu_0$ as a baseline, compute a preliminary estimate $\hat{\theta}_n$ and the corresponding moment estimates $\hat{\psi}_n(\hat{\theta}_n;\nu_0)$. In the absence of outliers, it can be shown that $|Q_n(\hat{\psi}_n(\hat{\theta}_n;\nu_0);\nu_j) - Q_n(\hat{\psi}_n(\hat{\theta}_n;\nu_0);\infty)| = O_p(\nu_j^{-1})$. This implies that, in the absence of outliers, the fit should be comparable accross different values of $\nu$: $|Q_n(\hat{\psi}_n(\hat{\theta}_n;\nu_0);\nu_j) - Q_n(\hat{\psi}_n(\hat{\theta}_n;\nu_0);\nu_0)| \leq O_p(\nu_0^{-1})$. 
The selection rule picks the largest value of $\nu_j$ such that the fit remains comparable:
\[ \hat{\nu}_n = \text{max} \left\{ \nu_j, \text{ s.t. } |Q_n(\hat{\psi}_n(\hat{\theta}_n;\nu_0);\nu_j) - Q_n(\hat{\psi}_n(\hat{\theta}_n;\nu_0);\nu_0)| \leq \frac{1+\log(n)}{\nu_0} \right\}. \]
The parameters and the moments are only estimated once, to reduce computation, at the smallest $\nu_0$ which produces the most robust estimate of the grid $\nu_0,\dots,\nu_J$. 

By design, $\hat{\nu}_n$ has rate $O(n^{s})$ or $O(n^{s} \log(n))$ which satisfies the conditions of Theorem \ref{th:asym} given restrictions on $n_o$. 
The following heuristic motivates the choice of criteria. As discussed above, the outliers have an asymptotic impact on non-robust estimates if $n_o n^{\alpha}/n \not \to 0$, and the estimator is robust as long as $n_o = o(\sqrt{n/\nu})$. Set $n_o = c \sqrt{n/\nu_0}$ then the sum over outliers in $Q_n(\cdot;\nu)$ increases proportionally to $\nu c \sqrt{\nu_0/n}\log(1+n^{2\alpha}/\nu) \sim c \nu \sqrt{\nu_0/n}\log(n)$. The change over the $n_P$ terms is a $O_p(\nu_0^{-1})$. For $\nu_0$ relatively small, the upper bound in the criteria above conservatively minors the sum of these two bounds.



\section{Simulated and Empirical Applications} \label{sec:apps}
All the estimations below use the same $\kappa_1 = \kappa_2 =10^{-2}$, giving wide bounds in Lemma \ref{lem:bounds}. With the data-driven choice of $\hat{\nu}_n$, the results are not too sensitive to this choice of penalty.
\subsection{Simulated Example} \label{sec:mc}
To illustrate the finite sample properties of the procedure, consider a linear regression $y_t = x_t^\prime \theta_0 + e_t$. There are three regressors $x_t = (1,x_{1t},x_{2t},x_{3t})$, each $x_{jt}$ and $e_t$ is drawn from $(\chi^2_5 - 5)/\sqrt{10}$ has mean zero and unit variance, $\theta_0 = (0,1,1,1)$. Sample size is $n=150$, several $n_o = 0,1,5,10$ are reported where each outlier has $x_{jt} = \sqrt{n}$ and $y_t = x_t^\prime \theta_\dagger$, $\theta_\dagger = (0,1/2,1/2,1/2)$. In this example, outliers are leveraged to mimic the motivating example. 

The simulations compares full sample $\hat{\theta}_{n}^{ols}$, an oracle which discards outliers $\hat{\theta}_{n_P}^{ols}$, R's robust regression estimates $\hat{\theta}_{n}^{rlm}$ with $\hat{\theta}_{n}$, $\tilde{\theta}_{n}$, $\vardbtilde{\theta}_{n}$ computed using $\hat{\nu}_n$ as described above. A further $\hat{\theta}^{un}_{n}$ is computed using $\hat{\nu}_n^2$ to illustrate undersmoothing as opposed to bias correction used in this paper. $\vardbtilde{\theta}_{n}$ applies the correction step twice as discussed at the end of Section \ref{sec:modsamest}.

\begin{table}[ht] \caption{Small sample properties of the estimators ($n=150$)} \label{tab:mc150}
  \centering
  \setlength\tabcolsep{4.5pt}
    \renewcommand{\arraystretch}{0.935} 
    {\small
  \begin{tabular}{l|cccaaac|cccaaac}
    \hline \hline
  & \multicolumn{7}{c|}{$100 \times \text{RMSE}$} & \multicolumn{7}{c}{Rejection Rate}\\\hline
  & \multicolumn{14}{c}{$n_o = 0$}\\ \hline
   & $\hat{\theta}_n^{ols}$ & $\hat{\theta}_{n_P}^{ols}$ & $\hat{\theta}_{n}^{rlm}$ & \mc{1}{$\hat{\theta}_{n}$} & \mc{1}{$\tilde{\theta}_{n}$} & \mc{1}{$\vardbtilde{\theta}_{n}$} & $\hat{\theta}_{n}^{un}$ & $\hat{\theta}_n^{ols}$ & $\hat{\theta}_{n_P}^{ols}$ & $\hat{\theta}_{n}^{rlm}$ & \mc{1}{$\hat{\theta}_{n}$} & \mc{1}{$\tilde{\theta}_{n}$} & \mc{1}{$\vardbtilde{\theta}_{n}$} & $\hat{\theta}_{n}^{un}$ \\ 
    \hline
    $\theta_0$ & 8.05 & 8.05 & 12.00 & 11.84 & 9.31 & 8.11 & 7.94 & 0.04 & 0.04 & 0.24 & 0.29 & 0.14 & 0.05 & 0.06 \\ 
    $\theta_1$ & 8.00 & 8.00 & 7.15 & 7.97 & 7.79 & 7.78 & 7.92 & 0.06 & 0.06 & 0.06 & 0.11 & 0.08 & 0.07 & 0.06 \\ 
    $\theta_2$ & 8.10 & 8.10 & 7.46 & 8.45 & 8.21 & 8.11 & 8.06 & 0.04 & 0.04 & 0.05 & 0.10 & 0.06 & 0.05 & 0.05 \\ 
    $\theta_3$ & 8.19 & 8.19 & 7.43 & 8.55 & 8.30 & 8.16 & 8.14 & 0.06 & 0.06 & 0.06 & 0.10 & 0.07 & 0.06 & 0.06 \\ \hline
    & \multicolumn{14}{c}{$n_o = 1$}\\ \hline
    & $\hat{\theta}_n^{ols}$ & $\hat{\theta}_{n_P}^{ols}$ & $\hat{\theta}_{n}^{rlm}$ & \mc{1}{$\hat{\theta}_{n}$} & \mc{1}{$\tilde{\theta}_{n}$} & \mc{1}{$\vardbtilde{\theta}_{n}$} & $\hat{\theta}_{n}^{un}$ & $\hat{\theta}_n^{ols}$ & $\hat{\theta}_{n_P}^{ols}$ & $\hat{\theta}_{n}^{rlm}$ & \mc{1}{$\hat{\theta}_{n}$} & \mc{1}{$\tilde{\theta}_{n}$} & \mc{1}{$\vardbtilde{\theta}_{n}$} & $\hat{\theta}_{n}^{un}$ \\ 
    \hline
    $\theta_0$ & 10.71 & 8.04 & 13.01 & 14.18 & 10.97 & 8.52 & 10.32 & 0.03 & 0.04 & 0.20 & 0.46 & 0.23 & 0.08 & 0.08 \\ 
    $\theta_1$ & 38.57 & 8.07 & 15.23 & 8.27 & 7.97 & 7.87 & 32.24 & 0.00 & 0.06 & 0.01 & 0.14 & 0.10 & 0.07 & 0.40 \\ 
    $\theta_2$ & 38.39 & 8.11 & 15.09 & 8.73 & 8.36 & 8.14 & 32.08 & 0.01 & 0.04 & 0.01 & 0.12 & 0.06 & 0.06 & 0.38 \\ 
    $\theta_3$ & 39.94 & 8.20 & 15.75 & 8.83 & 8.49 & 8.27 & 33.47 & 0.00 & 0.06 & 0.00 & 0.12 & 0.09 & 0.07 & 0.39 \\ 
   \hline
   & \multicolumn{14}{c}{$n_o = 5$}\\ \hline
    & $\hat{\theta}_n^{ols}$ & $\hat{\theta}_{n_P}^{ols}$ & $\hat{\theta}_{n}^{rlm}$ & \mc{1}{$\hat{\theta}_{n}$} & \mc{1}{$\tilde{\theta}_{n}$} & \mc{1}{$\vardbtilde{\theta}_{n}$} & $\hat{\theta}_{n}^{un}$ & $\hat{\theta}_n^{ols}$ & $\hat{\theta}_{n_P}^{ols}$ & $\hat{\theta}_{n}^{rlm}$ & \mc{1}{$\hat{\theta}_{n}$} & \mc{1}{$\tilde{\theta}_{n}$} & \mc{1}{$\vardbtilde{\theta}_{n}$} & $\hat{\theta}_{n}^{un}$ \\ 
    \hline
    $\theta_0$ & 11.98 & 8.14 & 16.57 & 16.98 & 13.38 & 9.82 & 13.45 & 0.10 & 0.04 & 0.24 & 0.59 & 0.38 & 0.13 & 0.16 \\ 
    $\theta_1$ & 47.57 & 8.40 & 47.17 & 9.02 & 8.62 & 8.40 & 46.72 & 0.99 & 0.06 & 0.99 & 0.12 & 0.08 & 0.06 & 0.99 \\ 
    $\theta_2$ & 47.48 & 8.26 & 48.25 & 9.28 & 8.80 & 8.53 & 47.14 & 0.99 & 0.04 & 1.00 & 0.12 & 0.05 & 0.03 & 1.00 \\ 
    $\theta_3$ & 49.17 & 8.28 & 49.48 & 9.33 & 8.94 & 8.72 & 48.65 & 0.98 & 0.06 & 0.98 & 0.10 & 0.08 & 0.04 & 0.98 \\ 
   \hline
   & \multicolumn{14}{c}{$n_o = 10$}\\ \hline
    & $\hat{\theta}_n^{ols}$ & $\hat{\theta}_{n_P}^{ols}$ & $\hat{\theta}_{n}^{rlm}$ & \mc{1}{$\hat{\theta}_{n}$} & \mc{1}{$\tilde{\theta}_{n}$} & \mc{1}{$\vardbtilde{\theta}_{n}$} & $\hat{\theta}_{n}^{un}$ & $\hat{\theta}_n^{ols}$ & $\hat{\theta}_{n_P}^{ols}$ & $\hat{\theta}_{n}^{rlm}$ & \mc{1}{$\hat{\theta}_{n}$} & \mc{1}{$\tilde{\theta}_{n}$} & \mc{1}{$\vardbtilde{\theta}_{n}$} & $\hat{\theta}_{n}^{un}$ \\ 
    \hline
    $\theta_0$ & 12.21 & 8.21 & 17.33 & 16.78 & 13.27 & 10.35 & 14.13 & 0.09 & 0.04 & 0.23 & 0.47 & 0.22 & 0.07 & 0.17 \\ 
    $\theta_1$ & 49.14 & 8.54 & 48.38 & 10.22 & 11.68 & 19.76 & 48.65 & 0.99 & 0.04 & 0.99 & 0.01 & 0.01 & 0.09 & 1.00 \\ 
    $\theta_2$ & 49.05 & 8.31 & 49.67 & 10.76 & 12.40 & 20.28 & 48.92 & 0.99 & 0.04 & 0.99 & 0.01 & 0.01 & 0.09 & 1.00 \\ 
    $\theta_3$ &50.52 & 8.51 & 50.70 & 11.04 & 13.00 & 20.96 & 50.19 & 0.98 & 0.06 & 0.98 & 0.00 & 0.01 & 0.09 & 0.99 \\ 
     \hline \hline
  \end{tabular}\\
  \notes{ \textbf{Legend:} $\hat{\theta}_n^{ols}$ full sample OLS, $\hat{\theta}_{n_P}^{ols}$ oracle OLS, $\hat{\theta}_n^{rlm}$ robust M-estimator, $\hat{\theta}_{n}$ robust estimates without bias correction, $\tilde{\theta}_{n}$ robust estimates with bias correction, $\vardbtilde{\theta}_{n}$ robust estimates with repeated bias correction, $\hat{\theta}^{un}_{n}$ undersmoothed robust estimates with $\hat{\nu}_n^2$. 200 Monte-Carlo replications. $n_o = $ number of outliers. Rejection rate for t-test at the $5\%$ significance level. Average $\hat{\nu}_n$: $35.85$, $16.00$, $11.00$, $10.71$ for $n_0 = 0$, $1$, $5$, $10$ respectively. Each $\hat{\nu}_n$ is selected on a grid $[\nu_0,\dots,\nu_J]$ where $\nu_0 = 8.77$, $\nu_J = 584.69$.
    } }
\end{table}

Table \ref{tab:mc150} shows that without outliers ($n_o=0$) the performance of bias-corrected and undersmoothed estimates is comparable to full sample OLS. The robust M-estimates of the intercept $\theta_0$ are biased, because the errors are skewed. The performance of OLS degrades as soon as $n_o = 1$, as expected. The undersmoothed and \textit{rlm} estimates are also less accurate. The non-corrected estimates $\hat{\theta}_n$ are more robust but biased. Bias correction, $\tilde{\theta}_n$ and $\vardbtilde{\theta}_n$, improves accuracy and rejection rates. The estimators still perform well for $n_o = 5$. Performance degrades for $n_o = 10$. This is perhaps not too surprising since $\log(n_o)/\log(n) \simeq 0.45 > 3/8$ for $n_o = 10$. Additional results for $n = 500$ are reported in Table \ref{tab:mc500}, Appendix \ref{apx:extra_mc}. Tables \ref{tab:mc150b}, \ref{tab:mc500b} has results with $\nu =O(n^{1/3})$ in the same Appendix.

\subsection{Empirical Applications} \label{sec:emp}

Two empirical applications further illustrate robust-GMM estimator in instrumental variable regression settings.  

\subsubsection{Trade Openness and Inflation} \label{sec:Romer}
The second empirical application is also inflation-related. \citet{romer1993} estimates the relationship between trade openness and inflation using country time averages between 1973 and 1993. Trade openness, measured by the share of imports to GDP, can be considered as endogenous given that monetary policy affects both inflation and exchange rates. He considers the following specification:
\[ \pi_t = \theta_0 + \theta_1 \text{open}_t + \theta_2 \log(\text{pcinc})_t + e_t, \]
where $\pi$ measures inflation, $\text{pcinc}$ is per-capita income in 1980, assumed exogenous. \citet{romer1993} further adds dummies in some specifications, these are not included here. The instrument for openness is $\log(\text{land})$ measuring the $\log$ of the square-mile surface of the country. The idea is that smaller land area economics should be more open to imports. \citet{romer1993} notes that ``A few countries in the sample have extremely high average inflation rates.'' and is concerned that ``the parameter estimates from a linear regression would be determined almost entirely by a handful of observations.'' As a remedy, he estimates the regression using the log of average inflation $\log(\pi/100)$. The influence of outliers in linear IV regressions is not intuitive because leverage can be either positive or negative (Lemma \ref{lem:lev_IV}). As a result, unlike OLS, the influence may not have the same sign as the residual: the impact of an outlier is less predictable than with OLS.

\begin{table}[ht] \caption{\citet{romer1993}: 10 Largest Contributors to $\hat{\theta}_{1n}^{IV}$, Sample Moments} \label{tab:Romer93_Influence}
  \centering
  \setlength\tabcolsep{4.5pt}
    \renewcommand{\arraystretch}{0.935} 
    { \small
  \begin{tabular}{l|acc||l|acc}
    \hline \hline
  \multicolumn{4}{c||}{Dependent variable: $y = \log(\frac{\pi}{100})$} & \multicolumn{4}{c}{Dependent variable: $y = \frac{\pi}{100}$}\\ \hline
  Country & \mc{1}{Contr.} & $\log(\frac{\pi}{100})$ & Open. & Country & \mc{1}{Contr.} & $\frac{\pi}{100}$ & Open. \\ 
    \hline
  Malta & -60.75 & -3.17 & 0.92 & Bolivia & -11.27 & 2.07 & 0.23\\ 
  Singapore & -56.77 & -3.32 & 1.64 & Argentina & -11.01 & 1.17 & 0.09\\ 
  Bahrain & -49.65 & -3.04 & 0.91 & Brazil & -9.40 & 0.74 & 0.07 \\ 
  Barbados & -40.74 & -2.23 & 0.73 & Israel & 4.28 & 0.75 & 0.57 \\ 
  United States & 39.32 & -2.78 & 0.09 & Peru & -3.18 & 0.49 & 0.20 \\ 
  Canada & 38.08 & -2.65 & 0.25 & Chile & -3.15 & 0.59 & 0.23 \\ 
  Hong Kong & -37.30 & -2.49 & 0.82 & Mexico & -2.73 & 0.33 & 0.11\\ 
  Luxembourg & -32.86 & -2.80 & 0.76 & Zaire & -2.57 & 0.43 & 0.40 \\ 
  Australia & 31.24 & -2.35 & 0.17 & Barbados & 1.95 & 0.11 & 0.73\\ 
  Mauritius & -29.07 & -2.02 & 0.57 & Mauritius & 1.92 & 0.13 & 0.57 \\
  \hline
  & \multicolumn{3}{c||}{Sample Moments} & & \multicolumn{3}{c}{Sample Moments}\\ \hline
  Mean & -1.25 & -2.10 & 0.37 & Mean & -0.34 & 0.17 & 0.37\\
  Stdev & 15.57 & 0.71 & 0.24 & Stdev & 1.93 & 0.24 & 0.24\\
  Skewness & -1.12 & 1.25 & 2.09 & Skewness & -3.91 & 5.34 & 2.09\\
  Kurtosis & 6.32 & 5.38 & 9.89 & Kurtosis & 22.22 & 38.10 & 9.89\\
   \hline \hline
  \end{tabular} }\\
  {\footnotesize \textbf{Note:} Contr.: Contribution = $(Z^\prime X/n)^{-1} z_i y_i$ to coefficient $\hat{\theta}_{1n}^{IV}$. Open.: Openness. $\pi$ = average inflation. Sample size $n = 114$. Countries sorted in decreasing order of contribution, in absolute values.}
  \end{table}
  Similar to the motivating example, Table \ref{tab:Romer93_Influence} provides diagnostics for both specifications. For $y = \pi/100$, the greatest contributors tend to be severely indebted countries that were particularly affected by the 1980s debt crisis. \citet{terra1998} argues that these countries overborrowed in the 1980s and had ``less pre-commitment in monetary policy'' resulting in higher inflation during the debt crisis.\footnote{\citet[p647]{terra1998} classifies Argentina, Bolivia, Brazil, Peru, Mexico, Zaire as severely indebted.} In contrast, for $y = \log(\pi/100)$, the greatest contributors are less indebted and other countries \citet[p647]{terra1998} which have low average inflation.\footnote{Singapore is the country with the lowest average inflation in the sample.} The log increases the influence of low-inflation countries, as one might expect. 
  
  \begin{table}[ht]
    \caption{\citet{romer1993}: IV, Robust and Bias-Corrected Estimates} \label{tab:Romer93_est}
  \centering
  \setlength\tabcolsep{4.5pt}
    \renewcommand{\arraystretch}{0.935} 
    { \small
    \begin{tabular}{l|ccc|cac|cac|cac}
      \hline \hline
      & \multicolumn{12}{c}{Dependent variable: $y = \log(\frac{\pi}{100})$}\\ \hline
     & $\hat{\theta}_{0n}^{\text{IV}}$ & $\hat{\theta}_{1n}^{\text{IV}}$& $\hat{\theta}_{2n}^{\text{IV}}$ & $\hat{\theta}_{0n}$ & \mc{1}{$\hat{\theta}_{1n}$} & $\hat{\theta}_{2n}$ & $\tilde{\theta}_{0n}$ & \mc{1}{$\tilde{\theta}_{1n}$} & $\tilde{\theta}_{2n}$ & $\vardbtilde{\theta}_{0n}$ & \mc{1}{$\vardbtilde{\theta}_{1n}$} & $\vardbtilde{\theta}_{2n}$ \\ 
      \hline
      est &  -1.21 & -1.25 & -5.64 & -1.19 & -1.13 & -6.82 & -1.18 & -1.21 & -6.42 & -1.19 & -1.29 & -5.70\\
      se & 0.42 & 0.40 & 5.60 & 0.37 & 0.36 & 5.01 & 0.40 & 0.38 & 5.41 & 0.43 & 0.41 & 5.70\\ \hline \hline
      & \multicolumn{12}{c}{Dependent variable: $y = \frac{\pi}{100}$}\\ \hline
      & $\hat{\theta}_{0n}^{\text{IV}}$ & $\hat{\theta}_{1n}^{\text{IV}}$& $\hat{\theta}_{2n}^{\text{IV}}$ & $\hat{\theta}_{0n}$ & \mc{1}{$\hat{\theta}_{1n}$} & $\hat{\theta}_{2n}$ & $\tilde{\theta}_{0n}$ & \mc{1}{$\tilde{\theta}_{1n}$} & $\tilde{\theta}_{2n}$ & $\vardbtilde{\theta}_{0n}$ & \mc{1}{$\vardbtilde{\theta}_{1n}$} & $\vardbtilde{\theta}_{2n}$ \\ 
      \hline
      est &  0.27 & -0.34 & 0.38 & 0.21 & -0.08 & -0.74 & 0.22 & -0.10 & -0.75 & 0.23 & -0.13 & -0.63 \\ 
      se & 0.11 & 0.16 & 1.36 & 0.04 & 0.04 & 0.53 & 0.05 & 0.05 & 0.65 & 0.06 & 0.06 & 0.81 \\ 
       \hline \hline
    \end{tabular} }\\
    {\footnotesize \textbf{Note:} $\hat{\theta}_n^{IV}$: IV estimates, $\hat{\theta}_n$: robust estimates, $\tilde{\theta}_n$: bias-corrected robust estimates, $\vardbtilde{\theta}_n$: repeated bias-corrected robust estimates. $\hat{\nu}_n = 38.33$, $14.10$ for $y = \log(\pi/100)$ and $\pi/100$, respectively. Estimates for $\theta_2$ reported using $\log(\text{pcinc})/100$ as a regressor. Sample size $n = 114.$}
  \end{table}

  The kurtosis indicates the $\log$-transformed regression is less prone to outliers, the standard deviation suggests the estimates will be significantly less accurate. This reflects the larger volatility of log-inflation compared to inflation. Also, the log transformation changes the interpretation of the coefficient $\theta_1$ which may not be desirable. The following replicates the original results and estimates the regression in levels, as in \citet[Ch16]{wooldridge2002}, to get the desired coefficient interpretation.

Table \ref{tab:Romer93_est} confirms that the log-transformed regression is less prone to outliers as the IV and robust estimates are very similar after bias-correction.\footnote{Estimates using a smaller $\nu = 12$ are nearly identical for the log regression (not reported here).} The non-transformed regression is, as \citet{romer1993} suspected, sensitive to some datapoints. Robust and bias-corrected estimates indicate IV overestimates the relationship between trade openness and inflation. Standard errors indicate the bias-corrected estimates are more accurate than the IV ones. The estimated effect is about one-third of the non-robust one. The bias correction adjusts the estimates by half to a full standard error. The full dataset of weights used to compute the estimates when $y = \pi/100$ are reported  in Tables \ref{tab:Romer_w1}, \ref{tab:Romer_w2}, Appendix \ref{apx:extra_emp}.

\subsubsection{Segregation and the Quality of Government} \label{sec:Alesina}
The third application considers the relationship between racial and religious discrimination and the quality of government. \citet{alesina2011} constructed a new dataset on ethnic, linguistic, and religious segregation and fractionalization for a large number of countries.  Mobility within a country, which determines segregation, can be endogenous to government quality. To address this particular issue, the authors predict segregation from neighboring country data. The main idea is that when a sub-population is at the border of the neighboring country, the same sub-group is more likely to be located near that border \citep[see][Figure 1, p1980]{alesina2011}. They illustrate using Switzerland as an example: most French speakers live near the French border, and Protestants are more commonly found near the German border. This is one of the papers surveyed in \citet{young2022}, which finds that published IV regressions tend to be highly leveraged and sensitive to a few observations. The following revisits some of the main results in the original paper. The regression specification is given by:
\[ \text{Rule of law}_i = \theta_0 + \theta_1 \text{Segregation}_i + \theta_2 \text{Fractionalization}_i + \text{Controls} + u_i, \]
where Segregation and Fractionalization are measured with respect to one of Ethnicity, Language, or Religion leading to three separate IV regressions. The controls are the same as in Table 6, Column 2 of \citet[p1897]{alesina2011}. Fractionalization controls for group heterogeneity in each dimension (ethnicity, language, and religion) as measured by a Herfindahl index. If there is only one group in the population, the index is zero. If there are many equal-sized groups, the measure is closer to $1$. See \citet[pp1779-1780]{alesina2011} for further details. 

\begin{table}[ht]
  \caption{\citet{alesina2011}: 10 Largest Contributors to $\hat{\theta}_{1n}$, $\hat{\theta}_{2n}$, Sample Moments, for each Measure of Segregation (ranked on coefficient $\theta_1$)} \label{tab:Alesina_Contr}
\centering
\setlength\tabcolsep{4.5pt}
  \renewcommand{\arraystretch}{0.935} 
  { \small
  \begin{tabular}{l|ac||l|ac||l|ac}
    \hline \hline
    \multicolumn{3}{c||}{Ethnicity} & \multicolumn{3}{c||}{Language} & \multicolumn{3}{c}{Religion} \\ \hline
    Country & \multicolumn{1}{c}{$\theta_1$} & $\theta_2$  & Country & \multicolumn{1}{c}{$\theta_1$} & $\theta_2$  & Country & \multicolumn{1}{c}{$\theta_1$} & $\theta_2$  \\ 
    \hline
    Zimbabwe & -99.76 & 19.98  & USA & 142.04 & -22.93 & Kazakhstan & -139.56 & -2.67\\
    Israel & 78.21 & -9.10  & Zimbabwe & -110.37 & 20.86 & Uzbekistan & 76.49 & -1.62\\
    Belgium & 61.05 & 6.76  & Austria & 106.92 & -7.21 & Cambodia & 53.48 & 7.31\\
    Cote d'Ivoire & -53.93 & 1.56 & Belgium & 76.80 & -0.48 & Indonesia & -51.20 & 6.27\\
    Guatemala & -32.69 & 3.02 & Canada & -58.79 & 17.34 & Switzerland & 46.39 & -4.31\\
    Ecuador & -27.86 & 1.57 & New Zealand & -58.21 & -0.22 & Netherlands & 45.01 & -1.37\\
    UK & -27.12 & -2.17 & Togo & -44.16 & 6.20 & CAR & -43.58 & 11.92\\
    Tajikistan & -26.56 & 6.03 & UK & -40.78 & -6.84 & Canada & -41.44 & 12.29\\
    France & -25.16 & 1.47 & Kyrgyzstan & -38.44 & 2.41 & Kenya & -41.28 & 4.92\\
    Spain & -25.12 & 9.11 & Rwanda & -36.45 & 11.63 & Israel & 41.10 & -4.39\\
    \hline
    \multicolumn{3}{c||}{Sample Moments} & \multicolumn{3}{c||}{Sample Moments} & \multicolumn{3}{c}{Sample Moments} \\ \hline
    Mean & -2.47 & 0.18 & Mean & -1.80 & 0.31 & Mean & -0.87 & 0.40\\
    Stdev & 19.11 & 5.08 & Stdev & 28.82 & 6.06 & Stdev &  30.53 & 5.85\\
  Skewness & -0.52 & 0.74 & Skewness & 1.31 & 0.67 & Skewness & -1.00 & 0.39\\
    Kurtosis & 12.63 & 5.11 & Kurtosis & 12.58 & 7.46 & Kurtosis & 7.17 & 4.21\\ \hline \hline
  \end{tabular} }\\
  {\footnotesize \textbf{Note:} CAR = Central African Republic.}
  \end{table}

  Table \ref{tab:Alesina_Contr} shows the 10 highest contributors for the coefficient $\theta_1$ in each regression as well as the sample moments of coefficient contribution. The regressions for ethnicity and language display somewhat heavy tails, as measured by the kurtosis. This indicates that the baseline results -- estimated coefficients, standard errors, or both --  may be sensitive to a few observations. Table \ref{tab:Alesina2011_est} reports standard IV and robust estimates with(out) bias correction. Robust estimates tend to produce more precise inferences, as measured by standard errors. The baseline results indicate that both ethnic and language segregation have a significant, negative impact on the rule of law in a given country. 
  
  Robust results indicate that ethnic segregation if the only significant determinant of the rule of law. Unlike the previous example, the estimate implies a larger effect than standard IV.  Notice that because the controls are correlated with the instrument, the direction of the change from standard to robust estimates does not necessarily coincide with the contribution to $\theta_1$ in Table \ref{tab:Alesina_Contr}.\footnote{If an outlier affects a coefficient on the controls and there is collinearity with the instrument, then robust estimates of $\theta_1$ will change with the coefficients on the controls, as they are correlated. Diagnostics may not fully reflect the multivariate effect of the outliers. In addition, when there are multiple outliers, the direction of change depends on the combined effect of the outliers.} Diagnostics can inform if the results are sensitive to some observations but, as explained in \citet{huber2011}, are not a substitute for robust estimation. Although non-significant, the coefficient $\theta_2$ for fractionalization does change from positive to negative in the first two regressions. Again, the full dataset of weights used in the three regressions is reported in Tables \ref{tab:Alesina_w1}, \ref{tab:Alesina_w2}, \ref{tab:Alesina_w3} of Appendix \ref{apx:extra_emp}.

  \begin{table}[ht]
    \caption{\citet{alesina2011}: IV, Robust and Bias-Corrected Estimates} \label{tab:Alesina2011_est}
  \centering
  \setlength\tabcolsep{4.5pt}
    \renewcommand{\arraystretch}{0.935} 
    { \small
    \begin{tabular}{l|cc|ab|ab|ab}
      \hline \hline
     & \multicolumn{8}{c}{Ethnicity} \\
      \hline
      & $\hat{\theta}_{1n}^{\text{IV}}$ & $\hat{\theta}_{2}^{\text{IV}}$ & \mc{1}{$\hat{\theta}_{1n}$} & \multicolumn{1}{c|}{$\hat{\theta}_{2n}$} & \multicolumn{1}{c}{$\tilde{\theta}_{1n}$} &\multicolumn{1}{c|}{$\tilde{\theta}_{2n}$} & \mc{1}{$\vardbtilde{\theta}_{1n}$} & \mc{1}{$\vardbtilde{\theta}_{2n}$} \\ \hline
      est & -2.47 & 0.18 & -3.18 & -0.13 & -3.19 & -0.13 & -2.75 & -0.09 \\ 
      se & 0.60 & 0.24 & 0.33 & 0.10 & 0.39 & 0.14 & 0.38 & 0.17 \\ \hline
      & \multicolumn{8}{c}{Language} \\
       \hline
       & $\hat{\theta}_{1n}^{\text{IV}}$ & $\hat{\theta}_{2}^{\text{IV}}$ & \mc{1}{$\hat{\theta}_{1n}$} & \multicolumn{1}{c|}{$\hat{\theta}_{2n}$} & \mc{1}{$\tilde{\theta}_{1n}$} &\multicolumn{1}{c|}{$\tilde{\theta}_{2n}$} & \mc{1}{$\vardbtilde{\theta}_{1n}$} & \mc{1}{$\vardbtilde{\theta}_{2n}$} \\ \hline
      est & -1.80 & 0.31 & -0.65 & -0.21 & -0.65 & -0.22 & -0.59 & -0.20 \\ 
      se & 0.80 & 0.24 & 0.23 & 0.06 & 0.35 & 0.08 & 0.54 & 0.11 \\ \hline
      & \multicolumn{8}{c}{Religion} \\
       \hline
       & $\hat{\theta}_{1n}^{\text{IV}}$ & $\hat{\theta}_{2}^{\text{IV}}$ & \mc{1}{$\hat{\theta}_{1n}$} & \multicolumn{1}{c|}{$\hat{\theta}_{2n}$} & \mc{1}{$\tilde{\theta}_{1n}$} & \multicolumn{1}{c|}{$\tilde{\theta}_{2n}$} & \mc{1}{$\vardbtilde{\theta}_{1n}$} & \mc{1}{$\vardbtilde{\theta}_{2n}$} \\ \hline
      est & -0.87 & 0.40 & -0.04 & -0.02 & 0.09 & 0.10 & 0.34 & 0.16 \\ 
      se & 1.82 & 0.23 & 0.52 & 0.12 & 0.69 & 0.16 & 0.86 & 0.16 \\
       \hline \hline
    \end{tabular} }\\
    {\footnotesize \textbf{Note:} $\hat{\theta}_n^{IV}$: IV estimates, $\hat{\theta}_n$: robust estimates, $\tilde{\theta}_n$: bias-corrected robust estimates, $\vardbtilde{\theta}_n$: repeated bias-corrected robust estimates. $\hat{\nu}_n = 10.71, 8.55, 11.80$ and sample size $n = 97, 92, 78$ for Ethnicity, Language and Religion, respectively. Sample sizes vary because of missing values. }
    \end{table}

\section{Conclusion} 
It is important to assess the robustness of empirical findings. Without symmetry restrictions, large differences between robust and non-robust estimates could be attributed to 1) improved resilience, or 2) significant asymmetry bias (or a combination of the two). This paper proposes a procedure with a simple asymptotic bias correction so that 2) is less likely. Reporting the implicit estimation weights makes the final results transparent and interpretable. This is illustrated in three empirical applications. 
\bibliographystyle{ecta}
\newpage
\bibliography{refs}

\begin{appendices}
  \renewcommand\thetable{\thesection\arabic{table}}
  \renewcommand\thefigure{\thesection\arabic{figure}}
  \renewcommand{\theequation}{\thesection.\arabic{equation}}
  \renewcommand\thelemma{\thesection\arabic{lemma}}
  \renewcommand\thetheorem{\thesection\arabic{theorem}}
  \renewcommand\thedefinition{\thesection\arabic{definition}}
    \renewcommand\theassumption{\thesection\arabic{assumption}}
  \renewcommand\theproposition{\thesection\arabic{proposition}}
    \renewcommand\theremark{\thesection\arabic{remark}}
    \renewcommand\thecorollary{\thesection\arabic{corollary}}
\setcounter{equation}{0}
\setcounter{lemma}{0}
\clearpage \baselineskip=18.0pt
\appendix

\makeatletter
\@addtoreset{assumption}{section}
\makeatother

\section{Preliminary Results} \label{apx:prelim}

\begin{lemma} \label{lem:Lip} Let $q_t (\psi;\theta) = (\nu+p)\log( 1+\|g(z_t;\theta) - \mu\|^2_{\Sigma^{-1}}/\nu )$. For all $\theta \in \Theta$: $\sup_{\psi \in \Psi_n} \|\partial_\mu q_t(\psi;\theta)\| \leq s_0^{-1/2}(1+p/\nu)\nu^{1/2}$, $\sup_{\psi \in \Psi_n} \|\partial_\Sigma q_t(\psi;\theta)\| \leq \nu \left(\frac{\nu^2 (1+p/\nu)}{\kappa_2} + \frac{\nu^4(1+p/\nu)^2}{4 \kappa_1\kappa_2} + \frac{p \nu}{\kappa_2}\right)^3$.
\end{lemma}

\begin{lemma} \label{lem:Concentration1}
  Suppose $z_t \sim P$ satisfying Assumption \ref{as:moments}, for $t\in \{1,\dots,n\}$, take $1 \leq \nu \leq n$. Let: \[ \overline{\Delta}_n(\psi;\theta) = \frac{1}{n}\sum_{t=1}^n \left(\log \left(1+\frac{\|g(z_t;\theta)-\mu\|^2_{\Sigma^{-1}}}{\nu} \right) - \mathbb{E}_P\left[\log \left(1+\frac{\|g(z_t;\theta)-\mu\|^2_{\Sigma^{-1}}}{\nu} \right)\right]\right)\] for any $\theta,\psi \in \Theta \times \Psi_n$. 
  
  1) If $z_t$ are iid, then there exists a constant $L>0$ which depends on $s_0,\kappa_1,\kappa_2,M_2,M_4$ such that for all $t \geq 0$:
  \begin{align}
    &\mathbb{P} \left( \sup_{\theta \in \Theta,\psi \in \Psi_n}|\overline{\Delta}_n(\psi,\theta)| \geq  L\log(1+p\nu)\left[  \sqrt{\frac{t}{n}} + \frac{t}{n} + \sqrt{\frac{C_n}{n}}+\frac{C_n}{n} \right]  \right)  \leq 4\exp(- t), \label{eq:Concentration1} 
  \end{align}
where $C_n = 1 + (k+2p^2)[ \log(p) + \log(\nu) + \log(n) ]$.

2) If $z_t$ is strictly stationary with mixing coefficient $\beta_m \leq a \exp( - b m)$ for $a,b > 0$, then for another constant $\tilde{L}>0$ which further depends on $a,b$ such that:
\begin{align}
  &\mathbb{P} \left( \sup_{\theta \in \Theta,\psi \in \Psi_n}|\overline{\Delta}_n(\psi,\theta)| \geq  \tilde{L}\log(1+p\nu)\left[  \sqrt{\frac{(t+C_n)t}{n}} + \frac{(t+C_n)t}{n} + \sqrt{\frac{C_n}{n}}+\frac{C_n}{n} \right]  \right) \tag{\ref{eq:Concentration1}'} \\  &\leq 12\exp(-t), \notag
\end{align}
for the same $C_n$ as 1).
\end{lemma}

\section{Proofs for the Main Results} \label{apx:proofs}
\paragraph{Proof of Lemma \ref{lem:bounds}.} Note that $Q_n(\psi) \to +\infty$ when $\text{trace}(\Sigma) \to +\infty$ so the solution is s.t. $\text{trace}(\hat{\Sigma}_n) < +\infty$, likewise $\|\hat{\mu}_n\| < \infty$. The first-order condition (foc) wrt $\mu$ implies:
\[ -\frac{\nu + p}{\nu n} \sum_{t=1}^n \frac{\hat{\Sigma}_n^{-1}(g(Z_t;\theta)-\hat{\mu}_n)}{1+\|g(Z_t;\theta)-\hat{\mu}_n\|^2_{\hat{\Sigma}_n^{-1}}/\nu} + \frac{\kappa_1}{\nu} \hat{\Sigma}_n^{-1}\hat{\mu}_n = 0. \]
Pre-multiply by $\Sigma_n^{1/2}$ and re-arrange terms to find:
\[ \|\hat{\Sigma}_n^{-1/2}\hat{\mu}_n\| \leq \frac{\nu}{\kappa_1}(1+p/\nu) \max_{t} \frac{\|g(Z_t;\theta)-\hat{\mu}_n\|_{\hat{\Sigma}_n^{-1}}}{1+\|g(Z_t;\theta)-\hat{\mu}_n\|^2_{\hat{\Sigma}_n^{-1}}/\nu}, \]
where $\max_{x \geq 0} \frac{x}{1+x^2/\nu} = \sqrt{\nu}/2$ yields the desired inequality. Take the foc wrt to $\Sigma^{-1}$:
\[ \frac{\nu + p}{\nu n} \sum_{t=1}^n \frac{(g(Z_t;\theta)-\hat{\mu}_n)(g(Z_t;\theta)-\hat{\mu}_n)^\prime}{1+\|g(Z_t;\theta)-\hat{\mu}_n\|^2_{\hat{\Sigma}_n^{-1}}/\nu} + \frac{\kappa_1}{\nu} \hat{\mu}_n\hat{\mu}_n^\prime - \frac{\kappa_2}{\nu}\Sigma^2_n - \Sigma_n = 0. \]
Pre and post-multiply by $\Sigma_n^{-1/2}$, re-arrange terms and compute the $\text{trace}$ to find:
\[ \text{trace}(\hat{\Sigma}_n) \leq \frac{\nu}{\kappa_2}\left( (1+p/\nu)\max_t \frac{\|g(Z_t;\theta)-\hat{\mu}_n\|^2_{\hat{\Sigma}_n^{-1}}}{1+\|g(Z_t;\theta)-\hat{\mu}_n\|^2_{\hat{\Sigma}_n^{-1}}/\nu} + \frac{\kappa_1}{\nu} \|\hat{\Sigma}_n^{-1/2}\hat{\mu}_n\|^2 + p \right). \]
The $\max$ is bounded above by $\sup_{x \geq 0} \frac{x^2}{1+x^2/\nu} = \nu$. Plug-in the bound for $\|\hat{\Sigma}_n^{-1/2}\hat{\mu}_n\|$ to get the desired inequality.
\qed

\paragraph{Proof of Proposition \ref{prop:unif}.} First, note that $\psi(\theta;\nu) \in \Psi_n$ for all $\theta \in \Theta$. By minimization, we have for all $\theta \in \Theta$:
\begin{align*}
  0 \leq Q_{\nu}(\hat\psi_n(\theta;\nu);\theta)-Q_{\nu}(\psi(\theta;\nu);\theta) &= \underbrace{Q_{n}(\hat\psi_n(\theta;\nu);\theta)-Q_{n}(\psi(\theta;\nu);\theta)}_{\leq 0} \\&+ (Q_{\nu} - Q_{n})(\hat\psi_n(\theta;\nu);\theta) - (Q_{\nu} - Q_{n})(\psi(\theta;\nu);\theta)\\
  &\leq 2\sup_{\theta \in \Theta,\psi \in \Psi_n } |(Q_{\nu} - Q_{n})(\psi;\theta)|,
\end{align*}
where $Q_{n} - Q_{\nu} = (\nu+p)\overline{\Delta}_n$ used in Lemma \ref{lem:Concentration1}. There are two bounds to derive: one for the $n_o$ outliers and another for the remaining $n_P$ observations. For any $z \in \mathcal{O}_n$, $\psi \in \Psi_n$, $1 \leq \nu \leq n$:
\begin{align*} 0 &\leq \log(1 + \|g(z;\theta)-\mu\|^2_{\Sigma^{-1}}/\nu)\leq \log(1 + 3s_0^{-1} A^2n^{2\alpha} /\nu  ) + \log(1 + 3/2 \kappa_1^{-1} \nu^{1/2}  ). \end{align*}
 We also have $Q_\nu = n_o/n Q_\nu + n_P/n Q_\nu$, the second is the centering term for well-behaved observations. We need to bound the first:
\[ 0 \leq (\nu + p)\mathbb{E}_P[\log(1+\|g(z_t;\theta)-\mu\|^2_{\Sigma^{-1}}/\nu)] \leq 3(1+p/\nu)s_0^{-1}M_2 + (\nu+p)\log(1 + 3/2 \kappa_1^{-1} \nu^{1/2}  ), \]
for any $(\theta,\psi) \in \Theta \times \Psi_n$, using $\log(1+x) \leq x$ for $x \geq 0$, Assumption \ref{as:moments} and Lemma \ref{lem:bounds}. Combine the two bounds to find:
\[ 2\Big|\frac{\nu+p}{n}\sum_{t=n_P+1}^{n} \log(1 + \|g(z_t;\theta)-\mu\|^2_{\Sigma^{-1}}/\nu) - \frac{n_o}{n}Q_{\nu}(\psi;\theta)\Big| \leq C_{\mathcal{O}}\frac{n_o(\nu + p)}{n} [1+\log(n)], \]
where $C_{\mathcal{O}}$ only depends on $s_0,M_2,\kappa_1,A,\alpha$. Define $Q_{n_P}$ to be the sample average over the $n_P$ uncontaminated observations, $Q_{n_P} - Q_{\nu} = (\nu+p)\overline{\Delta}_{n_P}$ which satisfies the conditions of Lemma \ref{lem:Concentration1}. Pre-multiply by $n_P/n$ to get the uncontaminated part of $Q_n - Q_\nu$ and multiply by $2$. Replace $L,\tilde{L}$ from Lemma \ref{lem:Concentration1} with $2L,2\tilde{L}$ to get the desired result.
\qed

\paragraph{Proof of Corollary \ref{corr:unif}.} Proceed in several steps: 1) show uniform convergence under the pseudo-distance $Q_\nu$ and that it implies some compactness restrictions, 2) derive a norm equivalence on compact sets, 3) combine these two steps with a uniform convergence for $\|\psi(\theta;\nu)-\psi(\theta;\infty)\|$ as $\nu \to \infty$.

\paragraph{Step 1.} Uniform convergence is implied by Proposition \ref{prop:unif} and the rate conditions.  The following shows that this implies: $\sup_{\theta \in \Theta}\|\hat{\psi}_n(\theta;\nu)\| \leq K$ with probability approaching 1 (wpa1), for some constant $K>0$. Then, all pairs $(\hat{\psi}_n(\theta;\nu),\psi(\theta;\nu))_{\theta \in \Theta}$ will be in a bounded compact subset of $\Psi$ wpa1.
First, note that for $1 \leq \nu$:
\[ Q_\nu(\psi;\theta) \leq (1+p)\mathbb{E}_P\left[\|g(z_t;\theta)\|^2_{\Sigma^{-1}} ) \right] + \log|\Sigma| + \kappa_1 \|\mu\|^2_{\Sigma^{-1}} + \kappa_2\text{trace}(\Sigma), \]
which implies that $\sup_{\theta \in \Theta,\nu \geq 1} \left( \inf_{\psi \in \Psi} Q_\nu(\psi;\theta) \right) \leq K_1$ for some constant $K_1$ which is less or equal to the largest (over $\theta$) minimal (over $\psi$) value of the upper bound which is finite by compactness, continuity and strict convexity, wrt $\psi$, of the upper bound.
\[ Q_\nu(\psi;\theta) \geq \mathbb{E}_P\left[ (\nu+p)\log( 1+\|g(z_t;\theta)\|^2_{\Sigma^{-1}}/\nu ) \right] + \log|\Sigma| \geq \log( \lambda_{\max}(\Sigma) ) \geq 2K_1, \]
for any $\theta,\nu,\mu$ as soon as $\lambda_{\max}(\Sigma) \geq \exp(2 K_1) := s_1$. Assumption \ref{as:moments} ii and compactness of $\Theta$ implies that:
\[ \|\mu(\theta;\infty)\| = \|\mathbb{E}_P[g(z_t;\theta)]\| \leq K_2, \]
for some constant $K_2$ which depends on $M_2$, $M_4$ and $\text{diam}(\Theta)$. In addition, for any $M >0$, Chebychev's inequality implies:
\[ \sup_{\theta \in \Theta} \mathbb{P} \left( \|g(z_t;\theta)-\mu(\theta;\infty)\| \geq M \right) \leq M_2/M := \varepsilon >0. \]
For $\lambda_{\max}(\Sigma) \leq s_1$ above, this implies for any $\theta \in \Theta$ and all $\|\mu\| \geq 2M + K_2$:
\[ Q_\nu(\psi;\theta) \geq (\nu+p) (1-\varepsilon)\log( 1 + s_1^{-1}M /\nu) + p \log(s_0) \geq (1+p) \frac{(1-\varepsilon)s_1^{-1}M}{1+s_1^{-1}M/\nu} + p \log(s_0) \geq 2K_1, \]
for $M$ and $\nu \geq \underline{\nu} \geq 1$ sufficiently large. 

The uniform convergence then implies that $\sup_{\theta \in \Theta} \|Q_{\nu}(\hat{\psi}_n(\theta;\nu);\theta)\| \leq \sup_{\theta \in \Theta} \|Q_{\nu}(\psi(\theta;\nu);\theta)\| + o_p(1) \leq 2K_1,$
wpa1. This implies that $\sup_{\theta \in \Theta}\|\hat{\mu}_n(\theta;\nu)\| \leq K_2 + 2M$ and $\sup_{\theta \in \Theta}\|\lambda_{\max}(\hat{\Sigma}_n(\theta;\nu))\| \leq \exp(2K_1)$ wpa1, which implies the desired result. The same holds for $\psi(\theta;\nu)$.

\paragraph{Step 2.} First, for any $x \geq 0$ we have $\frac{x}{1+x} \leq \log(1+x) \leq x$ which implies $|\log(1+x)-x| \leq \frac{x^2}{1+x}$. Take $(\theta,\psi) \in \Theta \times \Psi$, this implies:
\begin{align*} \Bigg| &\mathbb{E}_p\left[(\nu+p)\log( 1+\|g(z_t;\theta)-\mu\|^2_{\Sigma^{-1}}/\nu ) - \frac{\nu+p}{\nu}\|g(z_t;\theta)-\mu\|^2_{\Sigma^{-1}} \right]\Bigg| \\ &\leq \frac{\nu+p}{\nu^2} \mathbb{E}_P\left[ \|g(z_t;\theta)-\mu\|^4_{\Sigma^{-1}} \right] \leq 9s_0^{-2} \frac{\nu+p}{\nu^2}\left[ M_4 + \|\mu\|^4\right], \end{align*}
using Assumption \ref{as:moments} iii. to bound the 4th moment.
This implies that $\sup_{\theta \in \Theta}|Q_\nu(\psi;\theta) - Q_{\infty}(\psi;\theta)| \leq O(\nu^{-1})$ with respect to $\psi$ on bounded compact sets.

\paragraph{Step 3.} Given that $\sup_{\theta \in \Theta}( \|\hat{\psi}_n(\theta;\nu)\| + \|\psi(\theta;\nu)\| ) \leq 2K$ from Step 1, Step 2 and the triangular inequality imply:
\[ \sup_{\theta \in \Theta} |Q_\infty(\hat{\psi}_n(\theta;\nu);\theta) - Q_{\infty}(\psi(\theta;\nu);\theta)| = o_p(1). \]
Note that $Q_\infty$ is the Gaussian negative log-likelihood which is strictly convex for each $\theta \in \Theta$, so this also implies $\|\hat{\psi}_n(\theta;\nu)-\psi(\theta;\nu)\| = o_p(1)$ uniformly in $\theta$. Since we are actually interested in $\psi(\theta;\infty)$:
\begin{align*} 0 &\leq \sup_{\theta \in \Theta}\{Q_{\infty}(\psi(\theta;\nu);\theta) - Q_{\infty}(\psi(\theta;\infty);\theta)\}\leq \sup_{\theta \in \Theta}\{\underbrace{Q_{\nu}(\psi(\theta;\nu);\theta) - Q_{\nu}(\psi(\theta;\infty);\theta)}_{\leq 0}\} \\ &+ \sup_{\theta \in \Theta}[Q_{\infty}(\psi(\theta;\nu);\theta)-Q_{\nu}(\psi(\theta;\nu);\theta) - Q_{\infty}(\psi(\theta;\infty);\theta)+Q_{\nu}(\psi(\theta;\infty);\theta)]\\
&\leq O(\nu^{-1}), \end{align*}
using Step 2 and the compactness from Step 1. This implies the uniform convergence result $\|\hat{\psi}_n(\theta;\nu)-\psi(\theta;\infty)\| = o_p(1)$.
\qed

\newcommand{\Thalign}{ \frac{1}{n_P}\sum_{t=1}^{n_P} x_{t,\theta} \frac{1}{\nu n_{P}} \sum_{t=1}^{n_P} \frac{ (x_{t,\theta} - \hat{\mu}_n)\|x_{t,\theta} - \hat{\mu}_n\|_{\hat{\Sigma}_n^{-1}}^2 }{1+\|x_{t,\theta} - \hat{\mu}_n\|_{\hat{\Sigma}_n^{-1}}^2/\nu} \frac{ \kappa_1 n}{n_P}\frac{\hat{\mu}_n}{\nu} \frac{1}{n_P}\sum_{t > n_P}  \frac{  \Sigma_{n}^{1/2} \Sigma_{n}^{-1/2}(x_{t,\theta} - \hat{\mu}_n) }{1+\|x_{t,\theta} - \hat{\mu}_n\|_{\hat{\Sigma}_n^{-1}}^2/\nu}} 

\paragraph{Proof of Proposition \ref{prop:equiv}.} The foc wrt $\hat\mu_n(\theta;\nu)$ reads (the dependence on $\theta,\nu$ is omitted for brievety):
\[ \frac{1}{n} \sum_{t=1}^n \frac{ x_{t,\theta} - \hat{\mu}_n }{1+\|x_{t,\theta} - \hat{\mu}_n\|_{\hat{\Sigma}_n^{-1}}^2/\nu} + \kappa_1\frac{\hat{\mu}_n}{\nu} = 0, \] 
where $x_{t,\theta} = g(z_t;\theta)$ as in the proof of Lemma \ref{lem:Concentration1}. Re-arrange terms to find:
\[ \hat{\mu}_n = \underbrace{\frac{1}{n_P}\sum_{t=1}^{n_P} x_{t,\theta} \vphantom{\Thalign} }_{ (A)} - \underbrace{\frac{1}{\nu n_{P}} \sum_{t=1}^{n_P} \frac{ (x_{t,\theta} - \hat{\mu}_n)\|x_{t,\theta} - \hat{\mu}_n\|_{\hat{\Sigma}_n^{-1}}^2 }{1+\|x_{t,\theta} - \hat{\mu}_n\|_{\hat{\Sigma}_n^{-1}}^2/\nu} \vphantom{\Thalign}}_{ (B) } + \underbrace{\frac{ \kappa_1 n}{n_P}\frac{\hat{\mu}_n}{\nu} \vphantom{\Thalign}}_{(C)} + \underbrace{\frac{1}{n_P}\sum_{t > n_P}  \frac{  \Sigma_{n}^{1/2} \Sigma_{n}^{-1/2}(x_{t,\theta} - \hat{\mu}_n) }{1+\|x_{t,\theta} - \hat{\mu}_n\|_{\hat{\Sigma}_n^{-1}}^2/\nu} \vphantom{\Thalign}}_{(D)},\]
where $(A) = \overline{g}_{n_P}(\theta)$ and $\|(C)\| = O_p(\nu^{-1})$ uniformly in $\theta$ when $n_P/n \to 1$ using Corollary \ref{corr:unif}. Then, we have:
\begin{align*} \sup_{\theta \in \Theta}\|(B)\| &\leq ( \sup_{\theta \in \Theta} \lambda_{\max}(\Sigma(\theta;\infty)) + o_p(1) )^{-2} \frac{1}{\nu n_P} \sum_{t=1}^{n_P} 8\left( \sup_{\theta \in \Theta} \|x_{t,\theta}\|^3 + \sup_{\theta \in \Theta} \|\hat{\mu}_n\|^3 \right)\\
  &\leq ( \sup_{\theta \in \Theta} \lambda_{\max}(\Sigma(\theta;\infty)) + o_p(1) )^{-2} \frac{1}{\nu n_P} \sum_{t=1}^{n_P} \left( 64\|x_{t,\theta_0}\|^3 + 64 \text{diam}(\Theta)^3 G_t^3 + 8 \sup_{\theta \in \Theta}\|\hat{\mu}_n\|^3 \right)\\
  &=O_p(\nu^{-1}),  \end{align*}
by uniform consistency of $\hat{\mu}_n$ and a strong law of large numbers applied to the sample mean of $\|x_{t\theta_0}\|^3 + G_t^3$ \citep[][Cor3.48]{white2001}.
We also have:
\[ \sup_{\theta \in \Theta} \|(D)\| \leq \left[ \sup_{\theta \in \Theta} \lambda_{\max}(\Sigma(\theta;\infty)) + o_p(1) \right]^{1/2} \frac{\sqrt{\nu}n_o}{2n_P} = o(n^{-1/2}), \]
if $n_o = o(\sqrt{\nu/n})$. Corollary \ref{corr:unif} required $\nu = o(\sqrt{n})$, this yields the first result:
\[ \sup_{\theta \in \Theta} \|\hat{\mu}_n(\theta;\nu) - \overline{g}_{n_P}(\theta)\| = O_p\left(\max\left[\nu^{-1}, \frac{\sqrt{\nu}n_o}{n}\right]\right). \]
To derive results for the bias-corrected estimates, we additionally need convergence rates for $\hat{\Sigma}_n$, take the foc wrt $\Sigma^{-1}$ and re-arrange terms:
\begin{align*} \hat{\Sigma}_n &= 
  \frac{1}{n}\sum_{t=1}^{n_P} (x_{t,\theta}-\hat{\mu}_n)(x_{t,\theta}-\hat{\mu}_n)^\prime \tag{A}
  \\&- \frac{p}{\nu n}\sum_{t=1}^{n_P} (x_{t,\theta}-\hat{\mu}_n)(x_{t,\theta}-\hat{\mu}_n)^\prime \tag{B}
  \\ &-\frac{\nu + p}{\nu^2 n}\sum_{t=1}^{n_P} \frac{(x_{t,\theta}-\hat{\mu}_n)(x_{t,\theta}-\hat{\mu}_n)^\prime \|x_{t,\theta}-\hat{\mu}_n\|^2_{\hat{\Sigma}_n^{-1}}}{1+\|x_{t,\theta}-\hat{\mu}_n\|^2_{\hat{\Sigma}_n^{-1}}/\nu}  \tag{C}
  \\ &+ \frac{\nu + p}{\nu n}\sum_{t > n_P} \frac{\hat{\Sigma}_n^{1/2} \hat{\Sigma}_n^{-1/2}(x_{t,\theta}-\hat{\mu}_n)(x_{t,\theta}-\hat{\mu}_n)^\prime \hat{\Sigma}_n^{-1/2} \hat{\Sigma}_n^{1/2}}{1+\|x_{t,\theta}-\hat{\mu}_n\|^2_{\hat{\Sigma}_n^{-1}}/\nu} \tag{D}
  \\ &+ \kappa_1 \frac{\hat{\mu}_n \hat{\mu}_n^\prime}{\nu} - \kappa_2 \frac{\hat{\Sigma}_n^2}{\nu}, \tag{E}
\end{align*}
where $\sup_{\theta \in \Theta}\|(E)\| = O_p(\nu^{-1})$ by uniform convergence. $\sup_{\theta \in \Theta}\|(B)\| = O_p(\nu^{-1})$ by applying a uniform law of large numbers to $x_{t,\theta},x_{t,\theta}^2$ and uniform convergence of $\hat{\mu}_n$. Then, we have:
\[ \sup_{\theta \in \Theta} \|(C)\| \leq ( \sup_{\theta \in \Theta} \lambda_{\max}(\Sigma(\theta;\infty)) + o_p(1) )^{-2} \frac{1+p}{\nu n_P} \sum_{t=1}^{n_P} 16(\|x_{t,\theta}\|^4 + \|\hat{\mu}_n\|^4) = O_p(\nu^{-1}), \]
using a strong law of large numbers for $\|x_{t,\theta_0}\|^4,G_t^4$, as in the bound on $(B)$ for $\hat{\mu}_n$ above. Finally, $\sup_{\theta \in \Theta}\|(D)\| \leq ( \sup_{\theta \in \Theta} \lambda_{\max}(\Sigma(\theta;\infty)) + o_p(1) ) \nu (1+p) \frac{n_o}{n} = O_p\left( \frac{\nu n_o}{n} \right)$. Importantly, we also have:
\begin{align*} &\frac{1}{n}\sum_{t=1}^{n_P} \left[ (x_{t,\theta}-\hat{\mu}_n(\theta;\nu))(x_{t,\theta}-\hat{\mu}_n(\theta;\nu))^\prime -  (x_{t,\theta}-\hat{\mu}_n(\theta;\nu/2))(x_{t,\theta}-\hat{\mu}_n(\theta;\nu/2))^\prime \right]  \\ &= O_p\left(\max \left[\nu^{-1}, \frac{\sqrt{\nu}n_o}{n} \right]\right),\end{align*}
since $\hat{\mu}_n(\theta,\nu) - \hat{\mu}_n(\theta,\nu/2) = O_p(\max \left[\nu^{-1}, \frac{\sqrt{\nu}n_o}{n} \right])$ uniformly in $\theta$. This implies that $\hat{\Sigma}_n(\theta;\nu)-\hat{\Sigma}_n(\theta;\nu/2) = O_p(\max[\nu^{-1},\frac{\nu n_o}{n}])$ uniformly in $\theta$. 
We now have all the ingredients to expand the bias-corrected estimates $\tilde{\mu}_n(\theta;\nu) = 2\hat{\mu}_n(\theta;\nu) - \hat{\mu}_n(\theta;\nu/2)$, omit their dependence on $\theta$:
\begin{align*}
  &\tilde{\mu}_n(\nu)\\ &= \frac{2}{n_P}\sum_{t=1}^{n_P} x_{t,\theta} - \frac{1}{n_P}\sum_{t=1}^{n_P} x_{t,\theta} \tag{A}
  \\ & - \frac{2}{\nu n_P} \sum_{t=1}^{n_P} \left[ \frac{(x_{t,\theta}-\hat{\mu}_n(\nu))\|x_{t,\theta}-\hat{\mu}_n(\nu)\|^2_{\hat{\Sigma}^{-1}_n(\nu)}}{1+\|x_{t,\theta}-\hat{\mu}_n(\nu)\|^2_{\hat{\Sigma}^{-1}_n(\nu)}/\nu} - \frac{(x_{t,\theta}-\hat{\mu}_n(\frac{\nu}{2}))\|x_{t,\theta}-\hat{\mu}_n(\frac{\nu}{2})\|^2_{\hat{\Sigma}^{-1}_n(\frac{\nu}{2})}}{1+2\|x_{t,\theta}-\hat{\mu}_n(\frac{\nu}{2})\|^2_{\hat{\Sigma}^{-1}_n(\frac{\nu}{2})}/\nu} \right] \tag{B}\\
  &+ 2\kappa_1\frac{\hat{\mu}_n(\nu) - \hat{\mu}_n(\frac{\nu}{2})}{\nu} \tag{C}\\
  &+\frac{2}{n_P}\sum_{t > n_P}  \frac{  \Sigma_{n}^{1/2}(\nu) \Sigma_{n}^{-1/2}(\nu)(x_{t,\theta} - \hat{\mu}_n(\nu)) }{1+\|x_{t,\theta} - \hat{\mu}_n(\nu)\|_{\hat{\Sigma}_n^{-1}(\nu)}^2/\nu} - \frac{1}{n_P}\sum_{t > n_P}  \frac{  \Sigma_{n}^{1/2}(\frac{\nu}{2}) \Sigma_{n}^{-1/2}(\frac{\nu}{2})(x_{t,\theta} - \hat{\mu}_n(\frac{\nu}{2})) }{1+2\|x_{t,\theta} - \hat{\mu}_n(\frac{\nu}{2})\|_{\hat{\Sigma}_n^{-1}(\frac{\nu}{2})}^2/\nu}. \tag{D}
\end{align*}
Clearly $(A) = \overline{g}_{n_P}(\theta)$ and $\|(D)\| \leq O_p(\frac{\sqrt{\nu}n_o}{n})$ uniformly in $\theta \in \Theta$ as previously shown. Likewise, $\|(C)\| \leq O_p(\max[\nu^{-2} ,\frac{n_o}{\sqrt{\nu} n}]) \leq O_p(\max[\nu^{-2},\frac{\sqrt{\nu} n_o}{ n}])$, uniformly.

Remains to bound the longer term:
\begin{align*}
  (B) &= \frac{-2}{\nu n_p} \sum_{t=1}^{n_P} \frac{(x_{t,\theta}-\hat{\mu}_n(\nu))\|x_{t,\theta}-\hat{\mu}_n(\nu)\|^2_{\hat{\Sigma}^{-1}_n(\nu)} - (x_{t,\theta}-\hat{\mu}_n(\frac{\nu}{2}))\|x_{t,\theta}-\hat{\mu}_n(\frac{\nu}{2})\|^2_{\hat{\Sigma}^{-1}_n(\frac{\nu}{2})}}{(1+\|x_{t,\theta} - \hat{\mu}_n(\nu)\|_{\hat{\Sigma}_n^{-1}(\nu)}^2/\nu)(1+2\|x_{t,\theta} - \hat{\mu}_n(\frac{\nu}{2})\|_{\hat{\Sigma}_n^{-1}(\frac{\nu}{2})}^2/\nu)} \tag{B1}\\
  & +\frac{2}{\nu^2 n_p} \sum_{t=1}^{n_P} \frac{ (x_{t,\theta}-\hat{\mu}_n(\frac{\nu}{2}))\|x_{t,\theta}-\hat{\mu}_n(\frac{\nu}{2})\|^2_{\hat{\Sigma}^{-1}_n(\frac{\nu}{2})}\|x_{t,\theta} - \hat{\mu}_n(\nu)\|_{\hat{\Sigma}_n^{-1}(\nu)}^2 }{(1+\|x_{t,\theta} - \hat{\mu}_n(\nu)\|_{\hat{\Sigma}_n^{-1}(\nu)}^2/\nu)(1+2\|x_{t,\theta} - \hat{\mu}_n(\frac{\nu}{2})\|_{\hat{\Sigma}_n^{-1}(\frac{\nu}{2})}^2/\nu)} \tag{B2}\\
  & -\frac{2}{\nu^2 n_p} \sum_{t=1}^{n_P} \frac{ (x_{t,\theta}-\hat{\mu}_n(\nu))\|x_{t,\theta}-\hat{\mu}_n(\nu)\|^2_{\hat{\Sigma}^{-1}_n(\nu)}\|x_{t,\theta} - \hat{\mu}_n(\frac{\nu}{2})\|_{\hat{\Sigma}_n^{-1}(\frac{\nu}{2})}^2 }{(1+\|x_{t,\theta} - \hat{\mu}_n(\nu)\|_{\hat{\Sigma}_n^{-1}(\nu)}^2/\nu)(1+2\|x_{t,\theta} - \hat{\mu}_n(\frac{\nu}{2})\|_{\hat{\Sigma}_n^{-1}(\frac{\nu}{2})}^2/\nu)}, \tag{B3}
\end{align*}
where $\|(B2),(B3)\| = O_p(\nu^{-2})$ using a uniform of large numbers for $\|x_{t,\theta}\|^5$ and uniform convergence of $\hat{\mu}_n(\nu), \hat{\mu}_n(\nu/2)$. The last step is to show that the numerator in $(B1)$ is a $O_p(\max[\nu^{-1}, \frac{\sqrt{\nu}n_o}{n}])$, let $\delta_{n} = \nu^{-1} + \frac{\nu n_o}{n}$:
\begin{align*}
 &\|(x_{t,\theta}-\hat{\mu}_n(\nu))\|x_{t,\theta}-\hat{\mu}_n(\nu)\|^2_{\hat{\Sigma}^{-1}_n(\nu)} - (x_{t,\theta}-\hat{\mu}_n(\frac{\nu}{2}))\|x_{t,\theta}-\hat{\mu}_n(\frac{\nu}{2})\|^2_{\hat{\Sigma}^{-1}_n(\frac{\nu}{2})}\|\\
 &\leq O_p(\delta_n)\|x_{t,\theta}-\hat{\mu}_n(\nu)\|^2_{\hat{\Sigma}^{-1}_n(\nu)} + \|x_{t,\theta}-\hat{\mu}_n(\frac{\nu}{2})\|\left[ \|x_{t,\theta}-\hat{\mu}_n(\nu)\|^2_{\hat{\Sigma}^{-1}_n(\nu)} - \|x_{t,\theta}-\hat{\mu}_n(\frac{\nu}{2})\|^2_{\hat{\Sigma}^{-1}_n(\frac{\nu}{2})}\right]\\
 &\leq O_p(\delta_n)\|x_{t,\theta}-\hat{\mu}_n(\nu)\|^2_{\hat{\Sigma}^{-1}_n(\nu)} + O_p(\delta_n)s_0^{-2}\|x_{t,\theta}-\hat{\mu}_n(\frac{\nu}{2})\|^3 \\
 &+ \|x_{t,\theta}-\hat{\mu}_n(\frac{\nu}{2})\| s_0^{-2} \left[ \|\hat{\mu}_n(\nu)-\hat{\mu}_n(\frac{\nu}{2})\| \times \|2x_{t,\theta} - \hat{\mu}_n(\nu)-\hat{\mu}_n(\frac{\nu}{2})\| \right]\\
 &\leq O_p(\delta_n) \Bigg( \|x_{t,\theta}-\hat{\mu}_n(\nu)\|^2_{\hat{\Sigma}^{-1}_n(\nu)} +  s_0^{-2}\|x_{t,\theta}-\hat{\mu}_n(\frac{\nu}{2})\|^3 \\ &+ s_0^{-2} \|x_{t,\theta}-\hat{\mu}_n(\frac{\nu}{2})\|(\|x_{t,\theta}-\hat{\mu}_n(\nu)\|+\|x_{t,\theta}-\hat{\mu}_n(\frac{\nu}{2})\|) \Bigg).
\end{align*}
Apply a uniform law of large numbers to $\|x_{t,\theta}\|^2,\|x_{t,\theta}\|^3$, and invoke uniform convergence of $\hat{\mu}_n(\nu),\hat{\mu}_n(\nu/2)$ to get since the denominator in $(B1)$ is less or equal than $1$: $\sup_{\theta \in \Theta}\|(B1)\| \leq O_p(\nu^{-1}\delta_n) = O_p(\max[\nu^{-2},\frac{\sqrt{\nu}n_o}{n}])$ as desired. Putting everything together, we get the desired result:
\[ \sup_{\theta \in \Theta}\|\tilde{\mu}_n(\theta;\nu)-\overline{g}_{n_P}(\theta)\| \leq O_p\left(\max\left[\nu^{-2} , \frac{\sqrt{\nu}n_o}{n}\right]\right). \]
\qed

\paragraph{Proof of Theorem \ref{th:asym}.} By definition:
$\|\tilde{\mu}_n(\tilde{\theta}_n)\|_{W_n}^2 \leq \inf_{\theta \in \Theta} \|\tilde{\mu}_n(\theta)\|_{W_n}^2 +o_p(n^{-1}).$
Proposition \ref{prop:equiv} implies that, uniformly in $\theta \in \Theta$:
\[ \|\overline{g}_{n_P}(\theta)\|_{W_n} - o_p(n^{-1/2}) \leq \|\tilde{\mu}_n(\theta)\|_{W_n} \leq \|\overline{g}_{n_P}(\theta)\|_{W_n} + o_p(n^{-1/2}). \]
In particular the asymptotic equivalence and approximate minimization properties imply:
\[ \|\overline{g}_{n_P}(\tilde{\theta}_n)\|_{W_n}\leq \|\tilde{\mu}_n(\tilde{\theta}_n)\|_{W_n} + o_p(n^{-1/2}) \leq \|\tilde{\mu}_n(\hat{\theta}_n)\|_{W_n} +o_p(n^{-1/2}) \leq \|\overline{g}_{n_P}(\hat{\theta}_{n_P})\|_{W_n} + o_p(n^{-1/2}),\]
which implies that $\tilde{\theta}_n$ is an approximate minimizer of $\|\overline{g}_{n_P}(\cdot)\|_{W_n}$. Assumption \ref{as:momcond} then implies continuity and asymptotic normality for both $\tilde{\theta}_n$ and $\hat{\theta}_{n_P}$, e.g. \citet[Th2.6, Th7.2]{newey1994} in the iid setting. The results then follow from a first-order expansion of the two estimators, e.g.: $\sqrt{n_P}(\tilde{\theta}_n-\theta_0) = - (G^\prime W G)^{-1} G^\prime W \overline{g}_{n_P}(\theta_0) + o_p(1).$ \qed

\newcommand{\Pralign}{ \sum_{t=1}^{n_P} \frac{1+p/\nu}{n} \hat{\varepsilon}_t \hat{\varepsilon}_t^\prime \sum_{t=1}^{n_P} \frac{1+p/\nu}{n \nu} \frac{ \hat{\varepsilon}_t \hat{\varepsilon}_t^\prime \|\hat{\varepsilon}_t\|^2_{\hat{\Sigma}_n^{-1}} }{1 + \|\hat{\varepsilon}_t\|^2_{\hat{\Sigma}_n^{-1}}/\nu} \sum_{t=n_P+1}^{n} \frac{1+p/\nu}{n} \frac{ \hat{\varepsilon}_t \hat{\varepsilon}_t^\prime }{1 + \|\hat{\varepsilon}_t\|^2_{\hat{\Sigma}_n^{-1}}/\nu}} 

\paragraph{Proof of Proposition \ref{prop:HC}.} The weighted average representation follows from the first-order condition $\partial_\mu Q_n(\hat{\psi}_n;\theta) = 0$, which can be re-written as:
\[ 0 = \frac{1+p/\nu}{n} \sum_{t=1}^n \frac{ \hat{\mu}_n - g(z_t;\theta)  }{1+\|g(z_t;\theta) - \hat{\mu}_n\|^2_{\hat{\Sigma}_n^{-1}}/\nu} + \frac{\kappa_1}{\nu} \hat{\mu}_n.\]
Re-arrange terms to find $\hat{\mu}_n = \sum_{t=1}^n \omega_t(\theta;\nu) g(z_t;\theta)$ as in the Proposition. Since $\tilde{\mu}_n(\theta;\nu) = 2\hat{\mu}_n(\theta;\nu) - \hat{\mu}_n(\theta;\nu/2)$ we have also have $\tilde{\mu}_n(\theta;\nu) = \sum_{t=1}^n [2\omega_t(\theta;\nu)-\omega_t(\theta;\nu/2)] g(z_t;\theta)$.

Let $\bar{\omega}_n(\theta;\nu) = (1+p/\nu)/n\sum_{t=1}^n[1+q_t /\nu]^{-1} + \kappa_1/\nu$, we have:
\begin{align*} \bar{\omega}_n(\theta;\nu) \hat{\Sigma}_{n,\omega}(\theta) &= \sum_{t=1}^n \frac{1+p/\nu}{n} \frac{ \hat{\varepsilon}_t \hat{\varepsilon}_t^\prime }{1 + \|\hat{\varepsilon}_t\|^2_{\hat{\Sigma}_n^{-1}}/\nu} \\
&= \underbrace{\sum_{t=1}^{n_P} \frac{1+p/\nu}{n} \hat{\varepsilon}_t \hat{\varepsilon}_t^\prime \vphantom{\Pralign}}_{(A)} - \underbrace{\sum_{t=1}^{n_P} \frac{1+p/\nu}{n \nu} \frac{ \hat{\varepsilon}_t \hat{\varepsilon}_t^\prime \|\hat{\varepsilon}_t\|^2_{\hat{\Sigma}_n^{-1}} }{1 + \|\hat{\varepsilon}_t\|^2_{\hat{\Sigma}_n^{-1}}/\nu} \vphantom{\Pralign}}_{(B)} + \underbrace{\sum_{t=n_P+1}^{n} \frac{1+p/\nu}{n} \frac{ \hat{\varepsilon}_t \hat{\varepsilon}_t^\prime }{1 + \|\hat{\varepsilon}_t\|^2_{\hat{\Sigma}_n^{-1}}/\nu} \vphantom{\Pralign}}_{(C)}\\
&\overset{p}{\to} \Sigma(\theta).\end{align*} 
To get the result, note that $\|(C)\| \leq \lambda_{\max}(\hat{\Sigma}_n)(1+p/\nu)n_o \nu/n = o_p(1)$. Likewise, $\|(B)\| \leq \nu^{-1}(1+p/\nu)s_0^{-1} [1/n\sum_{t=1}^{n_P} \|\hat{\varepsilon}_t\|^4] = O_p(\nu^{-1})$ using $\hat{\mu}_n \overset{p}{\to} \mathbb{E}_P(g(z_t;\theta))$ and a law of large numbers for $\|g(z_t;\theta)\|^4$. Similarly, a law of large numbers implies $(A) \overset{p}{\to} \Sigma(\theta)$. Using $q_t \geq 0$, we have $\bar{\omega}_n(\theta;\nu) = (1+p/\nu)/n \sum_{t=1}^{n_P} 1 - (1+p/\nu)/[n\nu] \sum_{t=1}^{n_P} q_t/[1+q_t/\nu] + (1+p/\nu)/n \sum_{t=n_P+1}^n [1+q_t/\nu]^{-1}$. The first term converges to 1, the second term is a $O_p(\nu^{-1})$ using a law of large numbers, and the third term is less or equal than $n_o/n(1+p/\nu) = o(1)$. This implies $\bar{\omega}_n(\theta;\nu) \overset{p}{\to} 1.$ Combine the results to find $\hat{\Sigma}_{n,\omega}(\theta) \overset{p}{\to} \Sigma(\theta).$

To prove consistency for $\tilde{\Sigma}_{n,\omega}(\theta)$, we will first prove consistency for $\sum_{t=1}^n \omega_t(\theta;\nu) \tilde{\varepsilon}_t(\theta) \tilde{\varepsilon}_t(\theta)^\prime$ and $\sum_{t=1}^n \omega_t(\theta;\nu/2) \tilde{\varepsilon}_t(\theta) \tilde{\varepsilon}_t(\theta)^\prime$. Since $\tilde{\Sigma}_{n,\omega}(\theta)$ equals two times the first minus the second, consistency follows. First, note that $\tilde{\varepsilon}_t = \hat{\varepsilon}_t + \hat{\mu}_n - \tilde{\mu}_n$, where $\hat{\mu}_n - \tilde{\mu}_n = o_p(1)$ by Corollary \ref{corr:unif}.

\begin{align*}
  \sum_{t=1}^n \omega_t(\theta;\nu) \tilde{\varepsilon}_t(\theta) \tilde{\varepsilon}_t(\theta)^\prime &=  \sum_{t=1}^n \omega_t(\theta;\nu) \hat{\varepsilon}_t(\theta) \hat{\varepsilon}_t(\theta)^\prime \\ & + 2\sum_{t=1}^n \omega_t(\theta;\nu) \hat{\varepsilon}_t(\theta)(\tilde{\mu}_n - \hat{\mu}_n)+ \bar{\omega}_n(\theta;\nu) (\tilde{\mu}_n - \hat{\mu}_n) (\tilde{\mu}_n - \hat{\mu}_n)^\prime\\
  &\overset{p}{\to} \Sigma_0(\theta),
\end{align*}
because the first term is consistent for $\Sigma_0(\theta)$ from the previous result. The last term is a $o_p(1)$ since $\bar{\omega}_n(\theta) = 1 + o_p(1)$ is multiplied by a $o_p(1)$. The second term is equal to $2 \hat{\mu}_n o_p(1) - 2 \bar{\omega}_n(\theta) \hat{\mu}_n o_p(1) = o_p(1)$. Follow the same steps for $\sum_{t=1}^n \omega_t(\theta;\nu/2) \tilde{\varepsilon}_t(\theta) \tilde{\varepsilon}_t(\theta)^\prime$ using $\hat{\mu}_n(\theta;\nu/2)$ instead of $\hat{\mu}_n(\theta;\nu)$ to derive the result and conclude the proof.
\qed
\newpage
\section{Proofs for the Preliminary Results} \label{apx:proof_prelim}

\paragraph{Proof of Lemma \ref{lem:Lip}.} Take derivates wrt $\mu$:
\[ \partial_\mu q_t(\psi) = -2 \Sigma^{-1/2}\frac{\nu + p}{\nu} \frac{\Sigma^{-1/2}(x_t - \mu)}{1+\|x_t-\mu\|^2_{\Sigma^{-1}}/\nu }, \]
where $\lambda_{\max}(\Sigma^{-1/2}) \leq s_0^{-1/2}$. Use $\|\Sigma^{-1/2}(x_t - \mu)\|/(1+\|x_t-\mu\|^2_{\Sigma^{-1}}/\nu) \leq \sqrt{\nu}/2$ to get the first inequality. Take derivates wrt $\Sigma$:
\[ \partial_\Sigma q_t(\psi) = - \Sigma^{3/2}\frac{\nu + p}{\nu} \frac{\Sigma^{-1/2}(x_t - \mu)(x_t - \mu)^\prime \Sigma^{-1/2} }{1+\|x_t-\mu\|^2_{\Sigma^{-1}}/\nu } \Sigma^{3/2}. \]
This implies $\|\partial_\Sigma q_t(\psi)\| \leq \lambda_{\max}(\Sigma)^3  (1+p/\nu) \nu$ where $\lambda_{\max}(\Sigma) \leq \text{trace}(\Sigma)$, bounded in (\ref{eq:bounds}).
\qed

\paragraph{Proof of Lemma \ref{lem:Concentration1} - 1) IID Setting.} Let $x_{t,\theta} = g(z_t;\theta)$ and $\Delta_t(\psi;\theta) = \log(1+\|x_{t,\theta}-\mu\|^2_{\Sigma^{-1}}/\nu) - \mathbb{E}_P[\log(1+\|x_{t,\theta}-\mu\|^2_{\Sigma^{-1}}/\nu)]$, $\overline{\Delta}_n(\psi;\theta) = 1/n \sum_{t=1}^n \Delta_t(\psi,\theta)$. For any pair $(\psi_j,\theta_j)$, we have: 
\[ |\overline{\Delta}_n(\psi;\theta)| \leq \underbrace{|\overline{\Delta}_n(\psi;\theta) - \overline{\Delta}_n(\psi_j;\theta)|}_{(A)} + \underbrace{|\overline{\Delta}_n(\psi_j;\theta) - \overline{\Delta}_n(\psi_j;\theta_j)|}_{(B)} + \underbrace{|\overline{\Delta}_n(\psi_j;\theta_j)|}_{(C)}.\]
The following bounds each one of $(A)$, $(B)$, and $(C)$, either deterministically or in probability.

\paragraph{1. Bound for $(A)$.} Lemma \ref{lem:Lip} implies that for any $\psi = (\mu,\Sigma),\psi_j= (\mu_j,\Sigma_j)$ in $\Psi_n$: 
\[ |\log(1+\|x_{t,\theta}-\mu\|^2_{\Sigma^{-1}}/\nu)-\log(1+\|x_{t,\theta}-\mu_j\|^2_{\Sigma_j^{-1}}/\nu)| \leq  p^3 \nu^{12} L_1 \|\psi-\psi_j\|, \]
where $L_1$ depends on $s_0,\kappa_1,\kappa_2$. Taking either sample averages or expectations, yields:
\begin{align} (A) \leq 2 p^3 \nu^{12} L_1 \|\psi-\psi_j\|, \label{eq:boundA} \end{align}
since the bound is deterministic.
\paragraph{2. Bound for $(B)$.}  Suppose, without loss of generality that $\|x_{t,\theta}-\mu\|_{\Sigma^{-1}} \geq \|x_{t,\theta_j}-\mu\|_{\Sigma^{-1}}$, then:\footnote{For any $x \geq y \geq 0$, $0 \leq \log(1+x)-\log(1+y) = \log( 1 + (1+x)/(1+y) -1 ) = \log( 1 + (x-y)/(1+y) ) \leq \log( 1 + x-y )$.}
\begin{align*}  0 &\leq \log( 1 + \|x_{t,\theta}-\mu\|_{\Sigma^{-1}}^2/\nu ) - \log( 1 + \|x_{t,\theta_j}-\mu\|_{\Sigma^{-1}}^2/\nu ) \\ &\leq \log( 1 + \|x_{t,\theta}-\mu\|_{\Sigma^{-1}}^2/\nu - \|x_{t,\theta_j}-\mu\|_{\Sigma^{-1}}^2/\nu ). \end{align*}
Using properties of inner-products: $0 \leq \|x_{t,\theta}-\mu\|_{\Sigma^{-1}}^2/\nu - \|x_{t,\theta_j}-\mu\|_{\Sigma^{-1}}^2/\nu \leq \|x_{t,\theta}-x_{t,\theta_j}\|_{\Sigma^{-1}} \|x_{t,\theta}+x_{t,\theta_j} - 2 \mu\|_{\Sigma^{-1}} / \nu$.\footnote{For any two vectors $a,b$, we have $\langle a,a \rangle - \langle b,b \rangle = \langle a-b,a+b \rangle \leq \|a-b\| \times \|a+b\|$.} Assumption (\ref{as:moments}) implies $\|x_{t,\theta}-x_{t,\theta_j}\|_{\Sigma^{-1}} \leq s_0^{-1/2}G_t \|\theta-\theta_j\|$ and $\|x_{t,\theta} + x_{t,\theta_j}\|_{\Sigma^{-1}} \leq 2 s_0^{-1/2}G_t \text{diam}(\Theta)$. Also $\psi \in \Psi_n$ implies $\|2\mu\|_{\Sigma^{-1}} \leq \nu^{3/2} (1+p/\nu) \kappa_1^{-1}$. Hence, for some constant $L_2$ which depends on $s_0,\kappa_1$ and $\text{diam}(\Theta)$:
\[  |\log( 1 + \|x_{t,\theta}-\mu\|_{\Sigma^{-1}}^2/\nu ) - \log( 1 + \|x_{t,\theta_j}-\mu\|_{\Sigma^{-1}}^2/\nu )| \leq \log( 1 + \nu p L_2 (1+G_t)^2 \|\theta-\theta_j\| ), \]
and then taking expectations and using $\log(1+x) \leq x$ for $x \geq 0$:
\[ \mathbb{E}_P |\log( 1 + \|x_{t,\theta}-\mu\|_{\Sigma^{-1}}^2/\nu ) - \log( 1 + \|x_{t,\theta_j}-\mu\|_{\Sigma^{-1}}^2/\nu )| \leq 3 \nu p L_2 (1+M_2) \|\theta-\theta_j\|, \]
by taking expectations over $(1+G_t)^2 \leq 3(1+G_t^2)$.
Take $\varepsilon >0$ and $\|\theta-\theta_j\| \leq \varepsilon$, denote $\ell_{t,\varepsilon} = \log( 1 + \nu p L_2 (1+G_t)^2 \varepsilon )$, then:
\[ \sup_{\psi \in \Psi_n, \|\theta-\theta_j\| \leq \varepsilon} \underbrace{|\overline{\Delta}_n(\psi_j;\theta) - \overline{\Delta}_n(\psi_j;\theta_j)|}_{(B)} \leq |\overline{\ell}_{n,\varepsilon} - \mathbb{E}_P(\ell_{t,\varepsilon})| + 6 \nu p L_2 (1+M_2) \varepsilon. \]
Take $u_1 \geq 1$, we have:
\[ \mathbb{E}_P( \exp[\ell_{t,\varepsilon}/u_1] ) \leq \mathbb{E}_P([ 1+\nu \varepsilon p L_2(1+G_t)^2  ]^{1/u_1}) \leq [ 1+3\nu \varepsilon p L_2(1+M_2)  ]^{1/u_1} \leq 2, \]
if $u_1 = \max \left( 1, \log(1 + \nu \varepsilon p (1+M_4^{1/2})^2) \right)$, using $\mathbb{E}(X^{1/u_1}) \leq \mathbb{E}( X )^{1/u_1}$ for $u_1 \geq 1$ and $X \geq 0$. This implies that the sub-exponential norm of $\ell_{t,\varepsilon}$ is at most $u_1$. Because centering preserves sub-exponentiality, Bernstein's inequality \citep[][Cor2.8.3]{vershynin2018} implies:
\begin{align*}
  \mathbb{P} \left(|\overline{\ell}_{n,\varepsilon} - \mathbb{E}_P(\ell_{t,\varepsilon})| \geq  u_1\sqrt{\frac{t}{n}} + u_1\frac{t}{n}  \right) \leq 2\exp(-C t), 
\end{align*}
for some universal constant $C >0$. From this we deduce that:
\begin{align}
  &\mathbb{P} \left( \sup_{\psi \in \Psi_n, \|\theta-\theta_j\| \leq \varepsilon} |\overline{\Delta}_n(\psi_j;\theta) - \overline{\Delta}_n(\psi_j;\theta_j)| \geq  u_1\sqrt{\frac{t}{n}} + u_1\frac{t}{n} +  6 \nu p L_2 (1+M_2) \varepsilon \right) \notag \\ &\leq 2\exp(-C t). \label{eq:boundB}
\end{align}
\paragraph{3. Bound for $(C)$.} The first step is to show that $(C)$ is a sample average over a centered sub-exponential random variable. By Assumption \ref{as:moments}, $\sup_{\theta \in \Theta} \mathbb{E}_P( \|x_{t,\theta}\|^2) \leq M_2 < \infty$. For any $\theta \in \Theta,\psi \in \Psi_n$: $0 \leq \log(1+\|x_{t,\theta}-\mu\|^2_{\Sigma^{-1}}/\nu) \leq \log(1 + 3s_0^{-1}\|x_{t,\theta}\|^2/\nu) + \log(1+3/2\kappa_1^{-1}\nu(1+p/\nu))$. This inequality implies that for any $u_2 \geq 1$:
\[ \mathbb{E}_P( \exp[ \log(1+\|x_{t,\theta}-\mu\|^2_{\Sigma^{-1}}/\nu)/u_2 ] ) \leq \mathbb{E}_P[1 + 3s_0^{-1}\|x_{t,\theta}\|^2/\nu]^{1/u_2} \exp[ \log(1+3/2\kappa_1^{-1}\nu(1+p/\nu))/u_2 ]. \]
Take $u_2 = \max\left(1,\frac{\log(1+3/2 \kappa_1^{-1}\nu(1+p/\nu))}{1/2\log(2)},\frac{3M_2}{1/2\log(2) s_0 \nu}\right)$. We have $\mathbb{E}_P([1+3s_0^{-1}\|x_t\|^2/\nu]^{1/u_2}) \leq (\mathbb{E}_P[1+3s_0^{-1}\|x_t\|^2/\nu])^{1/u_2} \leq \sqrt{2}$ and $\exp(\log[1+3/2\kappa_1^{-1} \nu (1+p/\nu)]/u) \leq \sqrt{2}$, making the product less than $2$. This implies that the sub-exponential norm of $\log(1+\|x_{t,\theta}-\mu\|^2_{\Sigma^{-1}}/\nu)$ is at most $u_2$ for any $\psi,\theta$. Apply Bernstein's inequality to find:
\begin{align}
  \mathbb{P} \left( |\overline{\Delta}_n(\psi,\theta)| \geq  u_2\sqrt{\frac{t}{n}} + u_2\frac{t}{n}  \right) \leq 2\exp(-C t), \label{eq:boundC}
\end{align}
for the same universal constant $C >0$ as above, and for any $(\psi,\theta) \in \Psi_n \times \Theta$.

\paragraph{4. Overall Bound.} Take $\varepsilon >0$ and $N(\varepsilon)$ denote the smallest $N \geq 1$ such that there exists $(\psi_j,\theta_j) \in \Psi_n \times \Theta$ such that $\sup_{\psi,\theta \in \Psi_n \times \Theta} (\inf_{j=1,\dots,N} [\|\psi-\psi_j\| + \|\theta-\theta_j\|]) \leq \varepsilon$. Using this cover and a union bound, we have:
\begin{align}
  \mathbb{P} \left( \sup_{j=1,\dots,N(\varepsilon)}|\overline{\Delta}_n(\psi_j,\theta_j)| \geq  u_2\sqrt{\frac{t + \log[N(\varepsilon)]}{C n}} + u_2\frac{t + \log[N(\varepsilon)]}{C n}  \right) \leq 2\exp(-t). \tag{\ref{eq:boundC}'}
\end{align}
Take $u = u_1 + u_2$ and combine the bounds to find:
\begin{align}
  &\mathbb{P} \left( \sup_{\theta \in \Theta,\psi \in \Psi_n}|\overline{\Delta}_n(\psi,\theta)| \geq  2u\sqrt{\frac{t}{Cn}} + u\frac{t}{Cn} + u\left[\sqrt{\frac{\log[N(\varepsilon)]}{Cn}}+\frac{\log[N(\varepsilon)]}{Cn}\right]  + L_3 \nu^{12}p^3 \varepsilon  \right) \label{eq:boundABC}\\ &\leq 4\exp(-t). \notag 
\end{align}
Let $k = \text{dim}(\theta)$ and $p = \text{dim}(\mu)$. Lemma \ref{lem:bounds} implies that for some $L_4 >0$ which depends on $\kappa_1,\kappa_2$, we have for any $(\mu,\Sigma)\in\Psi_n$ that $\|\psi\| = \|\mu\| + \|\Sigma\| \leq L_4 p^2 \nu^4$. This yields the following bound $\log[N(\varepsilon)] \leq k \log(3 \text{diam}(\Theta)/\varepsilon) + 2p^2 \log(3 L_4 p^2 \nu^4 / \varepsilon )$. Pick $\varepsilon = \nu^{-12}p^{-2}n^{-1/2}$, then for some constant $L_5 > 0$ which depends on $L_4$ and $\text{diam}(\Theta)$: $\log[N(\varepsilon)] \leq L_5 (k+2p^2)[ \log(p) + \log(\nu) + \log(n) ].$ For the same choice of $\varepsilon$, we have $u \leq \log(1+\nu p)$, up to a constant that depends on $\kappa_1,\kappa_2,M_2,s_0$. This implies for some constant $L > 0$:
\begin{align}
  &\mathbb{P} \left( \sup_{\theta \in \Theta,\psi \in \Psi_n}|\overline{\Delta}_n(\psi,\theta)| \geq  L\log(1+p\nu)\left[  \sqrt{\frac{t}{n}} + \frac{t}{n} + \sqrt{\frac{C_n}{n}}+\frac{C_n}{n} \right]  \right)  \leq 4\exp(-t), \tag{\ref{eq:Concentration1}} 
\end{align}
where $C_n = 1 + (k+2p^2)[ \log(p) + \log(\nu) + \log(n) ]$.
\qed

\paragraph{Proof of Lemma \ref{lem:Concentration1} - 2) Dependent Setting.}
The core of the proof is similar to the iid setting, the main differences occur in the sub-exponential inequalities for $(B)$-$(C)$ in the inequality:
\[ |\overline{\Delta}_n(\psi;\theta)| \leq \underbrace{|\overline{\Delta}_n(\psi;\theta) - \overline{\Delta}_n(\psi_j;\theta)|}_{(A)} + \underbrace{|\overline{\Delta}_n(\psi_j;\theta) - \overline{\Delta}_n(\psi_j;\theta_j)|}_{(B)} + \underbrace{|\overline{\Delta}_n(\psi_j;\theta_j)|}_{(C)}.\]

\paragraph{1. Bound for $(A)$.} Same as iid setting.

\paragraph{2. Bound for $(B)$.} The following relies on a proof reduction technique by \citet{bosq1991}.\footnote{See also \citet{doukhan1994}, \citet{bosq1998}.}  Take an integer $q \geq 1$ and a real number $m \in (0,n)$ such that $m = \frac{n}{2 q}$.
Take $\varepsilon > 0$, $\ell_{t,\varepsilon} = \log(1+\nu p L_2(1+G_t)^2\varepsilon)$ from the iid setting, and, for $t \in [0,n]$, let $\mathcal{L}_{t,\varepsilon} = \ell_{[t+1],\varepsilon}$ be its continuous-time extension. By design, $\overline{\ell}_{n,\varepsilon} = \frac{1}{n}\int_{0}^n \mathcal{L}_{v,\varepsilon}dv$. Let $\mathcal{U}_i = \int_{2(i-1)m}^{(2i-1)m} \mathcal{L}_{v,\varepsilon}dv$, $\mathcal{V}_i = \int_{(2i-1)m}^{{2im}} \mathcal{L}_{v,\varepsilon}dv$ befine non-overlapping blocks; each contains $m$ consecutive discrete-time observations. By construction, $\overline{\ell}_{n,\varepsilon} = \frac{1}{n} \sum_{i=1}^q (\mathcal{U}_i + \mathcal{V}_i)$.

Both $\mathcal{U}_i$ and $\mathcal{V}_i$ are strictly stationary and $\beta$-mixing. Berbee's Lemma \citep[][Lem1.1]{bosq1998} implies that there exists $(\mathcal{U}^*_i,\mathcal{V}^*_i)_{i=1,\dots,q}$ iid such that $(\mathcal{U}^*_i,\mathcal{V}^*_i) \overset{d}{=} (\mathcal{U}_i,\mathcal{V}_i)$ and $\mathbb{P}( \mathcal{U}_i \neq \mathcal{U}^*_i ) \leq \beta_{[m]}$ (likewise for $\mathcal{V}_i,\mathcal{V}^*_i$). 
The next step is to compute the sub-exponential norm of $\mathcal{U}_i$, $\mathcal{V}_i$. For any $i \in \{1,\dots,q\}$ and $\tilde{u}_1 \geq m \geq 1$, Jensen's inequality and Fubini's Theorem imply:
\begin{align*}
  \mathbb{E}_P \left( \exp \left[ \int_{2(i-1)m}^{2im} \mathcal{L}_{v,\varepsilon}dv /\tilde{u}_1 \right] \right) &\leq \left[ \int_{2(i-1)m}^{2im}\mathbb{E}_P \left(  \exp \left[  \mathcal{L}_{v,\varepsilon} m/\tilde{u}_1   \right]  \right)dv  \right] /m,
\end{align*}
which is less than $2$ if the integrand itself is less than $2$ for all $v$. Following the proof in the iid setting, this is true whenever $\tilde{u}_1 \geq m \max\left(1,\log[1+3\nu\varepsilon p (1+M_2)]\right)$. Take $u_1 = \tilde{u}_1/m$ After recentering, Bernstein's inequality applied to the iid sequence $\mathcal{U}_i^*$ yields for the same choice of $u_1$ as the iid setting:
\[ \mathbb{P} \left( |\overline{\mathcal{U}}_i^* - \mathbb{E}_P(\mathcal{U}_i^*)| \geq m u_1 \sqrt{\frac{t}{q}} + m u_1 \frac{t}{q} \right) \leq 2 \exp(-C t), \]
for the same universal constant $C >0$ used in the iid setting, the same holds for $\mathcal{V}_i^*$. To get the bound for $(B)$, we need a tail inequality for $\overline{\ell}_{n,\varepsilon} - \mathbb{E}_P(\ell_{t,\varepsilon}) = \frac{1}{n} \sum_{i=1}^{q} (\mathcal{U}_i + \mathcal{V}_i - \mathbb{E}_P(\mathcal{U}_i) - \mathbb{E}_P(\mathcal{V}_i))$:\footnote{The derivation relies on the inequality: $\mathbb{P}( \sum_{i=1}^{q} (\mathcal{U}_i + \mathcal{V}_i - \mathbb{E}_P(\mathcal{U}_i) - \mathbb{E}_P(\mathcal{V}_i)) \geq 2t ) \leq \mathbb{P}( \sum_{i=1}^{q} (\mathcal{U}_i - \mathbb{E}_P(\mathcal{U}_i)  \geq t ) + \mathbb{P}( \sum_{i=1}^{q} (\mathcal{V}_i - \mathbb{E}_P(\mathcal{U}_i)) \geq t ) = 2 \mathbb{P}( \sum_{i=1}^{q} (\mathcal{U}_i - \mathbb{E}_P(\mathcal{U}_i)  \geq t )$.}
\begin{align*}
  \mathbb{P} \left( \frac{n}{2q}|\overline{\ell}_n,\varepsilon - \mathbb{E}_P(\ell_{t,\varepsilon})| \geq m u_1 \sqrt{\frac{t}{q}} + m u_1 \frac{t}{q} \right) 
  &\leq 2\mathbb{P} \left( |\overline{\mathcal{U}}_i - \mathbb{E}_P(\mathcal{U}_i)| \geq m u_1 \sqrt{\frac{t}{q}} + m u_1 \frac{t}{q} \right)\\
  &\leq 2\mathbb{P} \left( |\overline{\mathcal{U}}^*_i - \mathbb{E}_P(\mathcal{U}^*_i)| \geq m u_1 \sqrt{\frac{t}{q}} + m u_1 \frac{t}{q} \right) + 2q\beta_{[m]}.
\end{align*}
The mixing condition and the definition of $m$ imply that $2q\beta_{[m]} \leq \frac{n a}{m} \exp( - b [m])$. Then, we can re-write for $m \geq 1$:
\begin{align*}
  \mathbb{P} \left( |\overline{\ell}_{n,\varepsilon} - \mathbb{E}_P(\ell_{t,\varepsilon})| \geq 2 u_1 \sqrt{\frac{m t}{n}} + 2 u_1 \frac{ m t}{n} \right) 
  &\leq 4\exp \left(  - Ct \right) + \frac{n a }{\exp(b)}\exp( - b m ) = 6 \exp \left(  - Ct \right),
\end{align*}
for $m = 1 + [C t + \log(an)]/b.$ Note that the effect of $m$ on the tail inequality is comparable to the bounded case found in e.g. \citet[Ch1.4]{doukhan1994}, \citet[Ch6]{rio1999}. Going back to $(B)$ itself, following the same steps from the above inequality to the result yields:
\begin{align}
  &\mathbb{P} \left( \sup_{\psi \in \Psi_n, \|\theta-\theta_j\| \leq \varepsilon} |\overline{\Delta}_n(\psi_j;\theta) - \overline{\Delta}_n(\psi_j;\theta_j)| \geq  u_1\sqrt{\frac{m t}{n}} + u_1\frac{m t}{n} +  6 \nu p L_2 (1+M_2) \varepsilon \right) \notag \\ &\leq 6\exp(-C t), \tag{\ref{eq:boundB}}
\end{align}
where $m$ depends on $t$ and $n$ as stated above. 

\paragraph{3. Bound for $(C)$.} Using the same steps as above, we can take $m = 1 + [C t + \log(an)]/b$ and the same $u_2$ found in the iid setting to get the inequality:
\begin{align}
  &\mathbb{P} \left( |\overline{\Delta}_{n}(\psi;\theta)| \geq  u_2\sqrt{\frac{m t}{n}} + u_2\frac{m t}{n} \right) \leq 6\exp(-C t), \tag{\ref{eq:boundC}}
\end{align}
for the same universal constant $C >0$ and for any $(\psi,\theta) \in \Psi_n \times \Theta$.

\paragraph{4. Overall Bound.} Using the same collection $(\psi_j,\theta_j) \in \Psi_n \times \Theta$ as in the iid case, we have:
\begin{align}
  &\mathbb{P} \left( \sup_{j=1,\dots,N(\varepsilon)} |\overline{\Delta}_{n}(\psi_j;\theta_j)| \geq  u_2\sqrt{\frac{m_2 t + m_2\log[N(\varepsilon)]}{Cn}} + u_2\frac{m_2 t + m_2\log[N(\varepsilon)]}{Cn}  \right) \notag \\ &\leq 6\exp(-t), \tag{\ref{eq:boundC}'}
\end{align}
using $m_2 = 1 + [t + \log[N(\varepsilon)] + \log(an)]/b = m + \log[N(\varepsilon)]/b$.

Take $u = u_1 + u_2$ and combine these bounds:
\begin{align}
  &\mathbb{P} \Bigg( \sup_{\theta \in \Theta,\psi \in \Psi_n}|\overline{\Delta}_n(\psi,\theta)| \geq  2u\sqrt{\frac{(m+m_2)t}{Cn}} + u\frac{(m+m_2)t}{Cn} + u\left[\sqrt{\frac{\log[N(\varepsilon)]}{Cn}}+\frac{\log[N(\varepsilon)]}{Cn}\right] \tag{\ref{eq:boundABC}} \\& + L_3 \nu^{12}p^3 \varepsilon  \Bigg) \leq 12\exp(-t). \notag 
\end{align}
Take $\varepsilon = \nu^{-12}p^{-2}n^{-1/2}$ as in the iid case so that $\log[N(\varepsilon)] \leq L_5 (k+2p^2)[\log(p) + \log(\nu) + \log(n)]$. This implies that $ t \leq (m+m_2)t = t + t^2/b + \log[N(\varepsilon)]t/b + \log(an)t/b \leq t \tilde{L}_5 \left( t + C_n \right)$, for some constant $\tilde{L}_5$ which depends on $a,b$ and $L_5$. As in the iid setting $u \leq \log(1 + \nu p)$, up to a constant and for some constant $\tilde{L} > 0$:
\begin{align}
  &\mathbb{P} \left( \sup_{\theta \in \Theta,\psi \in \Psi_n}|\overline{\Delta}_n(\psi,\theta)| \geq  \tilde{L}\log(1+p\nu)\left[  \sqrt{\frac{(t+C_n)t}{n}} + \frac{(t+C_n)t}{n} + \sqrt{\frac{C_n}{n}}+\frac{C_n}{n} \right]  \right) \tag{\ref{eq:Concentration1}'} \\  &\leq 12\exp(-t), 
\end{align}
where $C_n = 1 + (k+2p^2)[ \log(p) + \log(\nu) + \log(n) ]$. \qed


\section{Leveraged outliers: an illustration} \label{apx:lev_outliers}
\begin{figure}[ht] \caption{Leveraged outlier: asymptotic size for 95\% and 66\% confidence intervals} \label{fig:RR}
  \includegraphics[scale = 0.5]{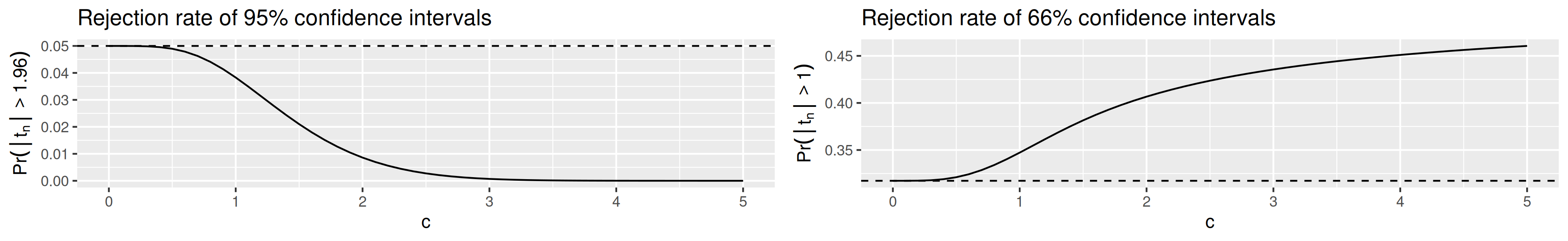}\\
  {\footnotesize \textbf{Note:} Solid line: rejection probability, dashed line: nominal size.}
\end{figure}

Before introducing the estimator, the following illustrates the asymptotic effect of excess leverage. Consider a single regressor linear model:
\[ y_t = \beta_0 + \beta_1 x_t + e_t, \]
for $t=1,\dots,n-1$ where $x_t \sim (0,\sigma_x^2),e_t \sim (0,\sigma_e^2)$ are iid with finite fourth moment. The last observation is $y_n = \beta_0 + (\beta_1 + c) x_n$. Here $c$ measures misspecification, and $x_n$ is such that $x_n^2 = \sqrt{n}\sigma_x^2$. Because of leverage, $(y_n,x_n)$ has some influence asymptotically, $\frac{x_n (y_n - \bar{y}_n)}{\sum_{t=1}^n (x_t - \bar{x}_n)^2} \approx \frac{x_n^2(\beta_1 + c)}{n\sigma_x^2} = \frac{\beta_1 + c}{\sqrt{n}}$, so that the estimator is asympotically biased:
\[ \sqrt{n}(\hat{\beta}_1 - \beta_1) \overset{d}{\to} \mathcal{N}(c,\sigma_e^2/\sigma_x^2), \]
with homoskedastic errors. The outlier further inflates heteroskedasticity-robust standard errors: $\hat{V}_{\hat{\beta}_1} \overset{p}{\to} c^2 + \sigma_e^2/\sigma_x^2$. The misspecification $c$ affects the t-statistic $t_n$ through both estimates and standard errors:
\[ t_n = \frac{\hat{\beta}_1-\beta_1}{\text{se}(\hat{\beta}_1)} \overset{d}{\to} \mathcal{N}\left( \frac{c}{\sqrt{c^2 +\sigma_e^2/\sigma_x^2}}, \frac{1}{\sqrt{c^2 \sigma_x^2 /\sigma_e^2 + 1}} \right). \]
Figure \ref{fig:RR} shows the coverage of 95\% and 66\% confidence intervals when $c$ increases. 

\section{Leverage in IV Regressions} \label{apx:lev_IV}
The following Lemma gives a measure of influence and leverage in just-identified linear instrumental variable regressions. The model is $y_t = x_t^\prime \theta + e_t$, let $\hat{\theta}_n$ be the IV estimates, $\hat{y}_t = x_t^\prime \hat{\theta}_n$ the predicted value and $\tilde{y}_t = x_t^\prime \hat{\theta}_{-t}$ the leave-one-out predicted value. Using standard notation, $Z$, $X$ and $y$ refer the to matrix of instruments, regressors and the vector of outcomes.
\begin{lemma} \label{lem:lev_IV} For each $t$, the difference between the full sample and the leave-one-out predicted value is:
\[ \hat{y}_t - \tilde{y}_t = x_t^\prime (Z^\prime X)^{-1} z_t \tilde{e}_t, \]
where $\tilde{e}_t = y_t -\tilde{y}_t$. Using the terminology from OLS, leverage is given by $h_{t} = x_t^\prime (Z^\prime X)^{-1} z_t$ and influence is $h_t \tilde{e}_t$. Leverage can be positive or negative. Unlike OLS, the sign of influence may not coincide with the sign of the residual $\tilde{e}_t$.
\end{lemma}
\paragraph{Proof of Lemma \ref{lem:lev_IV}.} 
The derivations are similar to OLS. The full sample $\hat{\theta}_n = (Z^\prime X)^{-1} Z^\prime y$, the leave-one-out $\hat{\theta}_{-t} = (Z^\prime X - z_tx_t^\prime)^{-1} (Z^\prime y - z_t y_t)$. Pre-multiply the latter by $(Z^\prime X)^{-1}(Z^\prime X-z_tx_t^\prime)$ to find:
\[ \hat{\theta}_{-t} - (Z^\prime X)^{-1} z_t \tilde{y}_t = \hat{\theta}_n - (Z^\prime X)^{-1}z_t y_t. \]
Re-arrange terms and pre-multiply by $x_t^\prime$ to find:
\[ \hat{y}_t - \tilde{y}_t = \underbrace{x_t^\prime (Z^\prime X)^{-1} z_t \tilde{e}_t}_{\text{Influence}}. \]
For OLS, $h_t = x_t^\prime (X^\prime X)^{-1} x_t \geq 0$, here $h_t = x_t^\prime (Z^\prime X)^{-1} z_t < 0$ can occur.
\qed

\section{Additional Simulation Results} \label{apx:extra_mc}

\begin{table}[H] \caption{Small sample properties of the estimators ($n=150$) -- $\nu = O(n^{1/3})$} \label{tab:mc150b}
  \centering
  \setlength\tabcolsep{4.5pt}
    \renewcommand{\arraystretch}{0.935} 
    {\small
  \begin{tabular}{l|cccaaac|cccaaac}
    \hline \hline
  & \multicolumn{7}{c|}{$100 \times \text{RMSE}$} & \multicolumn{7}{c}{Rejection Rate}\\\hline
  & \multicolumn{14}{c}{$n_o = 0$}\\ \hline
   & $\hat{\theta}_n^{ols}$ & $\hat{\theta}_{n_P}^{ols}$ & $\hat{\theta}_{n}^{rlm}$ & \mc{1}{$\hat{\theta}_{n}$} & \mc{1}{$\tilde{\theta}_{n}$} & \mc{1}{$\vardbtilde{\theta}_{n}$} & $\hat{\theta}_{n}^{un}$ & $\hat{\theta}_n^{ols}$ & $\hat{\theta}_{n_P}^{ols}$ & $\hat{\theta}_{n}^{rlm}$ & \mc{1}{$\hat{\theta}_{n}$} & \mc{1}{$\tilde{\theta}_{n}$} & \mc{1}{$\vardbtilde{\theta}_{n}$} & $\hat{\theta}_{n}^{un}$ \\ 
    \hline
    $\theta_0$ & 8.05 & 8.05 & 12.00 & 11.84 & 9.31 & 8.11 & 7.94 & 0.04 & 0.04 & 0.24 & 0.29 & 0.13 & 0.05 & 0.06 \\ 
    $\theta_1$ & 8.00 & 8.00 & 7.15 & 7.96 & 7.78 & 7.78 & 7.92 & 0.06 & 0.06 & 0.06 & 0.11 & 0.08 & 0.07 & 0.06 \\ 
    $\theta_2$ & 8.10 & 8.10 & 7.46 & 8.44 & 8.20 & 8.10 & 8.06 & 0.04 & 0.04 & 0.05 & 0.10 & 0.06 & 0.05 & 0.05 \\ 
    $\theta_3$ & 8.19 & 8.19 & 7.43 & 8.55 & 8.30 & 8.15 & 8.15 & 0.06 & 0.06 & 0.06 & 0.10 & 0.07 & 0.06 & 0.06 \\ \hline
    & \multicolumn{14}{c}{$n_o = 1$}\\ \hline
    & $\hat{\theta}_n^{ols}$ & $\hat{\theta}_{n_P}^{ols}$ & $\hat{\theta}_{n}^{rlm}$ & \mc{1}{$\hat{\theta}_{n}$} & \mc{1}{$\tilde{\theta}_{n}$} & \mc{1}{$\vardbtilde{\theta}_{n}$} & $\hat{\theta}_{n}^{un}$ & $\hat{\theta}_n^{ols}$ & $\hat{\theta}_{n_P}^{ols}$ & $\hat{\theta}_{n}^{rlm}$ & \mc{1}{$\hat{\theta}_{n}$} & \mc{1}{$\tilde{\theta}_{n}$} & \mc{1}{$\vardbtilde{\theta}_{n}$} & $\hat{\theta}_{n}^{un}$ \\ 
    \hline
    $\theta_0$ & 10.71 & 8.04 & 13.01 & 14.17 & 10.95 & 8.52 & 10.32 & 0.03 & 0.04 & 0.20 & 0.46 & 0.23 & 0.08 & 0.08 \\ 
    $\theta_1$ & 38.57 & 8.07 & 15.23 & 8.27 & 7.97 & 7.87 & 32.28 & 0.00 & 0.06 & 0.01 & 0.14 & 0.10 & 0.07 & 0.39 \\ 
    $\theta_2$ & 38.39 & 8.11 & 15.09 & 8.73 & 8.36 & 8.13 & 32.12 & 0.01 & 0.04 & 0.01 & 0.12 & 0.06 & 0.06 & 0.37 \\ 
    $\theta_3$ & 39.94 & 8.20 & 15.75 & 8.82 & 8.49 & 8.26 & 33.52 & 0.00 & 0.06 & 0.00 & 0.12 & 0.09 & 0.07 & 0.39 \\ 
   \hline
   & \multicolumn{14}{c}{$n_o = 5$}\\ \hline
    & $\hat{\theta}_n^{ols}$ & $\hat{\theta}_{n_P}^{ols}$ & $\hat{\theta}_{n}^{rlm}$ & \mc{1}{$\hat{\theta}_{n}$} & \mc{1}{$\tilde{\theta}_{n}$} & \mc{1}{$\vardbtilde{\theta}_{n}$} & $\hat{\theta}_{n}^{un}$ & $\hat{\theta}_n^{ols}$ & $\hat{\theta}_{n_P}^{ols}$ & $\hat{\theta}_{n}^{rlm}$ & \mc{1}{$\hat{\theta}_{n}$} & \mc{1}{$\tilde{\theta}_{n}$} & \mc{1}{$\vardbtilde{\theta}_{n}$} & $\hat{\theta}_{n}^{un}$ \\ 
    \hline
    $\theta_0$ & 11.98 & 8.14 & 16.57 & 16.96 & 13.36 & 9.79 & 13.44 & 0.10 & 0.04 & 0.24 & 0.59 & 0.38 & 0.13 & 0.16 \\ 
    $\theta_1$ & 47.57 & 8.40 & 47.17 & 9.03 & 8.63 & 8.41 & 46.72 & 0.99 & 0.06 & 0.99 & 0.12 & 0.08 & 0.06 & 0.99 \\ 
    $\theta_2$ & 47.48 & 8.26 & 48.25 & 9.26 & 8.78 & 8.51 & 47.14 & 0.99 & 0.04 & 1.00 & 0.11 & 0.04 & 0.03 & 1.00 \\ 
    $\theta_3$ & 49.17 & 8.28 & 49.48 & 9.34 & 8.95 & 8.72 & 48.64 & 0.98 & 0.06 & 0.98 & 0.10 & 0.08 & 0.04 & 0.98 \\ 
   \hline
   & \multicolumn{14}{c}{$n_o = 10$}\\ \hline
    & $\hat{\theta}_n^{ols}$ & $\hat{\theta}_{n_P}^{ols}$ & $\hat{\theta}_{n}^{rlm}$ & \mc{1}{$\hat{\theta}_{n}$} & \mc{1}{$\tilde{\theta}_{n}$} & \mc{1}{$\vardbtilde{\theta}_{n}$} & $\hat{\theta}_{n}^{un}$ & $\hat{\theta}_n^{ols}$ & $\hat{\theta}_{n_P}^{ols}$ & $\hat{\theta}_{n}^{rlm}$ & \mc{1}{$\hat{\theta}_{n}$} & \mc{1}{$\tilde{\theta}_{n}$} & \mc{1}{$\vardbtilde{\theta}_{n}$} & $\hat{\theta}_{n}^{un}$ \\ 
    \hline
    $\theta_0$ & 12.21 & 8.21 & 17.33 & 16.84 & 13.37 & 10.98 & 14.11 & 0.09 & 0.04 & 0.23 & 0.47 & 0.22 & 0.07 & 0.17 \\ 
    $\theta_1$ & 49.14 & 8.54 & 48.38 & 10.45 & 12.31 & 23.20 & 48.65 & 0.99 & 0.04 & 0.99 & 0.01 & 0.01 & 0.16 & 1.00 \\ 
    $\theta_2$ & 49.05 & 8.31 & 49.67 & 11.02 & 13.09 & 24.58 & 48.92 & 0.99 & 0.04 & 0.99 & 0.01 & 0.01 & 0.16 & 1.00 \\ 
    $\theta_3$ & 50.52 & 8.51 & 50.70 & 11.32 & 13.68 & 24.91 & 50.19 & 0.98 & 0.06 & 0.98 & 0.00 & 0.01 & 0.16 & 0.99 \\ 
     \hline \hline
  \end{tabular}\\
  \notes{ \textbf{Legend:} $\hat{\theta}_n^{ols}$ full sample OLS, $\hat{\theta}_{n_P}^{ols}$ oracle OLS, $\hat{\theta}_n^{rlm}$ robust M-estimator, $\hat{\theta}_{n}$ robust estimates without bias correction, $\tilde{\theta}_{n}$ robust estimates with bias correction, $\vardbtilde{\theta}_{n}$ robust estimates with repeated bias correction, $\hat{\theta}^{un}_{n}$ undersmoothed robust estimates with $\hat{\nu}_n^2$. 200 Monte-Carlo replications. $n_o = $ number of outliers. Rejection rate for t-test at the $5\%$ significance level. Average $\hat{\nu}_n$: $32.8$, $18.0$, $11.0$, $10.8$, $10.8$ for $n_0 = 0$, $1$, $5$, $10$, $20$ respectively. Each $\hat{\nu}_n$ is selected on a grid $[\nu_0,\dots,\nu_J]$ where $\nu_0 = 8.82$, $\nu_J = 177.16$.
    } }
\end{table}

\begin{table}[H] \caption{Small sample properties of the estimators ($n=500$) -- $\nu = O(n^{1/3})$} \label{tab:mc500b}
  \centering
  \setlength\tabcolsep{4.5pt}
    \renewcommand{\arraystretch}{0.935} 
    {\small
  \begin{tabular}{l|cccaaac|cccaaac}
    \hline \hline
  & \multicolumn{7}{c|}{$100 \times \text{RMSE}$} & \multicolumn{7}{c}{Rejection Rate}\\\hline
  & \multicolumn{14}{c}{$n_o = 0$}\\ \hline
   & $\hat{\theta}_n^{ols}$ & $\hat{\theta}_{n_P}^{ols}$ & $\hat{\theta}_{n}^{rlm}$ & \mc{1}{$\hat{\theta}_{n}$} & \mc{1}{$\tilde{\theta}_{n}$} & \mc{1}{$\vardbtilde{\theta}_{n}$} & $\hat{\theta}_{n}^{un}$ & $\hat{\theta}_n^{ols}$ & $\hat{\theta}_{n_P}^{ols}$ & $\hat{\theta}_{n}^{rlm}$ & \mc{1}{$\hat{\theta}_{n}$} & \mc{1}{$\tilde{\theta}_{n}$} & \mc{1}{$\vardbtilde{\theta}_{n}$} & $\hat{\theta}_{n}^{un}$ \\ 
    \hline
    $\theta_0$ & 4.59 & 4.59 & 10.67 & 9.29 & 6.29 & 4.70 & 4.56 & 0.07 & 0.07 & 0.65 & 0.51 & 0.21 & 0.07 & 0.07 \\ 
    $\theta_1$ & 4.21 & 4.21 & 3.93 & 4.57 & 4.50 & 4.44 & 4.21 & 0.04 & 0.04 & 0.05 & 0.09 & 0.07 & 0.07 & 0.04 \\ 
    $\theta_2$ & 4.76 & 4.76 & 4.21 & 4.65 & 4.61 & 4.60 & 4.72 & 0.06 & 0.06 & 0.07 & 0.09 & 0.09 & 0.07 & 0.07 \\ 
    $\theta_3$ & 4.51 & 4.51 & 4.09 & 4.66 & 4.56 & 4.52 & 4.48 & 0.09 & 0.09 & 0.07 & 0.13 & 0.10 & 0.09 & 0.09 \\ \hline
    & \multicolumn{14}{c}{$n_o = 1$}\\ \hline
    & $\hat{\theta}_n^{ols}$ & $\hat{\theta}_{n_P}^{ols}$ & $\hat{\theta}_{n}^{rlm}$ & \mc{1}{$\hat{\theta}_{n}$} & \mc{1}{$\tilde{\theta}_{n}$} & \mc{1}{$\vardbtilde{\theta}_{n}$} & $\hat{\theta}_{n}^{un}$ & $\hat{\theta}_n^{ols}$ & $\hat{\theta}_{n_P}^{ols}$ & $\hat{\theta}_{n}^{rlm}$ & \mc{1}{$\hat{\theta}_{n}$} & \mc{1}{$\tilde{\theta}_{n}$} & \mc{1}{$\vardbtilde{\theta}_{n}$} & $\hat{\theta}_{n}^{un}$ \\ 
    \hline
    $\theta_0$ & 5.40 & 4.58 & 10.87 & 10.92 & 7.41 & 4.87 & 4.88 & 0.03 & 0.07 & 0.64 & 0.67 & 0.30 & 0.09 & 0.04 \\ 
    $\theta_1$ & 38.16 & 4.22 & 7.98 & 4.67 & 4.60 & 4.56 & 22.53 & 0.00 & 0.04 & 0.01 & 0.10 & 0.07 & 0.07 & 0.02 \\ 
    $\theta_2$ & 38.00 & 4.77 & 7.95 & 4.74 & 4.68 & 4.67 & 22.66 & 0.00 & 0.07 & 0.00 & 0.10 & 0.09 & 0.09 & 0.04 \\ 
    $\theta_3$ & 37.38 & 4.50 & 7.42 & 4.73 & 4.59 & 4.52 & 21.92 & 0.00 & 0.08 & 0.01 & 0.13 & 0.09 & 0.08 & 0.03 \\ 
   \hline
   & \multicolumn{14}{c}{$n_o = 5$}\\ \hline
    & $\hat{\theta}_n^{ols}$ & $\hat{\theta}_{n_P}^{ols}$ & $\hat{\theta}_{n}^{rlm}$ & \mc{1}{$\hat{\theta}_{n}$} & \mc{1}{$\tilde{\theta}_{n}$} & \mc{1}{$\vardbtilde{\theta}_{n}$} & $\hat{\theta}_{n}^{un}$ & $\hat{\theta}_n^{ols}$ & $\hat{\theta}_{n_P}^{ols}$ & $\hat{\theta}_{n}^{rlm}$ & \mc{1}{$\hat{\theta}_{n}$} & \mc{1}{$\tilde{\theta}_{n}$} & \mc{1}{$\vardbtilde{\theta}_{n}$} & $\hat{\theta}_{n}^{un}$ \\ 
    \hline
    $\theta_0$ & 5.90 & 4.60 & 11.52 & 13.86 & 9.89 & 5.89 & 6.60 & 0.07 & 0.06 & 0.30 & 0.91 & 0.56 & 0.17 & 0.09 \\ 
    $\theta_1$ & 47.49 & 4.20 & 45.53 & 4.84 & 4.77 & 4.80 & 46.01 & 1.00 & 0.04 & 0.47 & 0.11 & 0.09 & 0.07 & 1.00 \\ 
    $\theta_2$ & 47.41 & 4.82 & 45.67 & 4.93 & 4.84 & 4.84 & 45.96 & 1.00 & 0.07 & 0.46 & 0.12 & 0.10 & 0.08 & 1.00 \\ 
    $\theta_3$ & 46.66 & 4.51 & 44.65 & 4.94 & 4.75 & 4.67 & 45.21 & 1.00 & 0.07 & 0.46 & 0.15 & 0.10 & 0.08 & 1.00 \\ 
   \hline
   & \multicolumn{14}{c}{$n_o = 10$}\\ \hline
    & $\hat{\theta}_n^{ols}$ & $\hat{\theta}_{n_P}^{ols}$ & $\hat{\theta}_{n}^{rlm}$ & \mc{1}{$\hat{\theta}_{n}$} & \mc{1}{$\tilde{\theta}_{n}$} & \mc{1}{$\vardbtilde{\theta}_{n}$} & $\hat{\theta}_{n}^{un}$ & $\hat{\theta}_n^{ols}$ & $\hat{\theta}_{n_P}^{ols}$ & $\hat{\theta}_{n}^{rlm}$ & \mc{1}{$\hat{\theta}_{n}$} & \mc{1}{$\tilde{\theta}_{n}$} & \mc{1}{$\vardbtilde{\theta}_{n}$} & $\hat{\theta}_{n}^{un}$ \\ 
    \hline
    $\theta_0$ & 5.95 & 4.60 & 11.69 & 15.11 & 11.06 & 6.62 & 7.43 & 0.07 & 0.05 & 0.42 & 0.94 & 0.71 & 0.23 & 0.13 \\ 
    $\theta_1$ & 48.95 & 4.19 & 49.02 & 4.89 & 4.81 & 4.86 & 48.44 & 1.00 & 0.03 & 1.00 & 0.12 & 0.07 & 0.04 & 1.00 \\ 
    $\theta_2$ & 48.98 & 4.85 & 49.27 & 5.05 & 4.95 & 4.96 & 48.58 & 1.00 & 0.07 & 1.00 & 0.11 & 0.10 & 0.08 & 1.00 \\ 
    $\theta_3$ & 48.15 & 4.56 & 48.22 & 5.12 & 4.89 & 4.80 & 47.69 & 1.00 & 0.07 & 1.00 & 0.16 & 0.10 & 0.07 & 1.00 \\ 
    \hline
   & \multicolumn{14}{c}{$n_o = 20$}\\ \hline
    & $\hat{\theta}_n^{ols}$ & $\hat{\theta}_{n_P}^{ols}$ & $\hat{\theta}_{n}^{rlm}$ & \mc{1}{$\hat{\theta}_{n}$} & \mc{1}{$\tilde{\theta}_{n}$} & \mc{1}{$\vardbtilde{\theta}_{n}$} & $\hat{\theta}_{n}^{un}$ & $\hat{\theta}_n^{ols}$ & $\hat{\theta}_{n_P}^{ols}$ & $\hat{\theta}_{n}^{rlm}$ & \mc{1}{$\hat{\theta}_{n}$} & \mc{1}{$\tilde{\theta}_{n}$} & \mc{1}{$\vardbtilde{\theta}_{n}$} & $\hat{\theta}_{n}^{un}$ \\ 
    \hline
    $\theta_0$ & 6.06 & 4.61 & 12.26 & 16.15 & 12.03 & 7.30 & 15.63 & 0.06 & 0.05 & 0.45 & 0.95 & 0.78 & 0.27 & 0.79 \\ 
    $\theta_1$ & 49.71 & 4.24 & 49.63 & 4.97 & 4.88 & 4.97 & 49.20 & 1.00 & 0.04 & 1.00 & 0.10 & 0.05 & 0.04 & 1.00 \\ 
    $\theta_2$ & 49.92 & 4.96 & 50.07 & 5.25 & 5.13 & 5.15 & 49.69 & 1.00 & 0.07 & 1.00 & 0.10 & 0.04 & 0.02 & 1.00 \\ 
    $\theta_3$ & 48.85 & 4.56 & 48.78 & 5.17 & 4.89 & 4.79 & 48.55 & 1.00 & 0.06 & 1.00 & 0.14 & 0.06 & 0.02 & 1.00 \\ 
     \hline \hline
  \end{tabular}\\
  \notes{ \textbf{Legend:} $\hat{\theta}_n^{ols}$ full sample OLS, $\hat{\theta}_{n_P}^{ols}$ oracle OLS, $\hat{\theta}_n^{rlm}$ robust M-estimator, $\hat{\theta}_{n}$ robust estimates without bias correction, $\tilde{\theta}_{n}$ robust estimates with bias correction, $\vardbtilde{\theta}_{n}$ robust estimates with repeated bias correction, $\hat{\theta}^{un}_{n}$ undersmoothed robust estimates with $\hat{\nu}_n^2$. 200 Monte-Carlo replications. $n_o = $ number of outliers. Rejection rate for t-test at the $5\%$ significance level. Average $\hat{\nu}_n$: $37.46$, $26.39$, $16.09$, $13.18$, $10.79$ for $n_0 = 0$, $1$, $5$, $10$, $20$ respectively. Each $\hat{\nu}_n$ is selected on a grid $[\nu_0,\dots,\nu_J]$ where $\nu_0 = 8.83$, $\nu_J = 264.64$.
    } }
\end{table}

\begin{table}[H] \caption{Small sample properties of the estimators ($n=500$), with $\nu = O(n^{1/4} \log(n))$} \label{tab:mc500}
  \centering
  \setlength\tabcolsep{4.5pt}
    \renewcommand{\arraystretch}{0.935} 
    {\small
  \begin{tabular}{l|cccaaac|cccaaac}
    \hline \hline
  & \multicolumn{7}{c|}{$100 \times \text{RMSE}$} & \multicolumn{7}{c}{Rejection Rate}\\\hline
  & \multicolumn{14}{c}{$n_o = 0$}\\ \hline
   & $\hat{\theta}_n^{ols}$ & $\hat{\theta}_{n_P}^{ols}$ & $\hat{\theta}_{n}^{rlm}$ & \mc{1}{$\hat{\theta}_{n}$} & \mc{1}{$\tilde{\theta}_{n}$} & \mc{1}{$\vardbtilde{\theta}_{n}$} & $\hat{\theta}_{n}^{un}$ & $\hat{\theta}_n^{ols}$ & $\hat{\theta}_{n_P}^{ols}$ & $\hat{\theta}_{n}^{rlm}$ & \mc{1}{$\hat{\theta}_{n}$} & \mc{1}{$\tilde{\theta}_{n}$} & \mc{1}{$\vardbtilde{\theta}_{n}$} & $\hat{\theta}_{n}^{un}$ \\ 
    \hline
    $\theta_0$ & 4.59 & 4.59 & 10.67 & 8.53 & 5.83 & 4.67 & 4.57 & 0.07 & 0.07 & 0.65 & 0.42 & 0.18 & 0.07 & 0.07 \\ 
    $\theta_1$ & 4.21 & 4.21 & 3.93 & 4.52 & 4.45 & 4.39 & 4.21 & 0.04 & 0.04 & 0.05 & 0.08 & 0.07 & 0.07 & 0.04 \\ 
    $\theta_2$ & 4.76 & 4.76 & 4.21 & 4.63 & 4.61 & 4.62 & 4.73 & 0.06 & 0.06 & 0.07 & 0.09 & 0.08 & 0.07 & 0.07 \\ 
    $\theta_3$ & 4.51 & 4.51 & 4.09 & 4.61 & 4.52 & 4.49 & 4.48 & 0.09 & 0.09 & 0.07 & 0.12 & 0.10 & 0.09 & 0.09 \\ \hline
    & \multicolumn{14}{c}{$n_o = 1$}\\ \hline
    & $\hat{\theta}_n^{ols}$ & $\hat{\theta}_{n_P}^{ols}$ & $\hat{\theta}_{n}^{rlm}$ & \mc{1}{$\hat{\theta}_{n}$} & \mc{1}{$\tilde{\theta}_{n}$} & \mc{1}{$\vardbtilde{\theta}_{n}$} & $\hat{\theta}_{n}^{un}$ & $\hat{\theta}_n^{ols}$ & $\hat{\theta}_{n_P}^{ols}$ & $\hat{\theta}_{n}^{rlm}$ & \mc{1}{$\hat{\theta}_{n}$} & \mc{1}{$\tilde{\theta}_{n}$} & \mc{1}{$\vardbtilde{\theta}_{n}$} & $\hat{\theta}_{n}^{un}$ \\ 
    \hline
    $\theta_0$ & 5.40 & 4.58 & 10.87 & 10.30 & 6.93 & 4.75 & 5.05 & 0.03 & 0.07 & 0.64 & 0.62 & 0.27 & 0.07 & 0.04 \\ 
    $\theta_1$ & 38.16 & 4.22 & 7.98 & 4.64 & 4.58 & 4.52 & 27.54 & 0.00 & 0.04 & 0.01 & 0.09 & 0.07 & 0.06 & 0.14 \\ 
    $\theta_2$ & 38.00 & 4.77 & 7.95 & 4.70 & 4.65 & 4.64 & 27.44 & 0.00 & 0.07 & 0.00 & 0.09 & 0.09 & 0.07 & 0.17 \\ 
    $\theta_3$ & 37.38 & 4.50 & 7.42 & 4.69 & 4.57 & 4.51 & 26.86 & 0.00 & 0.08 & 0.01 & 0.12 & 0.09 & 0.07 & 0.14 \\ 
   \hline
   & \multicolumn{14}{c}{$n_o = 5$}\\ \hline
    & $\hat{\theta}_n^{ols}$ & $\hat{\theta}_{n_P}^{ols}$ & $\hat{\theta}_{n}^{rlm}$ & \mc{1}{$\hat{\theta}_{n}$} & \mc{1}{$\tilde{\theta}_{n}$} & \mc{1}{$\vardbtilde{\theta}_{n}$} & $\hat{\theta}_{n}^{un}$ & $\hat{\theta}_n^{ols}$ & $\hat{\theta}_{n_P}^{ols}$ & $\hat{\theta}_{n}^{rlm}$ & \mc{1}{$\hat{\theta}_{n}$} & \mc{1}{$\tilde{\theta}_{n}$} & \mc{1}{$\vardbtilde{\theta}_{n}$} & $\hat{\theta}_{n}^{un}$ \\ 
    \hline
    $\theta_0$ & 5.90 & 4.60 & 11.52 & 12.93 & 9.02 & 5.41 & 6.44 & 0.07 & 0.06 & 0.30 & 0.89 & 0.49 & 0.14 & 0.09 \\ 
    $\theta_1$ & 47.49 & 4.20 & 45.53 & 4.78 & 4.72 & 4.72 & 46.45 & 1.00 & 0.04 & 0.47 & 0.10 & 0.09 & 0.06 & 1.00 \\ 
    $\theta_2$ & 47.41 & 4.82 & 45.67 & 4.89 & 4.81 & 4.80 & 46.41 & 1.00 & 0.07 & 0.46 & 0.10 & 0.09 & 0.06 & 1.00 \\ 
    $\theta_3$ & 46.66 & 4.51 & 44.65 & 4.87 & 4.70 & 4.62 & 45.64 & 1.00 & 0.07 & 0.46 & 0.14 & 0.10 & 0.07 & 1.00 \\ 
   \hline
   & \multicolumn{14}{c}{$n_o = 10$}\\ \hline
    & $\hat{\theta}_n^{ols}$ & $\hat{\theta}_{n_P}^{ols}$ & $\hat{\theta}_{n}^{rlm}$ & \mc{1}{$\hat{\theta}_{n}$} & \mc{1}{$\tilde{\theta}_{n}$} & \mc{1}{$\vardbtilde{\theta}_{n}$} & $\hat{\theta}_{n}^{un}$ & $\hat{\theta}_n^{ols}$ & $\hat{\theta}_{n_P}^{ols}$ & $\hat{\theta}_{n}^{rlm}$ & \mc{1}{$\hat{\theta}_{n}$} & \mc{1}{$\tilde{\theta}_{n}$} & \mc{1}{$\vardbtilde{\theta}_{n}$} & $\hat{\theta}_{n}^{un}$ \\ 
    \hline
    $\theta_0$ & 5.95 & 4.60 & 11.69 & 12.57 & 8.59 & 5.10 & 7.01 & 0.07 & 0.05 & 0.42 & 0.86 & 0.43 & 0.09 & 0.10 \\ 
    $\theta_1$ & 48.95 & 4.19 & 49.02 & 4.74 & 4.68 & 4.69 & 48.66 & 1.00 & 0.03 & 1.00 & 0.07 & 0.04 & 0.03 & 1.00 \\ 
    $\theta_2$ & 48.98 & 4.85 & 49.27 & 4.90 & 4.83 & 4.84 & 48.81 & 1.00 & 0.07 & 1.00 & 0.09 & 0.07 & 0.04 & 1.00 \\ 
    $\theta_3$ & 48.15 & 4.56 & 48.22 & 4.91 & 4.73 & 4.64 & 47.91 & 1.00 & 0.07 & 1.00 & 0.12 & 0.07 & 0.03 & 1.00 \\ 
    \hline
   & \multicolumn{14}{c}{$n_o = 20$}\\ \hline
    & $\hat{\theta}_n^{ols}$ & $\hat{\theta}_{n_P}^{ols}$ & $\hat{\theta}_{n}^{rlm}$ & \mc{1}{$\hat{\theta}_{n}$} & \mc{1}{$\tilde{\theta}_{n}$} & \mc{1}{$\vardbtilde{\theta}_{n}$} & $\hat{\theta}_{n}^{un}$ & $\hat{\theta}_n^{ols}$ & $\hat{\theta}_{n_P}^{ols}$ & $\hat{\theta}_{n}^{rlm}$ & \mc{1}{$\hat{\theta}_{n}$} & \mc{1}{$\tilde{\theta}_{n}$} & \mc{1}{$\vardbtilde{\theta}_{n}$} & $\hat{\theta}_{n}^{un}$ \\ 
    \hline
    $\theta_0$ & 6.06 & 4.61 & 12.26 & 13.32 & 9.14 & 5.22 & 15.91 & 0.06 & 0.05 & 0.45 & 0.89 & 0.47 & 0.10 & 0.78 \\ 
    $\theta_1$ & 49.71 & 4.24 & 49.63 & 4.80 & 4.77 & 4.86 & 49.23 & 1.00 & 0.04 & 1.00 & 0.03 & 0.02 & 0.01 & 1.00 \\ 
    $\theta_2$ & 49.92 & 4.96 & 50.07 & 5.09 & 5.02 & 5.09 & 49.71 & 1.00 & 0.07 & 1.00 & 0.03 & 0.02 & 0.00 & 1.00 \\ 
    $\theta_3$ & 48.85 & 4.56 & 48.78 & 4.91 & 4.71 & 4.67 & 48.58 & 1.00 & 0.06 & 1.00 & 0.03 & 0.01 & 0.00 & 1.00 \\ 
     \hline \hline
  \end{tabular}\\
  \notes{ \textbf{Legend:} $\hat{\theta}_n^{ols}$ full sample OLS, $\hat{\theta}_{n_P}^{ols}$ oracle OLS, $\hat{\theta}_n^{rlm}$ robust M-estimator, $\hat{\theta}_{n}$ robust estimates without bias correction, $\tilde{\theta}_{n}$ robust estimates with bias correction, $\vardbtilde{\theta}_{n}$ robust estimates with repeated bias correction, $\hat{\theta}^{un}_{n}$ undersmoothed robust estimates with $\hat{\nu}_n^2$. 200 Monte-Carlo replications. $n_o = $ number of outliers. Rejection rate for t-test at the $5\%$ significance level. Average $\hat{\nu}_n$: $48.78$, $28.88$, $18.34$, $17.95$, $14.69$ for $n_0 = 0$, $1$, $5$, $10$, $20$  respectively. Each $\hat{\nu}_n$ is selected on a grid $[\nu_0,\dots,\nu_J]$ where $\nu_0 = 14.69$, $\nu_J = 979.86$.
    } }
\end{table}

\section{Additional Empirical Results} \label{apx:extra_emp}

\subsection{ Additional Results for the Price Puzzle }

\begin{table}[H] \caption{Regression (\ref{eq:VAR1}): contribution to each coefficient (moments)} \label{tab:VAR1}
  \centering
  \setlength\tabcolsep{4.5pt}
    \renewcommand{\arraystretch}{0.935} 
    {\small
  \begin{tabular}{l|rrrrrrrrrrrrr}
    \hline \hline
   & $\hat{\beta}_0$ & $\hat{\beta}_1$ & $\hat{\beta}_2$ & $\hat{\beta}_3$ & $\hat{\beta}_4$ & $\hat{\beta}_5$ & $\hat{\beta}_6$ & $\hat{\beta}_7$ & $\hat{\beta}_8$ & $\hat{\beta}_9$ & $\hat{\beta}_{10}$ & $\hat{\beta}_{11}$ & $\hat{\beta}_{12}$ \\ 
    \hline
  skewness & -0.56 & 3.24 & -0.30 & 1.30 & -2.99 & 0.95 & 0.74 & -1.48 & -0.78 & 0.52 & -2.93 & -0.24 & 0.54 \\ 
  kurtosis & 4.42 & 27.81 & 8.98 & 7.88 & 36.70 & 9.77 & 8.41 & 27.78 & 6.95 & 7.76 & 32.66 & 9.29 & 7.24 \\ 
     \hline \hline
  \end{tabular} }
  \end{table}

\begin{figure}[H] \caption{Recursive VAR: OLS, Robust and Bias-Corrected Estimates ($\nu=10$)} \label{fig:VAR2b}
  \includegraphics[scale = 0.5]{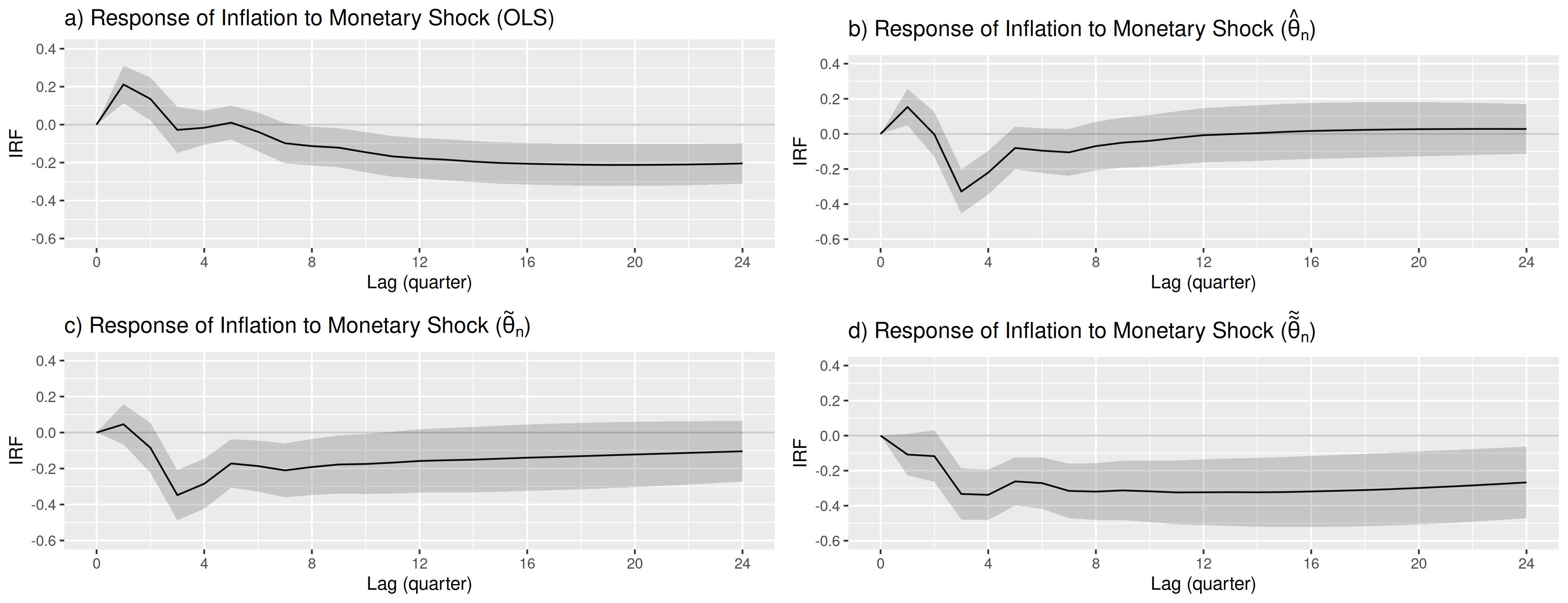}\\
  {\footnotesize \textbf{Note:} a) OLS estimates, b) $\hat{\theta}_{n}$ robust estimates without bias correction, c) $\tilde{\theta}_{n}$ robust estimates with bias correction, d) $\vardbtilde{\theta}_{n}$ robust estimates with repeated bias correction. Bands: estimates $\pm$ one standard error.}
\end{figure}

\begin{figure}[H] \caption{Recursive VAR: OLS, Robust and Bias-Corrected Estimates ($\nu=15$)} \label{fig:VAR2bb}
  \includegraphics[scale = 0.5]{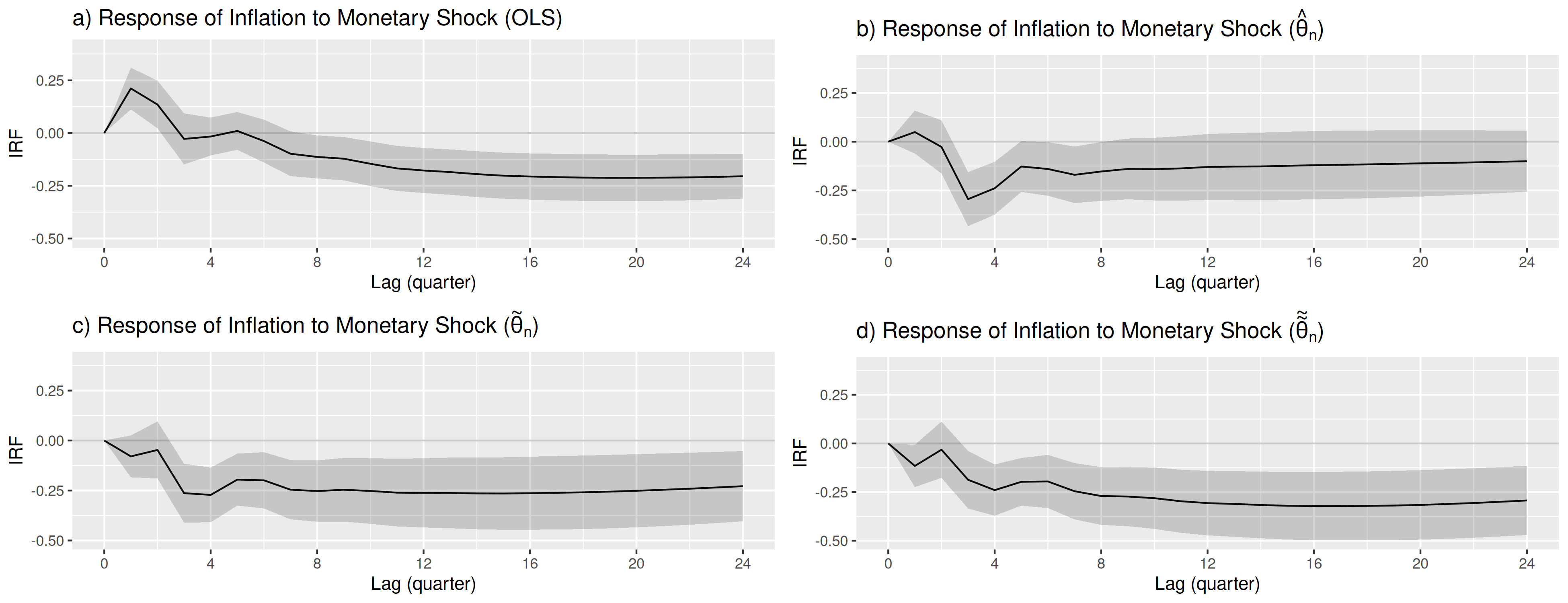}\\
  {\footnotesize \textbf{Note:} a) OLS estimates, b) $\hat{\theta}_{n}$ robust estimates without bias correction, c) $\tilde{\theta}_{n}$ robust estimates with bias correction, d) $\vardbtilde{\theta}_{n}$ robust estimates with repeated bias correction. Bands: estimates $\pm$ one standard error.}
\end{figure}

\begin{figure}[H] \caption{Recursive VAR: OLS, Robust and Bias-Corrected Estimates ($\nu=20$)} \label{fig:VAR2bbb}
  \includegraphics[scale = 0.5]{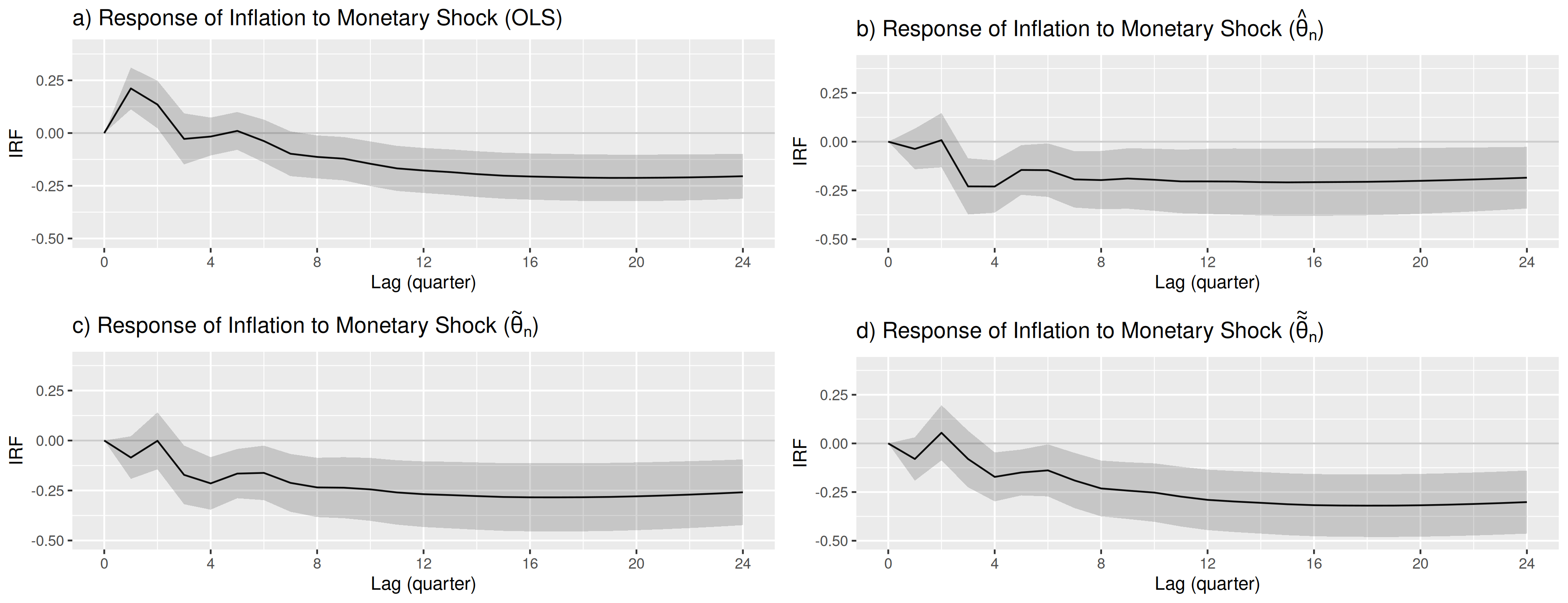}\\
  {\footnotesize \textbf{Note:} a) OLS estimates, b) $\hat{\theta}_{n}$ robust estimates without bias correction, c) $\tilde{\theta}_{n}$ robust estimates with bias correction, d) $\vardbtilde{\theta}_{n}$ robust estimates with repeated bias correction. Bands: estimates $\pm$ one standard error.}
\end{figure}

\begin{figure}[H] \caption{Recursive VAR, Estimation Weights: OLS, Robust, and Bias-Corrected Estimates ($\nu=10$) } \label{fig:VARwb}
  \includegraphics[scale = 0.5]{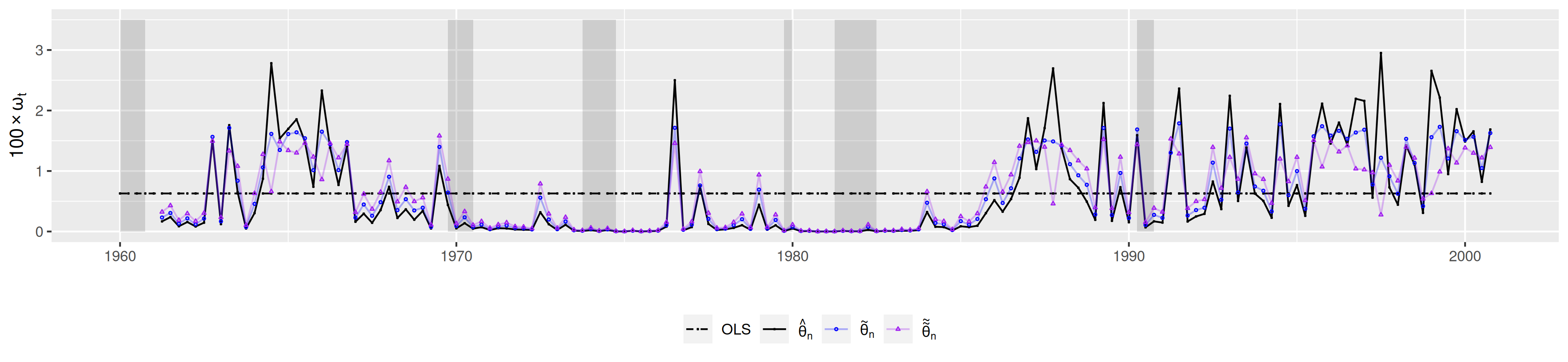}\\
  {\footnotesize \textbf{Note:} Estimation weights $\omega_t$ implicitely used to estimate $\theta$. OLS (dashed/black): $\omega_t = 1/n$. Robust estimates $\hat{\theta}_n$ (solid/black). Bias-corrected robust estimates $\tilde{\theta}_n$ (solid/circle/blue).  Repeated bias-corrected robust estimates $\vardbtilde{\theta}_n$ (solid/triangle/purple). Shaded vertical bars = NBER recession dates.}
\end{figure}

\begin{figure}[H] \caption{Recursive VAR, Estimation Weights: OLS, Robust, and Bias-Corrected Estimates ($\nu=15$) } \label{fig:VARwbb}
  \includegraphics[scale = 0.5]{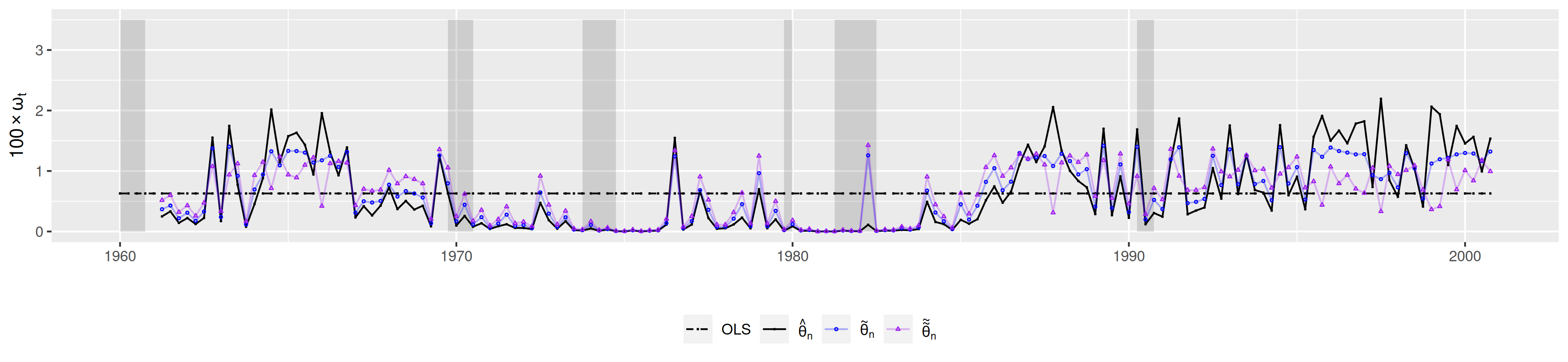}\\
  {\footnotesize \textbf{Note:} Estimation weights $\omega_t$ implicitely used to estimate $\theta$. OLS (dashed/black): $\omega_t = 1/n$. Robust estimates $\hat{\theta}_n$ (solid/black). Bias-corrected robust estimates $\tilde{\theta}_n$ (solid/circle/blue).  Repeated bias-corrected robust estimates $\vardbtilde{\theta}_n$ (solid/triangle/purple). Shaded vertical bars = NBER recession dates.}
\end{figure}

\begin{figure}[H] \caption{Recursive VAR, Estimation Weights: OLS, Robust, and Bias-Corrected Estimates ($\nu=20$) } \label{fig:VARwbbb}
  \includegraphics[scale = 0.5]{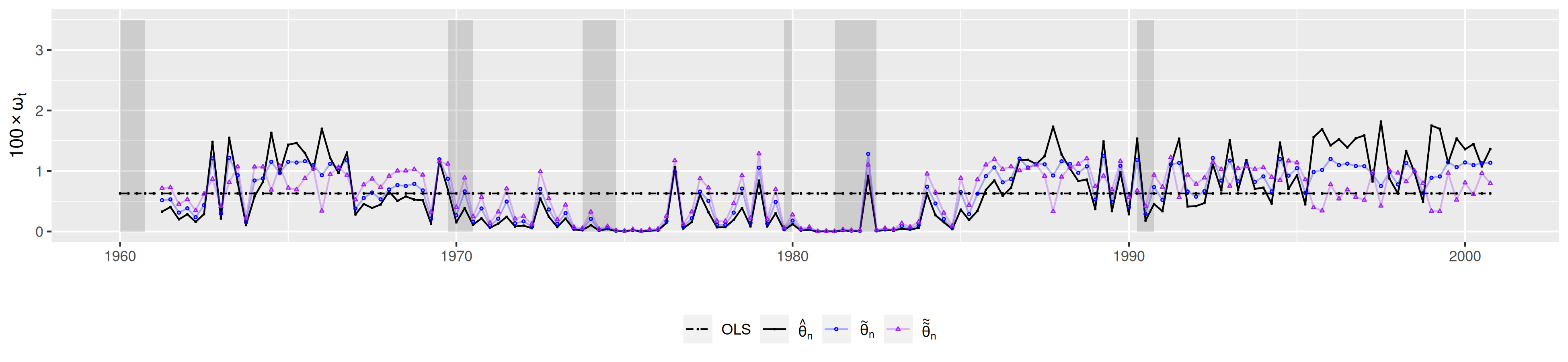}\\
  {\footnotesize \textbf{Note:} Estimation weights $\omega_t$ implicitely used to estimate $\theta$. OLS (dashed/black): $\omega_t = 1/n$. Robust estimates $\hat{\theta}_n$ (solid/black). Bias-corrected robust estimates $\tilde{\theta}_n$ (solid/circle/blue).  Repeated bias-corrected robust estimates $\vardbtilde{\theta}_n$ (solid/triangle/purple). Shaded vertical bars = NBER recession dates.}
\end{figure}

\begin{figure}[H] \caption{Recursive VAR, Estimation Weights: OLS, Robust, and Bias-Corrected Estimates ($\nu=10$, log scale) } \label{fig:VARwb_log}
  \includegraphics[scale = 0.5]{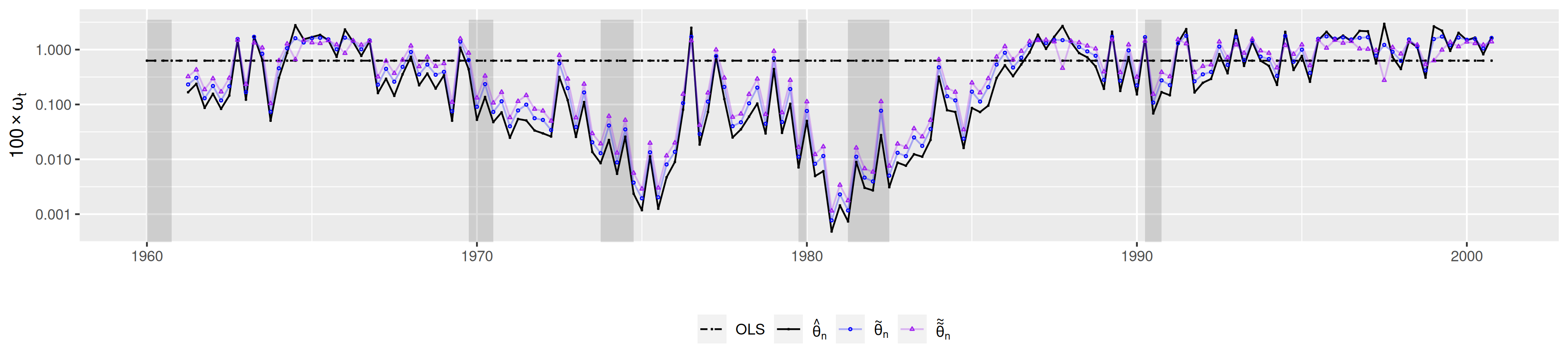}\\
  {\footnotesize \textbf{Note:} Estimation weights $\omega_t$ implicitely used to estimate $\theta$. OLS (dashed/black): $\omega_t = 1/n$. Robust estimates $\hat{\theta}_n$ (solid/black). Bias-corrected robust estimates $\tilde{\theta}_n$ (solid/circle/blue).  Repeated bias-corrected robust estimates $\vardbtilde{\theta}_n$ (solid/triangle/purple). Shaded vertical bars = NBER recession dates.}
\end{figure}

\begin{figure}[H] \caption{Recursive VAR, Estimation Weights: OLS, Robust, and Bias-Corrected Estimates ($\nu=15$, log scale) } \label{fig:VARwbb_log}
  \includegraphics[scale = 0.5]{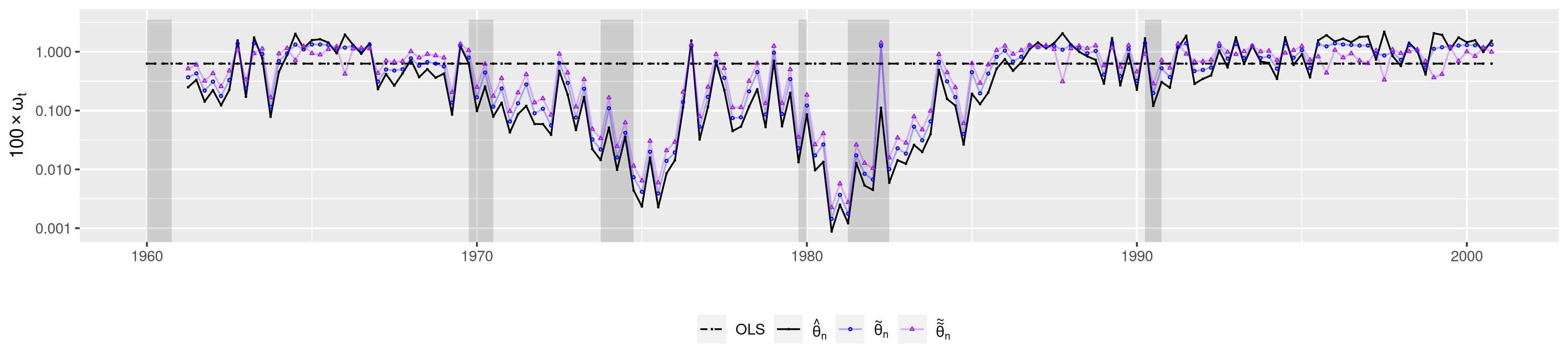}\\
  {\footnotesize \textbf{Note:} Estimation weights $\omega_t$ implicitely used to estimate $\theta$. OLS (dashed/black): $\omega_t = 1/n$. Robust estimates $\hat{\theta}_n$ (solid/black). Bias-corrected robust estimates $\tilde{\theta}_n$ (solid/circle/blue).  Repeated bias-corrected robust estimates $\vardbtilde{\theta}_n$ (solid/triangle/purple). Shaded vertical bars = NBER recession dates.}
\end{figure}

\begin{figure}[H] \caption{Recursive VAR, Estimation Weights: OLS, Robust, and Bias-Corrected Estimates ($\nu=20$, log scale) } \label{fig:VARwbbb_log}
  \includegraphics[scale = 0.5]{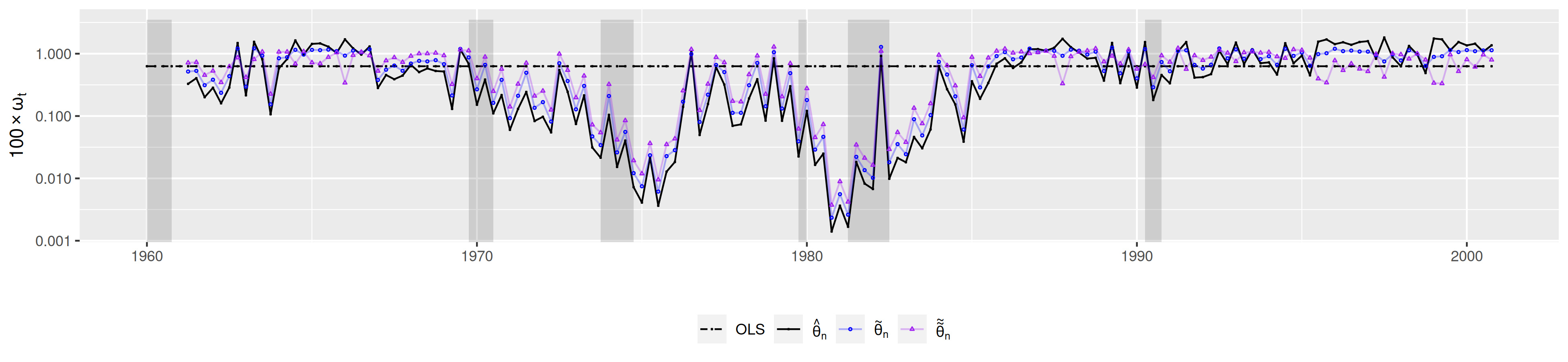}\\
  {\footnotesize \textbf{Note:} Estimation weights $\omega_t$ implicitely used to estimate $\theta$. OLS (dashed/black): $\omega_t = 1/n$. Robust estimates $\hat{\theta}_n$ (solid/black). Bias-corrected robust estimates $\tilde{\theta}_n$ (solid/circle/blue).  Repeated bias-corrected robust estimates $\vardbtilde{\theta}_n$ (solid/triangle/purple). Shaded vertical bars = NBER recession dates.}
\end{figure}

\subsection{ Additional Results for Inflation and Openness }
\begin{table}[H]
  \caption{Weights used in estimation ($y = \pi/100$) -- 1/2} \label{tab:Romer_w1}
  \centering
  \setlength\tabcolsep{4.5pt}
    \renewcommand{\arraystretch}{0.935} 
    {\small
    \begin{tabular}{l|caaa||l|caaa}
      \hline \hline
    Country & $\hat{\theta}_n^{\text{IV}}$ & \mc{1}{$\hat{\theta}_n$} & \mc{1}{$\tilde{\theta}_n$} & \mc{1}{$\vardbtilde{\theta}_n$} & Country & $\hat{\theta}_n^{\text{IV}}$ & \mc{1}{$\hat{\theta}_n$} & \mc{1}{$\tilde{\theta}_n$} & \mc{1}{$\vardbtilde{\theta}_n$} \\ 
      \hline
      Algeria & 0.88 & 1.04 & 0.95 & 0.78 & Ethiopia & 0.88 & 0.58 & 0.83 & 1.14 \\ 
      Argentina & 0.88 & 0.02 & 0.04 & 0.09 & Fiji & 0.88 & 1.05 & 0.90 & 0.81 \\ 
      Australia & 0.88 & 0.97 & 1.08 & 1.24 & Finland & 0.88 & 1.04 & 0.94 & 0.78 \\ 
      Austria & 0.88 & 0.87 & 1.02 & 1.09 & France & 0.88 & 1.01 & 1.01 & 0.92 \\ 
      Bahrain & 0.88 & 0.98 & 0.95 & 0.82 & Gabon & 0.88 & 1.06 & 0.89 & 0.86 \\ 
      Bangladesh & 0.88 & 1.05 & 0.92 & 0.81 & Gambia & 0.88 & 0.90 & 1.02 & 0.91 \\ 
      Barbados & 0.88 & 0.91 & 1.03 & 0.99 & Germany & 0.88 & 0.80 & 1.03 & 1.22 \\ 
      Belgium & 0.88 & 0.98 & 0.98 & 0.85 & Ghana & 0.88 & 0.25 & 0.45 & 0.82 \\ 
      Benin & 0.88 & 1.03 & 0.93 & 0.81 & Greece & 0.88 & 1.01 & 0.97 & 0.81 \\ 
      Bolivia & 0.88 & 0.01 & 0.02 & 0.05 & Guatemala & 0.88 & 1.06 & 0.89 & 0.83 \\ 
      Botswana & 0.88 & 1.06 & 0.89 & 0.85 & Guyana & 0.88 & 1.05 & 0.93 & 0.81 \\ 
      Brazil & 0.88 & 0.04 & 0.07 & 0.16 & Haiti & 0.88 & 0.75 & 0.95 & 1.13 \\ 
      Burkina Faso & 0.88 & 0.95 & 1.00 & 0.91 & Honduras & 0.88 & 0.96 & 0.97 & 0.86 \\ 
      Burma & 0.88 & 0.71 & 0.95 & 1.20 & Hong Kong & 0.88 & 1.04 & 0.94 & 0.82 \\ 
      Burundi & 0.88 & 0.70 & 0.91 & 1.14 & Iceland & 0.88 & 0.23 & 0.40 & 0.74 \\ 
      Cameroon & 0.88 & 1.03 & 0.94 & 0.80 & India & 0.88 & 0.79 & 1.03 & 1.31 \\ 
      Canada & 0.88 & 0.84 & 1.07 & 1.35 & Indonesia & 0.88 & 1.06 & 0.88 & 0.85 \\ 
      Central Afr. Rep. & 0.88 & 0.98 & 1.01 & 0.98 & Iran & 0.88 & 1.04 & 0.94 & 0.84 \\ 
      Chile & 0.88 & 0.16 & 0.30 & 0.59 & Ireland & 0.88 & 1.05 & 0.91 & 0.84 \\ 
      Colombia & 0.88 & 0.93 & 1.12 & 0.91 & Israel & 0.88 & 0.03 & 0.05 & 0.09 \\ 
      Congo & 0.88 & 1.06 & 0.88 & 0.87 & Italy & 0.88 & 1.05 & 0.90 & 0.86 \\ 
      Costa Rica & 0.88 & 0.77 & 1.00 & 1.10 & Ivory Coast & 0.88 & 1.06 & 0.89 & 0.84 \\ 
      Cyprus & 0.88 & 1.06 & 0.89 & 0.83 & Jamaica & 0.88 & 0.79 & 1.01 & 1.05 \\ 
      Denmark & 0.88 & 0.99 & 0.99 & 0.94 & Japan & 0.88 & 0.78 & 1.02 & 1.27 \\ 
      Dominican Republic & 0.88 & 1.06 & 0.89 & 0.81 & Jordan & 0.88 & 1.06 & 0.90 & 0.82 \\ 
      Ecuador & 0.88 & 0.96 & 1.05 & 0.87 & Kenya & 0.88 & 1.03 & 0.94 & 0.81 \\ 
      Egypt & 0.88 & 1.02 & 0.97 & 0.81 & Korea & 0.88 & 1.06 & 0.89 & 0.86 \\ 
      El Salvador & 0.88 & 1.06 & 0.88 & 0.85 & Kuwait & 0.88 & 1.06 & 0.91 & 0.81 \\
       \hline \hline
    \end{tabular}
     }\\
     {\footnotesize \textbf{Note:} $\hat{\theta}_n^{IV}$: IV estimates, $\hat{\theta}_n$: robust estimates, $\tilde{\theta}_n$: bias-corrected robust estimates, $\vardbtilde{\theta}_n$: repeated bias-corrected robust estimates. $\hat{\nu}_n = 12.62$. Estimates for $\theta_2$ reported using $\log(\text{pcinc})/100$ as a regressor. Sample size $n = 114.$ All weights were multiplied by $100$ for formatting.}
  \end{table}
  
  \begin{table}[H]
    \caption{Weights used in estimation ($y = \pi/100$) -- 2/2} \label{tab:Romer_w2}
    \centering
    \setlength\tabcolsep{4.5pt}
      \renewcommand{\arraystretch}{0.935} 
      {\small
      \begin{tabular}{l|caaa||l|caaa}
        \hline \hline
        Country & $\hat{\theta}_n^{\text{IV}}$ & \mc{1}{$\hat{\theta}_n$} & \mc{1}{$\tilde{\theta}_n$} & \mc{1}{$\vardbtilde{\theta}_n$} & Country & $\hat{\theta}_n^{\text{IV}}$ & \mc{1}{$\hat{\theta}_n$} & \mc{1}{$\tilde{\theta}_n$} & \mc{1}{$\vardbtilde{\theta}_n$} \\ 
        \hline
        Lesotho & 0.88 & 0.90 & 1.02 & 1.14 & Sierra Leone & 0.88 & 0.83 & 1.02 & 0.91 \\ 
        Liberia & 0.88 & 1.00 & 0.94 & 0.81 & Singapore & 0.88 & 0.86 & 0.93 & 1.02 \\ 
        Luxembourg & 0.88 & 1.02 & 0.93 & 0.80 & Somalia & 0.88 & 0.56 & 0.90 & 1.25 \\ 
        Madagascar & 0.88 & 1.06 & 0.89 & 0.82 & South Africa & 0.88 & 1.06 & 0.88 & 0.84 \\ 
        Malawi & 0.88 & 1.03 & 0.94 & 0.82 & Spain & 0.88 & 1.06 & 0.88 & 0.85 \\ 
        Malaysia & 0.88 & 1.00 & 0.97 & 0.77 & Sri Lanka & 0.88 & 1.06 & 0.88 & 0.86 \\ 
        Malta & 0.88 & 0.85 & 1.03 & 0.85 & Sudan & 0.88 & 0.85 & 1.10 & 0.94 \\ 
        Mauritania & 0.88 & 1.06 & 0.88 & 0.86 & Suriman & 0.88 & 1.06 & 0.88 & 0.87 \\ 
        Mauritius & 0.88 & 0.92 & 1.01 & 0.87 & Swaziland & 0.88 & 0.89 & 1.02 & 1.03 \\ 
        Mexico & 0.88 & 0.44 & 0.78 & 1.37 & Sweden & 0.88 & 1.03 & 0.96 & 0.80 \\ 
        Morocco & 0.88 & 1.02 & 0.94 & 0.79 & Switzerland & 0.88 & 0.74 & 0.96 & 1.18 \\ 
        Nepal & 0.88 & 0.90 & 1.00 & 1.03 & Syria & 0.88 & 1.06 & 0.89 & 0.87 \\ 
        Netherlands & 0.88 & 0.88 & 1.02 & 1.04 & Taiwan & 0.88 & 0.97 & 0.97 & 0.88 \\ 
        New Zealand & 0.88 & 1.06 & 0.88 & 0.83 & Tanzania & 0.88 & 1.04 & 0.90 & 0.86 \\ 
        Nicaragua & 0.88 & 0.34 & 0.55 & 0.92 & Thailand & 0.88 & 0.98 & 1.01 & 0.86 \\ 
        Niger & 0.88 & 1.06 & 0.88 & 0.85 & Togo & 0.88 & 1.00 & 0.95 & 0.81 \\ 
        Nigeria & 0.88 & 1.06 & 0.88 & 0.86 & Trinidad \& Tobago & 0.88 & 0.91 & 1.01 & 0.82 \\ 
        Norway & 0.88 & 1.03 & 0.96 & 0.79 & Tunisia & 0.88 & 1.03 & 0.91 & 0.81 \\ 
        Oman & 0.88 & 1.06 & 0.88 & 0.84 & Turkey & 0.88 & 0.66 & 1.04 & 1.47 \\ 
        Pakistan & 0.88 & 1.02 & 0.97 & 0.84 & Uganda & 0.88 & 0.17 & 0.32 & 0.63 \\ 
        Panama & 0.88 & 0.95 & 0.98 & 0.87 & U.A. Emirates & 0.88 & 1.06 & 0.89 & 0.79 \\ 
        Papua New Guinea & 0.88 & 1.03 & 0.92 & 0.79 & United Kingdom & 0.88 & 1.06 & 0.90 & 0.78 \\ 
        Paraguay & 0.88 & 1.05 & 0.91 & 0.85 & United States & 0.88 & 0.68 & 0.94 & 1.27 \\ 
        Peru & 0.88 & 0.26 & 0.48 & 0.90 & Uruguay & 0.88 & 0.28 & 0.48 & 0.86 \\ 
        Philippines & 0.88 & 1.06 & 0.88 & 0.87 & Venezuela & 0.88 & 1.06 & 0.90 & 0.86 \\ 
        Portugal & 0.88 & 0.89 & 1.04 & 0.96 & Yemen & 0.88 & 1.06 & 0.90 & 0.85 \\ 
        Rwanda & 0.88 & 0.85 & 0.99 & 1.13 & Zaire & 0.88 & 0.10 & 0.18 & 0.36 \\ 
        Saudi Arabia & 0.88 & 1.06 & 0.91 & 0.76 & Zambia & 0.88 & 0.97 & 1.02 & 0.83 \\ 
        Senegal & 0.88 & 1.06 & 0.89 & 0.84 & Zimbabwe & 0.88 & 1.04 & 0.92 & 0.80 \\
         \hline \hline
      \end{tabular}
       }\\
       {\footnotesize \textbf{Note:} $\hat{\theta}_n^{IV}$: IV estimates, $\hat{\theta}_n$: robust estimates, $\tilde{\theta}_n$: bias-corrected robust estimates, $\vardbtilde{\theta}_n$: repeated bias-corrected robust estimates. $\hat{\nu}_n = 12.62$. Estimates for $\theta_2$ reported using $\log(\text{pcinc})/100$ as a regressor. Sample size $n = 114.$ All weights were multiplied by $100$ for formatting.}
    \end{table}

\subsection{Additional Results for Segregation and Government Quality}

\begin{table}[H]
  \caption{Weights used in Estimation (Ethnicity)  } \label{tab:Alesina_w1}
  \centering
  \setlength\tabcolsep{4.5pt}
    \renewcommand{\arraystretch}{0.7} 
    {\small
  \begin{tabular}{l|caaa||l|caaa}
  \hline \hline
  Country & $\hat{\theta}_n^{\text{IV}}$ & \mc{1}{$\hat{\theta}_n$} & \mc{1}{$\tilde{\theta}_n$} & \mc{1}{$\vardbtilde{\theta}_n$} & Country & $\hat{\theta}_n^{\text{IV}}$ & \mc{1}{$\hat{\theta}_n$} & \mc{1}{$\tilde{\theta}_n$} & \mc{1}{$\vardbtilde{\theta}_n$} \\ 
  \hline
  Afghanistan & 1.03 & 0.22 & 0.59 & 2.1 & Kenya & 1.03 & 0.95 & 1.51 & 2.2 \\ 
  Argentina & 1.03 & 0.08 & 0.15 & 0.29 & Korea & 1.03 & 0.05 & 0.07 & 0.13 \\ 
  Armenia & 1.03 & 2.57 & 1.99 & 0.39 & Kyrgyzstan & 1.03 & 1.87 & 2.22 & 1.45 \\ 
  Australia & 1.03 & 0.93 & 0.97 & 1.29 & Latvia & 1.03 & 2.59 & 1.06 & 0.78 \\ 
  Austria & 1.03 & 0.23 & 2.55 & 1.62 & Lesotho & 1.03 & 0.11 & 0.17 & 0.23 \\ 
  Bahrain & 1.03 & 1.93 & 2.37 & 0.94 & Lithuania & 1.03 & 2.27 & 2.14 & 1.17 \\ 
  Bangladesh & 1.03 & 0.02 & 0.03 & 0.08 & Macedonia & 1.03 & 2.59 & 1.82 & 1.33 \\ 
  Belarus & 1.03 & 0.07 & 0.14 & 0.35 & Malawi & 1.03 & 2.59 & 1.69 & 0.63 \\ 
  Belgium & 1.03 & 0.06 & 0.08 & 0.17 & Mali & 1.03 & 0.6 & 1.12 & 1.58 \\ 
  Belize & 1.03 & 0.04 & 0.07 & 0.09 & Mexico & 1.03 & 2.59 & 1.04 & 0.58 \\ 
  Benin & 1.03 & 0.09 & 0.15 & 0.28 & Morocco & 1.03 & 0.08 & 0.13 & 0.23 \\ 
  Bolivia & 1.03 & 0.71 & 1.05 & 1.28 & Nepal & 1.03 & 2.1 & 1.89 & 0.44 \\ 
  Brazil & 1.03 & 0.28 & 0.52 & 1.07 & Netherlands & 1.03 & 2.58 & 1.97 & 0.38 \\ 
  Bulgaria & 1.03 & 0.15 & 0.22 & 0.35 & New Zealand & 1.03 & 0.29 & 0.47 & 1.33 \\ 
  Burkina\_faso & 1.03 & 0.04 & 0.19 & 0.3 & Niger & 1.03 & 1.98 & 2.21 & 1.86 \\ 
  Cambodia & 1.03 & 2.28 & 1.24 & 1.97 & Norway & 1.03 & 2.42 & 2.73 & 3.22 \\ 
  Cameroon & 1.03 & 0.17 & 0.3 & 0.57 & Pakistan & 1.03 & 2.48 & 1.51 & 1.79 \\ 
  Canada & 1.03 & 1.92 & 1.93 & 0.43 & Panama & 1.03 & 1.27 & 1.88 & 1.69 \\ 
  Central African Republic & 1.03 & 0.06 & 0.1 & 0.15 & Paraguay & 1.03 & 0.03 & 0.06 & 0.12 \\ 
  Chile & 1.03 & 0.1 & 0.17 & 0.26 & Peru & 1.03 & 2 & 2.26 & 1.93 \\ 
  China & 1.03 & 0.04 & 0.07 & 0.09 & Philippines & 1.03 & 2.56 & 1.17 & 0.95 \\ 
  Colombia & 1.03 & 1.09 & 1.59 & 1.27 & Portugal & 1.03 & 2.55 & 1.84 & 1.08 \\ 
  Costa Rica & 1.03 & 1.85 & 2.19 & 1.73 & Qatar & 1.03 & 0.05 & 0.09 & 0.29 \\ 
  Cote d'Ivoire & 1.03 & 1.55 & 2.22 & 0.92 & Romania & 1.03 & 2.56 & 1.86 & 1.45 \\ 
  Croatia & 1.03 & 2.59 & 1.15 & 0.96 & Russia & 1.03 & 0.59 & 1.16 & 1.47 \\ 
  Czech Republic & 1.03 & 1.19 & 1.57 & 1.61 & Rwanda & 1.03 & 1.18 & 1.67 & 1.8 \\ 
  Denmark & 1.03 & 0.94 & 2.37 & 2.85 & Saudi Arabia & 1.03 & 0.06 & 0.12 & 0.36 \\ 
  Ecuador & 1.03 & 0.25 & 0.48 & 1.63 & Senegal & 1.03 & 1.3 & 2.15 & 1.85 \\ 
  Estonia & 1.03 & 2.59 & 2.23 & 1.98 & Slovakia & 1.03 & 2.53 & 1.58 & 1.62 \\ 
  Ethiopia & 1.03 & 0.05 & 0.08 & 0.14 & Slovenia & 1.03 & 0.92 & 1.32 & 1.9 \\ 
  Finland & 1.03 & 0.11 & 0.05 & 0.26 & South Africa & 1.03 & 0.31 & 0.68 & 0.62 \\ 
  France & 1.03 & 0.76 & 1.11 & 1.41 & Spain & 1.03 & 0.08 & 0.14 & 0.3 \\ 
  Gabon & 1.03 & 0.05 & 0.1 & 0.22 & Sri Lanka & 1.03 & 0.37 & 0.86 & 1.69 \\ 
  Germany & 1.03 & 0.19 & 0.13 & 0.18 & Sweden & 1.03 & 0.02 & 0.01 & 0.07 \\ 
  Ghana & 1.03 & 2.16 & 2.19 & 1.38 & Switzerland & 1.03 & 0.04 & 0.24 & 1.44 \\ 
  Greece & 1.03 & 0.1 & 0.27 & 0.5 & Taiwan & 1.03 & 0.07 & 0.09 & 0.12 \\ 
  Guatemala & 1.03 & 0.34 & 0.65 & 1.75 & Tajikistan & 1.03 & 0.19 & 0.28 & 0.48 \\ 
  Guinea & 1.03 & 2.41 & 1.75 & 0.86 & Tanzania & 1.03 & 0.06 & 0.13 & 0.23 \\ 
  Honduras & 1.03 & 2.22 & 1.34 & 1.4 & Togo & 1.03 & 2.33 & 2.2 & 1.09 \\ 
  Hungary & 1.03 & 0.36 & 0.63 & 0.79 & Turkey & 1.03 & 0.06 & 0.13 & 0.72 \\ 
  Iceland & 1.03 & 0.01 & 0.01 & 0.06 & Uganda & 1.03 & 0.01 & 0.03 & 0.05 \\ 
  India & 1.03 & 0.41 & 0.56 & 0.93 & Ukraine & 1.03 & 0.2 & 0.42 & 0.92 \\ 
  Indonesia & 1.03 & 2.5 & 1.03 & 0.87 & United Kingdom & 1.03 & 1.82 & 1.62 & 1.72 \\ 
  Ireland & 1.03 & 0.85 & 1.64 & 2.16 & USA & 1.03 & 0.79 & 1.65 & 1.91 \\ 
  Israel & 1.03 & 0.34 & 1.69 & 1.84 & Uzbekistan & 1.03 & 0.16 & 0.28 & 0.8 \\ 
  Italy & 1.03 & 0.52 & 1.01 & 1.6 & Vietnam & 1.03 & 0.11 & 0.17 & 0.19 \\ 
  Japan & 1.03 & 2.59 & 1.29 & 2.54 & Zambia & 1.03 & 2.52 & 2.1 & 1.18 \\ 
  Jordan & 1.03 & 2.45 & 1.29 & 1.92 & Zimbabwe & 1.03 & 2.45 & 2.25 & 2.01 \\ 
  Kazakhstan & 1.03 & 0.18 & 0.32 & 0.79 &  & \mc{1}{} & \mc{1}{} & \mc{1}{} & \mc{1}{} \\ 
     \hline \hline
  \end{tabular} }
  \end{table}
  
  \begin{table}[H]
    \caption{Weights used in Estimation (Language)  } \label{tab:Alesina_w2}
    \centering
    \setlength\tabcolsep{4.5pt}
      \renewcommand{\arraystretch}{0.7} 
      {\small
    \begin{tabular}{l|caaa||l|caaa}
    \hline \hline
    Country & $\hat{\theta}_n^{\text{IV}}$ & \mc{1}{$\hat{\theta}_n$} & \mc{1}{$\tilde{\theta}_n$} & \mc{1}{$\vardbtilde{\theta}_n$} & Country & $\hat{\theta}_n^{\text{IV}}$ & \mc{1}{$\hat{\theta}_n$} & \mc{1}{$\tilde{\theta}_n$} & \mc{1}{$\vardbtilde{\theta}_n$} \\ 
    \hline
    Afghanistan & 1.09 & 2.5 & 2.66 & 1.32 & Lesotho & 1.09 & 2.94 & 2.41 & 0.95 \\ 
    Armenia & 1.09 & 0.26 & 0.4 & 0.52 & Lithuania & 1.09 & 0.03 & 0.07 & 0.22 \\ 
    Australia & 1.09 & 2.28 & 2.58 & 2.14 & Macedonia & 1.09 & 3.1 & 1.38 & 1.45 \\ 
    Austria & 1.09 & 0.01 & 0.02 & 0.62 & Malawi & 1.09 & 1.37 & 1.66 & 2.15 \\ 
    Bangladesh & 1.09 & 0.01 & 0.01 & 0.04 & Mali & 1.09 & 0.12 & 0.23 & 0.45 \\ 
    Belarus & 1.09 & 0.01 & 0.02 & 0.04 & Mauritius & 1.09 & 1.83 & 2.4 & 2.34 \\ 
    Belgium & 1.09 & 2.98 & 1.61 & 0.43 & Mexico & 1.09 & 1.7 & 2.53 & 2.74 \\ 
    Belize & 1.09 & 0.01 & 0.03 & 0.07 & Morocco & 1.09 & 0.01 & 0.02 & 0.06 \\ 
    Benin & 1.09 & 0.05 & 0.11 & 0.26 & Mozambique & 1.09 & 0.12 & 0.24 & 0.55 \\ 
    Bolivia & 1.09 & 0.24 & 0.57 & 1.65 & Namibia & 1.09 & 0.04 & 0.09 & 0.2 \\ 
    Brazil & 1.09 & 0.12 & 0.26 & 0.91 & Nepal & 1.09 & 0.23 & 0.4 & 0.49 \\ 
    Bulgaria & 1.09 & 0.79 & 1.65 & 2.51 & New Zealand & 1.09 & 0.05 & 0.11 & 0.29 \\ 
    Burkina Faso & 1.09 & 0.07 & 0.1 & 4.06 & Nicaragua & 1.09 & 0.54 & 1.14 & 2.76 \\ 
    Cambodia & 1.09 & 2.62 & 2.63 & 0.95 & Niger & 1.09 & 3.12 & 1.85 & 2.65 \\ 
    Cameroon & 1.09 & 1.25 & 2.19 & 2.76 & Nigeria & 1.09 & 0.01 & 0.03 & 0.08 \\ 
    Canada & 1.09 & 1.63 & 2.55 & 2.43 & Norway & 1.09 & 0.02 & 0.05 & 2.95 \\ 
    Central African Republic & 1.09 & 0.04 & 0.09 & 0.2 & Pakistan & 1.09 & 0.02 & 0.04 & 0.1 \\ 
    Chile & 1.09 & 0.01 & 0.03 & 0.06 & Panama & 1.09 & 0.47 & 0.77 & 1.2 \\ 
    China & 1.09 & 0.03 & 0.08 & 0.22 & Paraguay & 1.09 & 0.06 & 0.12 & 0.35 \\ 
    Colombia & 1.09 & 0.06 & 0.12 & 0.33 & Peru & 1.09 & 3.12 & 1.97 & 1.88 \\ 
    Costa Rica & 1.09 & 3.12 & 1.73 & 2.34 & Philippines & 1.09 & 2.34 & 2.53 & 2.62 \\ 
    Cote d'Ivoire & 1.09 & 3.1 & 1.19 & 0.42 & Portugal & 1.09 & 1.81 & 2.53 & 2.67 \\ 
    Croatia & 1.09 & 0.61 & 1.13 & 1.39 & Romania & 1.09 & 2.9 & 2.58 & 2.17 \\ 
    Czech Republic & 1.09 & 1.49 & 2.47 & 2.6 & Russia & 1.09 & 0.01 & 0.02 & 0.05 \\ 
    Denmark & 1.09 & 0.07 & 0.16 & 3.39 & Rwanda & 1.09 & 0.03 & 0.05 & 0.12 \\ 
    Ecuador & 1.09 & 0.25 & 0.56 & 1.78 & Saudi Arabia & 1.09 & 2.57 & 2.6 & 1.65 \\ 
    Estonia & 1.09 & 2.77 & 2.54 & 0.54 & Senegal & 1.09 & 0.07 & 0.14 & 0.27 \\ 
    Ethiopia & 1.09 & 0.02 & 0.04 & 0.1 & Slovakia & 1.09 & 1.83 & 2.54 & 1.45 \\ 
    Finland & 1.09 & 0.25 & 0.51 & 0.11 & Slovenia & 1.09 & 3.11 & 1.71 & 1.56 \\ 
    Gabon & 1.09 & 0.35 & 0.82 & 1.46 & South Africa & 1.09 & 0.01 & 0.02 & 0.03 \\ 
    Ghana & 1.09 & 2.34 & 2.61 & 2.73 & Spain & 1.09 & 0.07 & 0.14 & 0.3 \\ 
    Guatemala & 1.09 & 3.01 & 2.12 & 0.62 & Sweden & 1.09 & 0.01 & 0.02 & 0.13 \\ 
    Guinea & 1.09 & 0.88 & 1.57 & 2.27 & Switzerland & 1.09 & 0 & 0.01 & 0.06 \\ 
    Haiti & 1.09 & 0.02 & 0.04 & 0.1 & Tajikistan & 1.09 & 0.04 & 0.08 & 0.23 \\ 
    Honduras & 1.09 & 2.01 & 2.7 & 0.27 & Tanzania & 1.09 & 0.01 & 0.02 & 0.05 \\ 
    Hungary & 1.09 & 0.22 & 0.43 & 0.67 & Thailand & 1.09 & 0.02 & 0.03 & 0.07 \\ 
    Iceland & 1.09 & 3.13 & 3.81 & 0.16 & Togo & 1.09 & 2.99 & 1.91 & 0.89 \\ 
    India & 1.09 & 2.8 & 2.62 & 2.13 & Turkey & 1.09 & 3.07 & 1.12 & 0.68 \\ 
    Indonesia & 1.09 & 3.08 & 1.65 & 1.76 & Uganda & 1.09 & 0 & 0.01 & 0.02 \\ 
    Italy & 1.09 & 0.38 & 0.68 & 1.28 & Ukraine & 1.09 & 0.02 & 0.04 & 0.09 \\ 
    Japan & 1.09 & 3 & 3.8 & 0.09 & United Kingdom & 1.09 & 2.44 & 2.48 & 2.49 \\ 
    Kazakhstan & 1.09 & 0.06 & 0.1 & 0.24 & USA & 1.09 & 1.75 & 2.72 & 1.14 \\ 
    Kenya & 1.09 & 0.26 & 0.61 & 1.52 & Uzbekistan & 1.09 & 0.02 & 0.04 & 0.12 \\ 
    Korea & 1.09 & 0 & 0.01 & 0.01 & Vietnam & 1.09 & 0.06 & 0.14 & 0.35 \\ 
    Kyrgyzstan & 1.09 & 3.01 & 2.6 & 2.74 & Zambia & 1.09 & 3.12 & 2.36 & 2.67 \\ 
    Latvia & 1.09 & 1.56 & 2.42 & 1.96 & Zimbabwe & 1.09 & 0.01 & 0.01 & 0.02 \\   
       \hline \hline
    \end{tabular} }
    \end{table}

    \begin{table}[H]
      \caption{Weights used in Estimation (Religion)  } \label{tab:Alesina_w3}
      \centering
      \setlength\tabcolsep{4.5pt}
        \renewcommand{\arraystretch}{0.7} 
        {\small
      \begin{tabular}{l|caaa||l|caaa}
      \hline \hline
      Country & $\hat{\theta}_n^{\text{IV}}$ & \mc{1}{$\hat{\theta}_n$} & \mc{1}{$\tilde{\theta}_n$} & \mc{1}{$\vardbtilde{\theta}_n$} & Country & $\hat{\theta}_n^{\text{IV}}$ & \mc{1}{$\hat{\theta}_n$} & \mc{1}{$\tilde{\theta}_n$} & \mc{1}{$\vardbtilde{\theta}_n$} \\ 
      \hline
Armenia & 1.09 & 2.5 & 2.66 & 1.32 & Malawi & 1.09 & 0.38 & 0.68 & 1.28 \\ 
Australia & 1.09 & 0.26 & 0.4 & 0.52 & Mali & 1.09 & 3 & 3.8 & 0.09 \\ 
Austria & 1.09 & 2.28 & 2.58 & 2.14 & Mauritius & 1.09 & 0.06 & 0.1 & 0.24 \\ 
Bangladesh & 1.09 & 0.01 & 0.02 & 0.62 & Mexico & 1.09 & 0.26 & 0.61 & 1.52 \\ 
Belize & 1.09 & 0.01 & 0.01 & 0.04 & Mozambique & 1.09 & 0 & 0.01 & 0.01 \\ 
Benin & 1.09 & 0.01 & 0.02 & 0.04 & Namibia & 1.09 & 3.01 & 2.6 & 2.74 \\ 
Brazil & 1.09 & 2.98 & 1.61 & 0.43 & Nepal & 1.09 & 1.56 & 2.42 & 1.96 \\ 
Bulgaria & 1.09 & 0.01 & 0.03 & 0.07 & Netherlands & 1.09 & 2.94 & 2.41 & 0.95 \\ 
Burkina Faso & 1.09 & 0.05 & 0.11 & 0.26 & New Zealand & 1.09 & 0.03 & 0.07 & 0.22 \\ 
Cambodia & 1.09 & 0.24 & 0.57 & 1.65 & Nicaragua & 1.09 & 3.1 & 1.38 & 1.45 \\ 
Cameroon & 1.09 & 0.12 & 0.26 & 0.91 & Niger & 1.09 & 1.37 & 1.66 & 2.15 \\ 
Canada & 1.09 & 0.79 & 1.65 & 2.51 & Nigeria & 1.09 & 0.12 & 0.23 & 0.45 \\ 
Central African Republic & 1.09 & 0.07 & 0.1 & 4.06 & Pakistan & 1.09 & 1.83 & 2.4 & 2.34 \\ 
Chile & 1.09 & 2.62 & 2.63 & 0.95 & Paraguay & 1.09 & 1.7 & 2.53 & 2.74 \\ 
Cote d'Ivoire & 1.09 & 1.25 & 2.19 & 2.76 & Peru & 1.09 & 0.01 & 0.02 & 0.06 \\ 
Croatia & 1.09 & 1.63 & 2.55 & 2.43 & Philippines & 1.09 & 0.12 & 0.24 & 0.55 \\ 
Czech Republic & 1.09 & 0.04 & 0.09 & 0.2 & Portugal & 1.09 & 0.04 & 0.09 & 0.2 \\ 
Dominican Republic & 1.09 & 0.01 & 0.03 & 0.06 & qatar & 1.09 & 0.23 & 0.4 & 0.49 \\ 
Egypt & 1.09 & 0.03 & 0.08 & 0.22 & Romania & 1.09 & 0.05 & 0.11 & 0.29 \\ 
Estonia & 1.09 & 0.06 & 0.12 & 0.33 & Russia & 1.09 & 0.54 & 1.14 & 2.76 \\ 
Ethiopia & 1.09 & 3.12 & 1.73 & 2.34 & Rwanda & 1.09 & 3.12 & 1.85 & 2.65 \\ 
Gabon & 1.09 & 3.1 & 1.19 & 0.42 & Sao Tome & 1.09 & 0.01 & 0.03 & 0.08 \\ 
Ghana & 1.09 & 0.61 & 1.13 & 1.39 & Senegal & 1.09 & 0.02 & 0.05 & 2.95 \\ 
Guatemala & 1.09 & 1.49 & 2.47 & 2.6 & Slovakia & 1.09 & 0.02 & 0.04 & 0.1 \\ 
Guinea & 1.09 & 0.07 & 0.16 & 3.39 & Slovenia & 1.09 & 0.47 & 0.77 & 1.2 \\ 
Haiti & 1.09 & 0.25 & 0.56 & 1.78 & South Africa & 1.09 & 0.06 & 0.12 & 0.35 \\ 
Hungary & 1.09 & 2.77 & 2.54 & 0.54 & Sri Lanka & 1.09 & 3.12 & 1.97 & 1.88 \\ 
India & 1.09 & 0.02 & 0.04 & 0.1 & Switzerland & 1.09 & 2.34 & 2.53 & 2.62 \\ 
Indonesia & 1.09 & 0.25 & 0.51 & 0.11 & Tanzania & 1.09 & 1.81 & 2.53 & 2.67 \\ 
Iran & 1.09 & 0.35 & 0.82 & 1.46 & Thailand & 1.09 & 2.9 & 2.58 & 2.17 \\ 
Ireland & 1.09 & 2.34 & 2.61 & 2.73 & Togo & 1.09 & 0.01 & 0.02 & 0.05 \\ 
Israel & 1.09 & 3.01 & 2.12 & 0.62 & Turkey & 1.09 & 0.03 & 0.05 & 0.12 \\ 
Japan & 1.09 & 0.88 & 1.57 & 2.27 & Uganda & 1.09 & 2.57 & 2.6 & 1.65 \\ 
Kazakhstan & 1.09 & 0.02 & 0.04 & 0.1 & United Kingdom & 1.09 & 0.07 & 0.14 & 0.27 \\ 
Kenya & 1.09 & 2.01 & 2.7 & 0.27 & USA & 1.09 & 1.83 & 2.54 & 1.45 \\ 
Korea & 1.09 & 0.22 & 0.43 & 0.67 & Uzbekistan & 1.09 & 3.11 & 1.71 & 1.56 \\ 
Kyrgyzstan & 1.09 & 3.13 & 3.81 & 0.16 & Vietnam & 1.09 & 0.01 & 0.02 & 0.03 \\ 
Lithuania & 1.09 & 2.8 & 2.62 & 2.13 & Zambia & 1.09 & 0.07 & 0.14 & 0.3 \\ 
Madagascar & 1.09 & 3.08 & 1.65 & 1.76 & Zimbabwe & 1.09 & 0.01 & 0.02 & 0.13 \\  
   \hline \hline
      \end{tabular} }
      \end{table}

\section{Algorithms for computing $\hat{\psi}_n(\theta;\nu)$, $\hat{\theta}_n$, $\tilde{\theta}_n$} \label{apx:algos}
The following describes the algorithm used to compute $\hat{\psi}_n$ in the simulated and empirical examples. Algorithm \ref{algo:minQn} relies on explicit gradient calculations with respect to $\mu$ and $\Sigma$. The updates preserve symmetry and positive definiteness for $\Sigma$ which makes the iterations more stable than a direct implementation of gradient-descent for instance. A line search is used to update $\psi_b \to \psi_{b+1}$, in practice searching over $\gamma \in \{0.1,1\}$ provides good results more quickly. The initial $\mu_0 = 0$ is chosen specifically because $\hat{\mu}_n(\hat{\theta}_n;\nu) = 0$ is eventually the solution so that Algorithm \ref{algo:minQn} tends to speed up as $\theta$ gets closer to $\hat{\theta}_n$.  

\begin{algorithm}[h] 
  \caption{Computing $\hat{\psi}_n(\theta;\nu)$} \label{algo:minQn} 
        \begin{algorithmic}
          \State 1) \textbf{Inputs} (a) $\kappa_1,\kappa_2 > 0$, $\nu \geq 1$ (b) $\text{tol} > 0$, $\text{maxit} \geq 1$, (c) $\mu_0 = 0$, $\Sigma_0 = I_d$. 
        \State 2) \textbf{Iterations} 
        \State set $b=0$, $\psi_0 = (\mu_0,\Sigma_0)$
        \Repeat
              \State compute $\delta_t = \|g(z_t;\theta)-\mu_b\|_{\Sigma_b^{-1}}^{2}$, $w_t = (1+p/\nu)(1+\delta_t/\nu)$,
              \State normalize $w_t = \frac{w_t}{ \kappa_1/\nu + \sum_t w_t }$, compute $\bar{\mu}_{b+1} = \sum_t w_t g(z_t;\theta)$
              \State compute $\bar{S} = (I_d + \kappa_2 \Sigma_b/\nu)^{-1}$, center $\bar{x}_t = g(z_t;\theta)-\mu$ 
              \State compute $\bar{\Sigma}_{b+1} = \bar{S}\left( \sum_t w_t \bar{x}_t\bar{x}_t^\prime  + \kappa_1 \mu \mu^\prime / \nu\right)\bar{S}$
              \State minimize $Q_n( \gamma \psi_b + (1-\gamma)\bar{\psi}_{b+1};\nu )$ over $\gamma \in [0,1)$, $\bar{\psi}_{b+1} = (\bar{\mu}_{b+1},\bar{\Sigma}_{b+1})$
              \State compute $\psi_{b+1} = \gamma^\star \psi_b + (1-\gamma^\star)\bar{\psi}_{b+1}$, $\gamma^\star$ is the arg-minimizer of $Q_n$ above
              \State increment $b := b+1$
        \Until{ $|Q_n(\psi_{b}) - Q_n(\psi_{b+1})| < \text{tol}$, or $b > \text{maxit}$ }
        \State 3) \textbf{Output} estimates $\hat{\psi}_n(\theta;\nu) = \psi_{b+1}$, weights $w_t$
        \end{algorithmic}
\end{algorithm}

Algorithm \ref{algo:minGMM} describes more specifically the steps used to minimize $\|\tilde{\mu}_n(\theta)\|_{W_n}^2$. It is a Gauss-Newton algorithm where the Jacobian is approximated using the weighted average representation rather than a more costly computation based on the implicit function Theorem. For OLS, $\tilde{G}_n(\theta) = - \sum_{t} \tilde{w}_t(\theta;\nu) x_t x_t^\prime$, and IV $\tilde{G}_n(\theta) = - \sum_{t} \tilde{w}_t(\theta;\nu) z_t x_t^\prime$. Although the Jacobian $\tilde{G}_n(\theta)$ is inexact, the Gauss-Newton algorithm performed well in the simulated and empirical applications. The Algorithm is essentially the same when computing $\hat{\theta}_n$ or $\vardbtilde{\theta}_n$.

\begin{algorithm}[h] 
  \caption{Computing $\tilde{\theta}_n$} \label{algo:minGMM} 
        \begin{algorithmic}
          \State 1) \textbf{Inputs} (a) $\kappa_1,\kappa_2 > 0$, $\nu \geq 1$ (b) $\text{tol} > 0$, $\text{maxit} \geq 1$, $\gamma \in (0,1)$ (c) inital guess $\theta_0$. 
        \State 2) \textbf{Iterations} 
        \State set $b=0$,
        \Repeat
              \State compute $\hat{\psi}_n(\theta_b;\nu), \hat{\psi}_n(\theta_b;\nu/2)$
              \State compute $\tilde{\mu}_n(\theta) = 2\hat{\mu}_n(\theta;\nu) - \hat{\mu}_n(\theta;\nu/2)$ and $\tilde{w}_t(\theta;\nu) = 2 w_t(\theta;\nu) - w_t(\theta;\nu/2)$
              \State compute $\tilde{G}_n(\theta) = \sum_{t} \tilde{w}_t(\theta;\nu) \partial_\theta g(z_t;\theta)$ 
              \State update $\theta_{b+1} = \theta_b - \gamma \left( \tilde{G}_n(\theta)^\prime W_n \tilde{G}_n(\theta) \right)^{-1} \tilde{G}_n(\theta)^\prime W_n \tilde{\mu}_n(\theta_b)$
              \State increment $b := b+1$
        \Until{ $\|\tilde{\mu}_n(\theta_{b+1})\|_{W_n} < \text{tol}$, or $b > \text{maxit}$ }
        \State 3) \textbf{Output} estimates $\tilde{\theta}_n = \theta_{b+1}$, weights $\tilde{w}_t$
        \end{algorithmic}
\end{algorithm}

\end{appendices}
\end{document}